\documentclass[longauth]{aa}

\usepackage{txfonts}
\usepackage{graphicx}
\usepackage{natbib}
\bibpunct{(}{)}{;}{a}{}{,}

\begin{document}

\title{The dominant role of mergers in the size evolution of massive early-type galaxies since $z \sim 1$}

\author{C. L\'opez-Sanjuan\inst{1,2}\fnmsep
\thanks{Based on observations made at the European Southern Observatory
(ESO) Very Large Telescope (VLT) under Large Program 175.A-0839.}
 \and O.~Le F\`evre  \inst{1}
 \and O.~Ilbert \inst{1}
 \and L.~A.~M. Tasca\inst{1}
 \and C.~Bridge \inst{3}
 \and O.~Cucciati \inst{4}
 \and P.~Kampczyk  \inst{5}
 \and L.~Pozzetti \inst{6}
 \and C.~K. Xu \inst{7}
 \and C. M. Carollo \inst{8}
 \and T.~Contini \inst{8,9}
 \and J.~-P. Kneib \inst{1}
 \and S.~J. Lilly  \inst{5}
 \and V.~Mainieri \inst{10}
 \and A.~Renzini \inst{11}
 \and D.~Sanders \inst{12}
 \and M.~Scodeggio \inst{13}
 \and N.~Z. Scoville \inst{3}
 \and Y. Taniguchi \inst{14}
 \and G.~Zamorani \inst{6}
 \and H.~Aussel \inst{15}
 \and S.~Bardelli \inst{6}
 \and M.~Bolzonella \inst{6}
 \and A.~Bongiorno \inst{16}
 \and P.~Capak \inst{3}
 \and K.~Caputi \inst{17}
 \and S.~de la Torre \inst{18}
 \and L.~de Ravel  \inst{18}
 \and P.~Franzetti \inst{13}
 \and B.~Garilli \inst{13}
 \and A.~Iovino \inst{19}
 \and C.~Knobel \inst{5}
 \and K.~Kova\v{c} \inst{5,20}
 \and F.~Lamareille \inst{8,9}
 \and J.~-F.~Le Borgne \inst{8,9}
 \and V.~Le Brun \inst{1}
 \and E.~Le Floc'h \inst{12,15}
 \and C.~Maier \inst{5,21}
 \and H.~J. McCracken \inst{22}
 \and M.~Mignoli \inst{6}
 \and R.~Pell\'o \inst{8}
 \and Y.~Peng \inst{5}
 \and E.~P\'erez-Montero \inst{8,9,23}
 \and V.~Presotto \inst{24,18}
 \and E.~Ricciardelli \inst{25,26}
 \and M.~Salvato  \inst{3}
 \and J.~D. Silverman \inst{27}
 \and M.~Tanaka \inst{27}
 \and L.~Tresse  \inst{1}
 \and D.~Vergani \inst{6,28}
 \and E.~Zucca \inst{6}
 \and L.~Barnes \inst{5}
 \and R.~Bordoloi \inst{5}
 \and A.~Cappi \inst{6}
 \and A.~Cimatti \inst{29}
 \and G.~Coppa \inst{16,6}
 \and A.~Koekemoer \inst{30}
 \and C.~T. Liu \inst{31}
 \and M.~Moresco \inst{29}
 \and P.~Nair \inst{30,6}
 \and P.~Oesch \inst{5,32}
 \and K.~Schawinski \inst{33,34}
 \and N.~Welikala \inst{35}
 }

\institute{Laboratoire d'Astrophysique de Marseille - LAM, Universit\'e d'Aix-Marseille \& CNRS, UMR7326, 38 rue F. Joliot-Curie, 13388 Marseille Cedex 13, France
\and
Centro de Estudios de F\'{\i}sica del Cosmos de Arag\'on, Plaza San Juan 1, planta 2, E-44001, Teruel, Spain\\ 
\email{clsj@cefca.es}
\and 
 California Institute of Technology, MC 105-24, 1200 East California Boulevard, Pasadena, CA 91125 USA 
\and
 INAF Osservatorio Astronomico di Trieste, Via Tiepolo, 11, I-34143 Trieste, Italy 
\and
 Institute of Astronomy, ETH Zurich, CH-8093, Zurich, Switzerland 
\and
 INAF Osservatorio Astronomico di Bologna, via Ranzani 1, I-40127, Bologna, Italy 
\and
 Infrared Processing and Analysis Center, California Institute of Technology 100-22, Pasadena, CA 91125, USA 
\and
Institut de Recherche en Astrophysique et Plan\'etologie (IRAP), CNRS, 14, avenue Edouard Belin, F-31400 Toulouse, France 
 \and
IRAP, Universit\'e de Toulouse, UPS-OMP, Toulouse, France 
\and
 European Southern Observatory, Karl-Schwarzschild-Strasse 2, Garching, D-85748, Germany 
\and
  Dipartimento di Astronomia, Universit\'a di Padova, vicolo Osservatorio 3, I-35122 Padova, Italy 
\and
 Institute for Astronomy, 2680 Woodlawn Drive, University of Hawaii, Honolulu, HI 96822, USA 
\and
 INAF-IASF, Via Bassini 15, I-20133, Milano, Italy 
\and
 Research Center for Space and Cosmic Evolution, Ehime University, Bunkyo-cho 2-5, Matsuyama 790-8577, Japan 
\and
 CNRS, AIM-Unite Mixte de Recherche CEA-CNRS-Universit\'e Paris VII-UMR 7158, F-91191 Gif-sur-Yvette, France 
\and
 Max-Planck-Institut f\"{u}r Extraterrestrische Physik, D-84571 Garching b. Muenchen, Germany 
 \and
 Kapteyn Astronomical Institute, University of Groningen, P.O. Box 800, 9700
AV Groningen, The Netherlands 
 \and
 SUPA, Institute for Astronomy, University of Edinburgh, Royal Observatory, Edinburgh EH9 3HJ 
\and
 INAF Osservatorio Astronomico di Brera, Via Brera 28, I-20121 Milano, Italy 
\and
MPA - Max Planck Institut f\"{u}r Astrophysik, Karl-Schwarzschild-Str. 1, 85741 Garching, Germany 
\and
University of Vienna, Department of Astronomy, Tuerkenschanzstrasse 17, 1180 Vienna, Austria 
 \and
 Institut d'Astrophysique de Paris, UMR 7095 CNRS, Universit\'e Pierre et Marie Curie, 98 bis Boulevard Arago, F-75014 Paris, France 
 \and
  Instituto de Astrof\'{\i}sica de Andaluc\'{\i}a, CSIC, Apdo. 3004, 18080, Granada, Spain 
\and
 Universit\'a degli Studi dell’Insubria, Via Valleggio 11, 22100 Como, Italy 
\and
Instituto de Astrof\'{\i}sica de Canarias, v\'{\i}a Lactea s/n, 38200 La Laguna, Tenerife, Spain 
\and
Departamento de Astrof\'{\i}sica, Universidad de La Laguna, 38205 Tenerife, Spain 
\and
IPMU, Institute for the Physics and Mathematics of the Universe, 5-1-5 Kashiwanoha, Kashiwa, 277-8583, Japan 
\and
INAF - IASF Bologna, Via P. Gobetti 101, I-40129 Bologna, Italy 
\and
 Dipartimento di Astronomia, Universit\'a di Bologna, via Ranzani 1, I-40127, Bologna, Italy 
\and
 Space Telescope Science Institute, 3700 San Martin Drive, Baltimore, MD 21218 
\and
 Astrophysical Observatory, City University of New York, College of Staten Island, 2800 Victory Blvd, Staten Island, NY 10314, USA 
\and
UCO/Lick Observatory, Department of Astronomy and Astrophysics, University of California, Santa Cruz, CA 95064 
\and
 Department of Physics, Yale University, New Haven, CT 06511, USA 
 \and
 Yale Center for Astronomy and Astrophysics, Yale University, PO Box 208121, New Haven, CT 06520, USA 
\and
 Institut d'Astrophysique Spatiale, Batiment 121, CNRS \& Universit\'e Paris Sud XI, 91405 Orsay Cedex, France 
}

\date{Submitted February 21, 2012}

\abstract
{}{The role of galaxy mergers in massive galaxy evolution, and in particular to mass assembly and size growth, remains an open question. In this paper we measure the merger fraction and rate, both minor and major, of massive early-type galaxies ($M_{\star} \geq 10^{11}\ M_{\odot}$) in the COSMOS field, and study their role in mass and size evolution.}{We use the 30-band photometric catalogue in COSMOS, complemented with the spectroscopy of the zCOSMOS survey, to define close pairs with a separation on the sky plane $10h^{-1}$ kpc $\leq r_{\rm p} \leq 30h^{-1}$ kpc and a relative velocity $\Delta v \leq 500$ km $s^{-1}$ in redshift space. We measure both major (stellar mass ratio $\mu \equiv M_{\star,2}/M_{\star,1} \geq 1/4$) and minor ($1/10 \leq \mu < 1/4$) merger fractions of massive galaxies, and study their dependence on redshift and on morphology (early types vs late types).}{The merger fraction and rate of massive galaxies evolves as a power-law $(1+z)^{n}$, with major mergers increasing with redshift, $n_{\rm MM} = 1.4$, and minor mergers showing little evolution, $n_{\rm mm} \sim 0$. When split by their morphology, the minor merger fraction for early-type galaxies (ETGs) is higher by a factor of three than that for late-type galaxies (LTGs), and both are nearly constant with redshift. The fraction of major mergers for massive LTGs evolves faster ($n_{\rm MM}^{\rm LT} \sim 4$) than for ETGs ($n_{\rm MM}^{\rm ET}  = 1.8$).}{Our results show that massive ETGs have undergone 0.89 mergers (0.43 major and 0.46 minor) since $z \sim 1$, leading to a mass growth of $\sim30$\%. We find that $\mu \geq 1/10$ mergers can explain $\sim 55$\% of the observed size evolution of these galaxies since $z \sim 1$. Another $\sim20$\% is due to the progenitor bias (younger galaxies are more extended) and we estimate that very minor mergers ($\mu < 1/10$) could contribute with an extra $\sim20$\%. The remaining $\sim5$\% should come from other processes (e.g., adiabatic expansion or observational effects). This picture also reproduces the mass growth and the velocity dispersion evolution of these galaxies. We conclude from these results, and after exploring all the possible uncertainties in our picture, that merging is the main contributor to the size evolution of massive ETGs at $z \lesssim 1$, accounting for $\sim 50-75$\% of that evolution in the last 8 Gyr. Nearly half of the evolution due to mergers is related to minor ($\mu < 1/4$) events.}


\titlerunning{Mergers and the size evolution of massive ETGs since $z \sim 1$}

\maketitle

\section{Introduction}\label{intro}
The history of mass assembly is a major component of the galaxy formation and evolution scenario. The evolution in the number of galaxies of a given mass, as well as the size and shapes of galaxies building the Hubble sequence, provides strong input to this scenario. The optical colour –- magnitude diagram of local galaxies shows two distinct populations: the ''red sequence'', consisting primarily of old, spheroid-dominated, quiescent galaxies, and the ''blue cloud'', formed primarily by spiral and irregular star-forming galaxies \citep[e.g.,][]{strateva01,baldry04}. This bimodality has been traced at increasingly higher redshifts \citep[e.g.,][]{ilbert10}, showing that the most massive galaxies were the first to populate the red sequence as a result of the so-called ''downsizing'' \citep[e.g.,][]{bundy06,pgon08,pozzetti10}. These properties result from several physical mechanisms for which it is necessary to evaluate the relative impact. In this paper we examine the contribution of major and minor mergers to the mass growth and size evolution of massive early-type galaxies (ETGs), based on new measurements of the pair fraction from the COSMOS\footnote{http://cosmos.astro.caltech.edu/} (Cosmological Evolution Survey, \citealt{cosmos}) and zCOSMOS\footnote{http://www.astro.phys.ethz.ch/zCOSMOS/} \citep{lilly07} surveys. 

The number density of massive ETGs galaxies with $M_{\star} \gtrsim 10^{11}\ M_{\odot}$ is roughly constant since $z \sim 0.8$ \citep[][and references therein]{pozzetti10}, with major mergers (mass or luminosity ratio higher than 1/4) common enough to explain their number evolution since $z = 1$ \citep{eliche10I,robaina10,oesch10}. However, and despite that they seem ''dead'' since $z \sim 0.8$, two observational facts rule out the passive evolution of these massive ETGs after they have reached the red sequence: the presence of Recent Star Formation (RSF) episodes and their size evolution. In the former, the study of red sequence galaxies in the $NUV$--optical colour vs magnitude diagram reveals that $\sim$30\% have undergone RSF, as seen from their blue $NUV-r$ colours, both locally \citep{kaviraj07} and at higher redshifts \citep[$z \sim 0.6$,][]{kaviraj11}. This RSF typically involves $5-15$\% of the galaxy stellar mass \citep{scarlata07ee,kaviraj08,kaviraj11}. Some authors suggest that minor mergers, i.e., the merger of a massive red sequence galaxy with a less massive (mass or luminosity ratio lower than 1/4), gas-rich satellite, could explain the observed properties of galaxies with RSF \citep{kaviraj09,onti11,desai11}.

Regarding size evolution, it is now well established that massive ETGs have, on average, lower effective radius ($r_{\rm e}$) at high redshift than locally, being $\sim 2$ and $\sim 4$ times smaller at $z \sim 1$ and $z \sim 2$, respectively \citep[][but see \citealt{saracco10}; and \citealt{vale10b} for a different point of view]{daddi05, trujillo06, trujillo07, trujillo11, buitrago08, vandokkum08, vandokkum10, vanderwel08esize, toft09, williams10, newman10, newman12, damjanov11,weinzirl11,cassata11}. Massive ETGs as compact as observed at high redshifts are rare in the local universe \citep{trujillo09,taylor10,cassata11}, suggesting that they must evolve since $z \sim 2$ to the present. It has been proposed that high redshift compact galaxies are the cores of present day ellipticals, and that they increased their size by adding stellar mass in the outskirts of the galaxy \citep{bezanson09,hopkins09core,vandokkum10}. Several studies suggest that merging, especially the minor one, could explain the observed size evolution \citep{naab09,bezanson09,hopkins10size,feldmann10,shankar11,oser12}, while other processes, as adiabatic expansion due to AGNs or to the passive evolution of the stellar population, should have a mild role at $z \lesssim 1$ \citep{fan10,ragone11,trujillo11}. In addition, a significant fraction of local ellipticals present signs of recent interactions \citep{vandokkum05,tal09}.

While minor mergers are expected to contribute significantly to the evolution of massive ETGs, there is no direct observational measurement of their contribution yet. As a first effort, \citet{jogee09} estimate the minor merger fraction in massive galaxies out to $z \sim 0.8$ using morphological criteria, and
find that the minor merger fraction has a lower limit which is $\sim$3 times larger than the corresponding major merger fraction. The minor merger fraction of the global population of $L_{B} \gtrsim L_{B}^{*}$ galaxies has been studied quantitatively for the first time by \citet[][LS11 hereafter]{clsj11mmvvds} in the VVDS-Deep\footnote{http://cesam.oamp.fr/vvdsproject/vvds.htm} (VIMOS VLT Deep Spectroscopic Survey, \citealt{lefevre05}). They show that minor mergers are quite common, that their importance decrease with redshift \citep[see also][]{lotz11}, and that they participate to about 25\% of the mass growth by merging of such galaxies. Focusing on massive galaxies, \citet{williams11}, \citet{marmol12}, or \citet{newman12} study their total (major + minor) merger fraction to $z \sim 2$, finding also that it is nearly constant with redshift. In this paper we present the detailed merger history, both minor and major, of massive ($M_{\star} \geq 10^{11}\ M_{\odot}$) ETGs since $z \sim 1$ using close pair statistics in the COSMOS field, and use it to infer the role of major and minor mergers in the mass assembly and in the size evolution of these systems in the last $\sim 8$ Gyr.

The paper is organised as follow. In Sect.~\ref{data} we present our photometric catalogue in the COSMOS field, while in Sect.~\ref{method} we review the methodology used to measure close pair merger fractions when photometric redshifts are used. We present our merger fractions of massive galaxies in Sect.~\ref{ffcosmos}, and the inferred merger rates for ETGs in Sect.~\ref{mrcosmos}. The role of mergers in the mass assembly and in the size evolution of massive ETGs is discussed in Sect.~\ref{discussion}, and in Sect.~\ref{conclusion} we present our conclusions. Throughout this paper we use a standard cosmology with $\Omega_{m} = 0.3$, $\Omega_{\Lambda} = 0.7$, $H_{0}= 100h$ Km s$^{-1}$ Mpc$^{-1}$ and $h = 0.7$. Magnitudes are given in the AB system.

\section{The COSMOS photometric catalogue}\label{data}
We use the COSMOS catalogue with photometric redshifts derived from 30 broad and medium bands described in \citet{ilbert09} and \citet{capak07}, version 1.8. We restrict ourselves to objects with $i^+ \leq 25$. The detection completeness at this limit is higher than 90\% \citep{capak07}. In order to obtain accurate colours, all the images were degraded to the same point spread function (PSF) of $1.5^{\prime\prime}$. At $i^+ \sim 25$, the rms accuracy of the photometric redshifts ($z_{\rm phot}$) at $z \lesssim 1$ is $\sim0.04$ in $(z_{\rm spec} - z_{\rm phot})/(1 + z_{\rm spec}$), where $z_{\rm spec}$ is the spectroscopic redshift of the sources \citep[Fig. 9 in ][]{ilbert09}. At $z > 1$ the quality of the photometric redshifts quickly deteriorates. Additionally, and because we are interested on minor companions, we require a detection in the $K_{\rm s}$ band to ensure that the stellar mass estimates are reliable, thus we add the constraint $K_{\rm s} \leq 24$.

Stellar masses of the photometric catalogue have been derived following the same approach than in \citet{ilbert10}. We used stellar population synthesis models to convert luminosity into stellar mass \citep[e.g.,][]{bell03,fontana04}. The stellar mass is the factor needed to rescale the best-fit template (normalised at one solar mass) for the intrinsic luminosities. The Spectral Energy Distribution (SED) templates were generated with the stellar population synthesis package developed by \citet[BC03]{bc03}. We assumed a universal initial mass function (IMF) from \citet{chabrier03} and an exponentially declining star formation rate, $SFR \propto {\rm e}^{−t/\tau}$ ($\tau$ in the range 0.1 Gyr to 30 Gyr). The SEDs were generated for a grid of 51 ages (in the range 0.1 Gyr to 14.5 Gyr). Dust extinction was applied to the templates using the \citet{calzetti00} law, with $E(B - V)$ in the range 0 to 0.5. We used models with two different metallicities. Following \citet{fontana06} and \citet{pozzetti07}, we imposed the prior $E(B - V ) < 0.15$ if age/$\tau > 4$ (a significant extinction is only allowed for galaxies with a high $SFR$). The stellar masses derived in this way have a systematic uncertainty of $\sim$0.3 dex \citep[e.g.,][]{pozzetti07,barro11mass}.

We supplement the previous photometric catalogue with the spectroscopic information from zCOSMOS survey, a large spectroscopic redshift survey in the central area of the COSMOS field. In this analysis we use the final release of the bright part of this survey, called the zCOSMOS-bright 20k sample. This is a pure magnitude selected sample with $I_{\rm AB} \leq 22.5$. For a detailed description and relevant results of the previous 10k release, see \citet{zcosmos10k,tasca09,pozzetti10} or \citet{peng10}. A total of 20604 galaxies have been observed with the VIMOS spectrograph \citep{lefevre03} in multi-slit mode, and the data have been processed using the VIPGI data processing pipeline \citep{scodeggio05}. A spectroscopic flag has been assigned to each galaxy providing an estimate of the robustness of the redshift measurement \citep{lilly07}. If a redshift has been measured, the corresponding spectroscopic flag value can be 1, 2, 3, 4 or 9. Flag = 1 means that the redshift is $\sim$ 70\% secure and flag = 4 that the redshift is $\sim 99$\% secure. Flag = 9 means that the redshift measurement relies on one single narrow emission line ($\ion{O}{ii}$ or H$\alpha$ mainly). The information about the consistency between photometric and spectroscopic redshifts has also been included as a decimal in the spectroscopic flag. In this study we select the highest reliable redshifts, i.e., with confidence class 4.5, 4.4, 3.5, 3.4, 9.5, 9.3, and 2.5. This flag selection ensures that 99\% of redshifts are believed to be reliable based on duplicate objects \citep{zcosmos10k}.

Our final COSMOS catalogue comprises 134028 galaxies at $0.1 \leq z < 1.1$, our range of interest (see Sect.~\ref{samples}). Nearly 35\% of the galaxies with $i^+ \lesssim 22.5$ have a high reliable spectroscopic redshift. For consistency and to avoid systematics, we always use the stellar masses and other derived quantities from the photometric catalogue. We checked that the dispersion when comparing stellar masses from $z_{\rm phot}$ and $z_{\rm spec}$ is $\sim0.15$ dex, lower than the typical error in the measured stellar masses ($\sim 0.3$ dex). Thanks to the methodology developed in \citet{clsj10pargoods} we are able to obtain reliable merger fractions from photometric catalogues under some quality conditions (Sect.~\ref{method}). We check that the COSMOS catalogue is adequate for our purposes in Sects.~\ref{20ktest} and \ref{cosvar}.

\begin{figure}[t!]
\resizebox{\hsize}{!}{\includegraphics{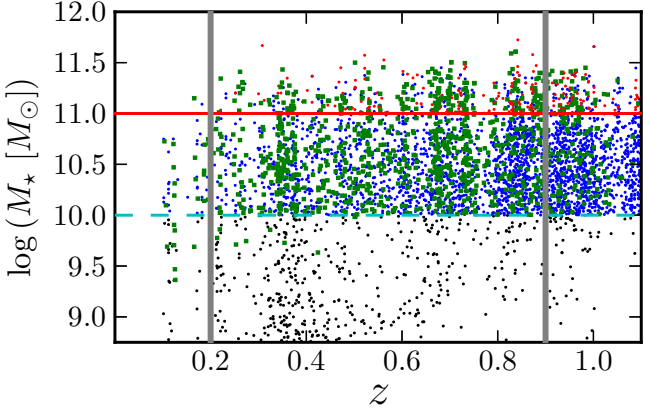}}
\caption{Stellar mass as a function of redshift in the COSMOS field. Red dots are principal galaxies ($M_{\star} \geq 10^{11}\ M_{\odot}$) with $z_{\rm phot}$ in the zCOSMOS area, blue dots are companion galaxies ($M_{\star} \geq 10^{10}\ M_{\odot}$) with $z_{\rm phot}$ in the COSMOS area, and black dots are the red galaxies ($NUV-r^{+} \geq 3.5$) with $z_{\rm phot}$ in the COSMOS area. We only show a random 15\% of the total populations for visualisation purposes. Green squares mark those galaxies in previous populations with a spectroscopic resdhift. The vertical lines mark the lower and upper redshift in our study, while the horizontal ones the mass selection of the principal (solid) and the companion (dashed) samples.}
\label{msvsz}
\end{figure}

\subsection{Definition of the mass-selected samples}\label{samples}
We define two samples selected in stellar mass. The first one comprises 2047 principal massive galaxies in the zCOSMOS area, where spectroscopic information is available, with $M_{\star} \geq 10^{11}\ M_{\odot}$ ($M_{\star} \gtrsim M_{\star}^*$, \citealt{ilbert10}) at $0.1 \leq z < 1.1$. The second sample comprises the 23992 companion galaxies with $M_{\star} \geq 10^{10}\ M_{\odot}$ in the full COSMOS area and in the same redshift range. The mass limit of the companion sample ensures completeness for red galaxies up to $z \sim 0.9$ \citep{drory09, ilbert10}. Because of that, we set $z_{\rm up} = 0.9$ as the upper redshift in our study, while $z_{\rm down} = 0.2$ to probe enough cosmological volume. However, our methodology takes into account the photometric redshift errors (see Sect.~\ref{method}, for details), so we must include in the samples not only the sources with $z < z_{\rm up}$, but also those sources with $z - 2\sigma_{\rm phot} < z_{\rm up}$ in order to ensure completeness in redshift space. Because of this, we set the maximum and minimum redshift of the catalogues to $z_{\rm min} = 0.1$ and $z_{\rm max} = 1.1$. We show the mass distribution of our samples as a function of $z$ in Fig.~\ref{msvsz}, and we assume our samples as volume-limited mass-selected in the following.

Our final goal is to measure the merger fraction and rate of massive ETGs, but our principal sample comprises ETGs, spirals and irregulars. We segregate morphologically our principal sample thanks to the morphological classification defined in \citet{tasca09}. Their method use as morphological indicator the distance of the galaxies in the multi-space $C-A-G$ (Concentration, Asymmetry and Gini coefficient) to the position in this space of a training sample of $\sim$500 eye-ball classified galaxies. These morphological indices were measured in the HST/ACS images of the COSMOS field, taken through the wide {\it F814W} filter \citep{koekemoer07}. The galaxies in the training sample were classified into ellipticals, lenticulars, spirals of all types (Sa, Sb, Sc, Sd), irregulars, point-like and undefined sources, and then these classes were grouped into early-type (E,S0), spirals (Sa, Sb, Sc, Sd) and irregular galaxies. It is this coarser classification that was considered in building the training set. The unclassified objects were not used for the training. We refer the reader to \citet{tasca09} for further details. The morphological classification in the COSMOS field is reliable for galaxies brighter than $i^+ < 24$, and all our principal galaxies are brighter than $i^+ < 23.5$ up to $z = 1$. According to the classification presented in \citet{tasca09} our principal sample comprises 1285 (63\%) ETGs (E/S0) and 632 (31\%) spiral galaxies. The remaining 6\% sources are half irregulars (65 sources) and half massive galaxies without morphological classification (65 sources). We stress that the  classification of the principal sample is exclusively morphological, without taking into account any additional colour information, i.e., some of our ETGs could be star-forming. We checked that $\sim$95\% of our massive ETGs are also quiescent (they have a rest-frame, dust reddening corrected colour $NUV-r^{+} \geq 3.5$, \citealt{ilbert10}). Regarding the companion sample, we do not attempt to segregate it morphologically because the morphological classification is not reliable for all companion galaxies (see Sect~\ref{colsec}, for details).

We used $C$, $A$ and $G$ automatic indices to classify morphologically the principal galaxy of the close pair systems. However, these indices are affected by interactions, e.g., the asymmetry increases, and we could misclassify ETGs and spirals as irregular galaxies. \citet{htoledo05,htoledo06} study how these morphological indices vary on major interactions in the local universe. They find that ETGs are slightly affected by interactions and that interacting ETGs do not reach the loci of irregular galaxies in the $C-A$ space. However, spiral galaxies are strongly affected by interactions and they can be classified as irregulars by automatic methods. Thus, we do not expect misclassifications in our ETGs sample, while some of our irregular galaxies can be interacting spirals. This is in fact observed by \citet{pawel12} in the 10k zCOSMOS sample. They find that the fraction of ETGs in close pairs is similar to that in the underlying non-interacting population, while the fraction of spirals/irregulars in close pairs is lower/higher than expected. However, the sum of spirals and irregulars is similar to that in the underlying population, suggesting a spiral to irregular transformation due to interactions.

\begin{figure}[t!]
\resizebox{\hsize}{!}{\includegraphics{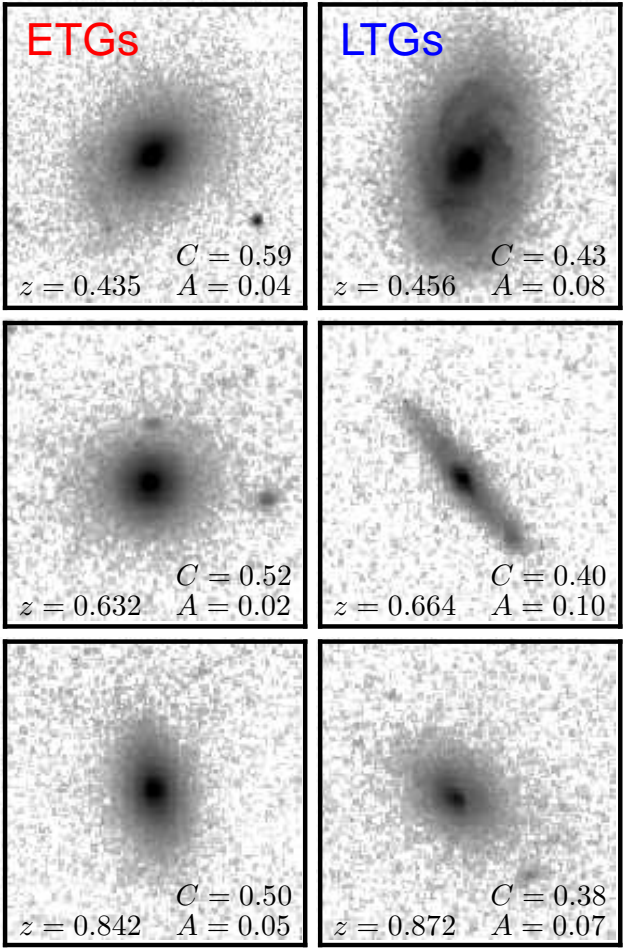}}
\caption{Examples of the typical ETGs (left) and LTGs (right) with $M_{\star} \geq 10^{11}\ M_{\odot}$ in the COSMOS field. The postage stamps show a $30h^{-1}$ kpc x $30h^{-1}$ kpc area of the HST/ACS {\it F814W} image at the redshift of the source, with the North on the top and the East on the left. The pixel scale of the HST/ACS image is 0.05\arcsec. The grey scale ranges from 0.5$\sigma_{\rm sky}$ to 150$\sigma_{\rm sky}$, where $\sigma_{\rm sky}$ is the dispersion of the sky around the source. The redshift, the concentration ($C$) and the asymmetry ($A$) of the sources are labelled in the panels.}
\label{etgfig}
\end{figure}

In summary, the morphology of ETGs is slightly affected by interactions, while some spirals could be classified as irregulars during a merger. Because of this, we define late-type galaxies (LTGs) as spirals + irregulars, thus avoiding any bias due to morphological transformations during the merger process. We show some representative examples of our massive ETGs and LTGs in Fig.~\ref{etgfig}. The mean mass of both ETGs and LTGs is similar, $\overline{M_{\star}} \sim 10^{11.2}\ M_{\odot}$.

\subsection{Dependence of the photometric errors on stellar mass}\label{deltazphot}
\begin{figure}[t!]
\resizebox{\hsize}{!}{\includegraphics{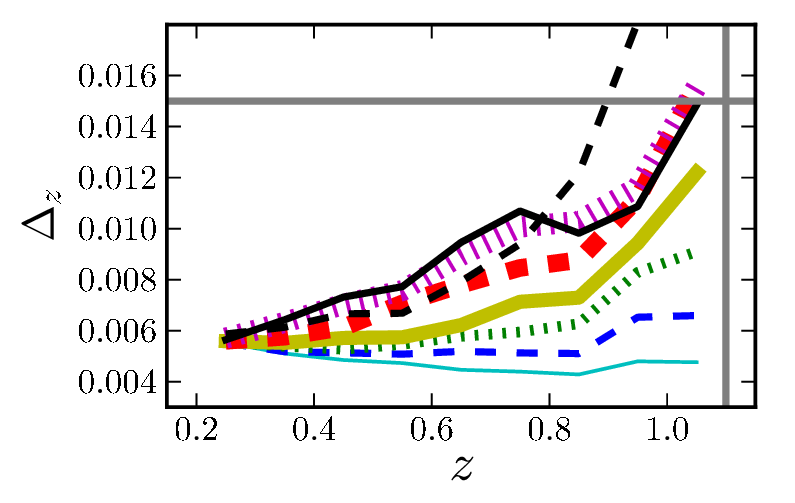}}
\caption{$\Delta_{z}$ as a function of redshift in the mass-selected sample, from $M_{\star} \geq 10^{11}\ M_{\odot}$ (thiner line) to $10^{10}\ M_{\odot} \leq M_{\star} < 10^{10.2}\ M_{\odot}$ (thicker line) galaxies in bins of 0.2 dex. The black solid line marks the photometric errors of blue galaxies in the lower mass bin, while the black dashed line is for red galaxies in the same mass bin. The vertical line marks the higher redshift in our samples, $z_{\rm max} = 1.1$. The horizontal line marks the median $\Delta_{z}$ for low-mass galaxies at the high redshift end of our sample $1.0 \leq z < 1.1$, $\Delta_{z} = 0.015$.}
\label{deltazfig}
\end{figure}

The quality of the photometric redshifts in COSMOS decreases for faint objects in the $i^+$ band \citep{ilbert09}. In this section we study in details how redshift errors depend on the mass of the sources, since this imposes limits on our ability of measure reliable merger fractions in photometric catalogues (Sect.~\ref{20ktest}). As shown by \citet{ilbert10}, we can estimate the photometric redshift error ($\sigma_{z_{\rm phot}}$) from the Probability Distribution Function of the photometric redshift fit. In Fig.~\ref{deltazfig} we show the median $\Delta_{z} \equiv \sigma_{z_{\rm phot}}/(1 + z_{\rm phot})$ of galaxies with different stellar masses, from $M_{\star} \geq 10^{11}\ M_{\odot}$ (massive galaxies) to $10^{10}\ M_{\odot} \leq M_{\star} < 10^{10.2}\ M_{\odot}$ (low-mass galaxies) in bins of 0.2 dex.

Massive galaxies are bright in the whole redshift range under study. Thus, their photometric errors are small up to $z \sim 1$, $\Delta_{z} \sim 0.005$. On the other hand, low-mass galaxies are fainter at high redshift than their local counterparts, so their their photometric errors increase with $z$ and reach $\Delta_{z} \sim 0.015$ at $z \sim 1$. We study separately the photometric errors of low-mass red and blue galaxies. We took as red galaxies those with SED (rest-frame, dust reddening corrected) colour $NUV-r^{+} \geq 3.5$, while as blue those with $NUV-r^{+} < 3.5$ \citep[see][for details]{ilbert10}. Blue galaxies also have $\Delta_{z} \sim 0.015$ up to $z \sim 1$, while red galaxies have higher photometric redshift errors, with $\Delta_{z} \sim 0.020$ at $z = 0.95$ and $\Delta_{z} \sim 0.040$ at $z = 1.05$. This different behaviour can be explained by the different mass-to-light ratio ($M_{\star}/L$) of both populations. Faint ($i^+ \sim 25$) blue galaxies, whose photometric errors are higher, reach masses as low as $M_{\star} \sim 10^{8.5}\ M_{\odot}$ at $z \sim 1$. On the other hand, we are in the detection limit for red galaxies at these redshifts (red galaxies have $i^+ \sim 25$ at $z \sim 1$, Sect.~\ref{samples}), explaining their high photometric redshift errors. Similar trends in the COSMOS photometric redshift errors were found by \citet{george11}. In Sect.~\ref{20ktest} we prove that our methodology is able to recover reliable merger fractions in COSMOS samples with $\Delta_{z} \lesssim 0.040$, as those in our study.

\section{Close pairs using photometric redshifts}\label{method}
The linear distance between two sources can be obtained from their projected separation, $r_{\rm p} = \theta d_A(z_i)$, and their rest-frame relative velocity along the line of sight, $\Delta v = {c\, |z_j - z_i|}/(1+z_i)$, where $z_i$ and $z_j$ are the redshift of the principal (more luminous/massive galaxy in the pair) and companion galaxy, respectively; $\theta$ is the angular separation, in arcsec, of the two galaxies on the sky plane; and $d_A(z)$ is the angular scale, in kpc/arcsec, at redshift $z$. Two galaxies are defined as a close pair if $r_{\rm p}^{\rm min} \leq r_{\rm p} \leq r_{\rm p}^{\rm max}$ and $\Delta v \leq \Delta v^{\rm max}$. The lower limit in $r_{\rm p}$ is imposed to avoid seeing effects. We used $r_{\rm p}^{\rm min} = 10h^{-1}$ kpc, $r_{\rm p}^{\rm max} = 30h^{-1}$ kpc, and $\Delta v^{\rm max} = 500$ km s$^{-1}$. With these constraints 50\%-70\% of the selected close pairs will finally merge \citep{patton00,patton08,lin04,bell06}. The PSF of the COSMOS ground-based images is $1.5\arcsec$ \citep{capak07}, which corresponds to $\sim 8h^{-1}$ kpc in our cosmology at $z \sim 0.9$. To ensure well deblended sources and to minimise colour contamination, we fixed $r_{\rm p}^{\rm min}$ to $10h^{-1}$ kpc ($\theta \gtrsim 2\arcsec$). On the other hand, we set $r_{\rm p}^{\rm max}$ to $30h^{-1}$ kpc to ensure reliable merger fractions in our study (see Sect.~\ref{20ktest}, for details).

To compute close pairs we defined a principal and a companion sample (Sect.~\ref{samples}). The principal sample comprises the more massive galaxy of the pair, and we looked for those galaxies in the companion sample that fulfil the close pair criterion for each galaxy of the principal sample. If one principal galaxy has more than one close companion, we took each possible pair separately (i.e., if the companion galaxies B and C are close to the principal galaxy A, we studied the pairs A-B and A-C as independent). In addition, we imposed a mass difference between the pair members. We denote the ratio between the mass of the principal galaxy, $M_{\star,1}$, and the companion galaxy, $M_{\star,2}$, as 
\begin{equation}
\mu \equiv \frac{M_{\star,2}}{M_{\star,1}}
\end{equation}
and looked for those systems with $M_{\star,2} \geq \mu M_{\star,1}$. We define as major companions those close pairs with $\mu \geq 1/4$, while minor companions those with $1/10 \leq \mu < 1/4$.

With the previous definitions the merger fraction is
\begin{equation}
f_{\rm m}\,(\geq \mu) = \frac{N_{\rm p}\,(\geq \mu)}{N_{1}},\label{ncspec}
\end{equation}
where $N_1$ is the number of sources in the principal sample, and $N_{\rm p}$ the number of principal galaxies with a companion that fulfil the close pair criterion for a given $\mu$. This definition applies to spectroscopic volume-limited samples. Our samples are volume-limited, but combine spectroscopic and photometric redshifts. In a previous work, \citet{clsj10pargoods} developed a statistical method to obtain reliable merger fractions from photometric catalogues. We recall the main points of this methodology below, while we study its limits when applied to our COSMOS photometric catalogue in Sect.~\ref{20ktest}.

We used the following procedure to define a close pair system in our photometric catalogue \citep[see][for details]{clsj10pargoods}: first we search for close spatial companions of a principal galaxy, with redshift $z_1$ and uncertainty $\sigma_{z_1}$, assuming that the galaxy is located at $z_1 - 2\sigma_{z_1}$. This defines the maximum $\theta$ possible for a given $r_{\rm p}^{\rm max}$ in the first instance. If we find a companion galaxy with redshift $z_2$ and uncertainty $\sigma_{z_2}$ in the range $r_{\rm p} \leq r_{\rm p}^{\rm max}$ and with a given mass with respect to the principal galaxy, then we study both galaxies in redshift space. For convenience, we assume below that every principal galaxy has, at most, one close companion. In this case, our two galaxies could be a close pair in the redshift range
\begin{equation}
[z^{-},z^{+}] = [z_1 - 2\sigma_{z_1}, z_1 + 2\sigma_{z_1}] \cap [z_2- 2\sigma_{z_2}, z_2 + 2\sigma_{z_2}].
\end{equation}
Because of variation in the range $[z^{-},z^{+}]$ of the function $d_A(z)$, a sky pair at $z_1 - 2\sigma_{z_1}$ might not be a pair at $z_1 + 2\sigma_{z_1}$. We thus impose the condition $r_{\rm p}^{\rm min} \leq r_{\rm p} \leq r_{\rm p}^{\rm max}$ at all $z \in [z^{-},z^{+}]$, and redefine this redshift interval if the sky pair condition is not satisfied at every redshift. After this, our two galaxies define the close pair system $k$ in the redshift interval $[z^{-}_k,z^{+}_k]$, where the index $k$ covers all the close pair systems in the sample.

The next step is to define the number of pairs associated at each close pair system $k$. For this, we suppose in the following that a galaxy $i$ in whatever sample is described in redshift space by a probability distribution $P_i\, (z_i\, |\, \eta_i)$, where $z_i$ is the source's redshift and $\eta_i$ are the parameters that define the distribution. If the source $i$ has a photometric redshift, we assume that

\begin{eqnarray}
P_i\, (z_i\, |\, \eta_i) = P_G\, (z_i\, |\, z_{{\rm phot},i},\sigma_{z_{{\rm phot},i}}) \nonumber\\
= \frac{1}{\sqrt{2\pi}\sigma_{z_{{\rm phot},i}}}\exp\left\{{-\frac{(z_i-z_{{\rm phot},i})^2}{2\sigma_{z_{{\rm phot},i}}^2}}\right\}\label{zgauss},
\end{eqnarray}
while if the source has a spectroscopic redshift
\begin{equation}
P_i\, (z_i\, |\, \eta_i) = P_D\, (z_i\, |\, z_{{\rm spec},i}) = \delta(z_i - z_{{\rm spec},i}),
\end{equation}
where $\delta(x)$ is delta's Dirac function. With this distribution we are able to statistically treat all the available information in $z$ space and define the number of pairs at redshift $z_1$ in system $k$ as
\begin{equation}
\nu_{k}\,(z_1) = {\rm C}_k\, P_1 (z_1\, |\, \eta_1) \int_{z_{\rm m}^{-}}^{z_{\rm m}^{+}} P_2 (z_2\, |\, \eta_2)\, {\rm d}z_2,\label{nuj}
\end{equation}
where $z_1 \in [z^{-}_k,z^{+}_k]$, the integration limits are
\begin{eqnarray}
z_{\rm m}^{-} = z_1(1-\Delta v^{\rm max}/c) - \Delta v^{\rm max}/c,\\
z_{\rm m}^{+} = z_1(1+\Delta v^{\rm max}/c) + \Delta v^{\rm max}/c,
\end{eqnarray}
the subindex 1 [2] refers to the principal [companion] galaxy in $k$ system, and the constant ${\rm C}_k$ normalises the function to the total number of pairs in the interest range
\begin{equation}
2 N_{\rm p}^k = \int_{z_k^{-}}^{z_k^{+}} P_1 (z_1\, |\, \eta_i)\, {\rm d}z_1  + \int_{z_k^{-}}^{z_k^{+}} P_2 (z_2\, |\, \eta_2)\, {\rm d}z_2.
\end{equation}
Note that $\nu_k = 0$ if $z_1 < z_k^-$ or  $z_1 > z_k^+$. The function $\nu_k$ (Eq.~[\ref{nuj}]) tells us how the number of pairs in the system $k$, $N_{\rm p}^k$, are distributed in redshift space. The integral in Eq.~(\ref{nuj}) spans those redshifts in which the companion galaxy has $\Delta v \leq \Delta v^{\rm max}$ for a given redshift of the principal galaxy.

With previous definitions, the merger fraction in the interval $z_{r,l} = [z_l,z_{l+1})$ is
\begin{equation}
f_{{\rm m},l} = \frac{\sum_k \int_{z_l}^{z_{l+1}}{\nu_k(z_1)}\, {\rm d}z_1}{\sum_i \int_{z_l}^{z_{l+1}} P_i\, (z_i\, |\, \eta_i)\, {\rm d}z_i},\label{ncphot}
\end{equation}
where the index $l$ spans the redshift bins defined over the redshift range under study. If we integrate over the whole redshift space, $z_{r} = [0,\infty)$, Eq.~(\ref{ncphot}) becomes
\begin{equation}
f_{{\rm m}}\,(\geq \mu) = \frac{\sum_k N_{\rm p}^k\,(\geq \mu)}{N_1},\label{ncphot2}
\end{equation}
where $\sum_k N_{\rm p}^k$ is analogous to $N_{\rm p}$ in Eq.~(\ref{ncspec}). In order to estimate the statistical error of $f_{{\rm m},l}$, denoted $\sigma_{{\rm stat},l}$, we used the jackknife technique \citep{efron82}. We computed partial standard deviations, $\delta_k$, for each system $k$ by taking the difference between the measured $f_{{\rm m},l}$ and the same quantity with the $k$th pair removed for the sample, $f_{{\rm m},l}^k$, such that $\delta_k = f_{{\rm m},l} - f_{{\rm m},l}^k$. For a redshift range with $N_{\rm p}$ systems, the variance is given by $\sigma_{{\rm stat},l}^2 = [(N_{\rm p}-1) \sum_k \delta_k^2]/N_{\rm p}$. When $N_{\rm p} \leq 5$ we used instead the Bayesian approach of \citet{cameron11}, that provides accurate asymmetric confidence intervals in these low statistical cases. We checked that for $N_{\rm p} > 5$ both jackknife and Bayesian methods provide similar statistical errors within 10\%.

\subsection{Dealing with border effects}\label{borders}
When we search for close companions near to the edges of the images it may happen that a fraction of the search volume is outside of the surveyed area, lowering artificially the number of companions. To deal with this we selected as principal galaxies those in the zCOSMOS area, i.e., in the central 1.6 deg$^{2}$, while we selected as companions those in the whole photometric COSMOS area. This maximise the spectroscopic fraction of the principal sample and ensures that we have companions inside all the searching volume.

\subsection{Testing the methodology with 20k spectroscopic sources}\label{20ktest}
Following \citet{clsj10pargoods}, we test in this section if we are able to obtain reliable merger fractions from our COSMOS photometric catalogue. For this, we study the merger fraction $f_{\rm m}$ in the zCOSMOS-bright 20k sample. The merger fraction in the 10k sample was studied in details by \citet{deravel11} and \citet{pawel12}. We define $f_{\rm spec}$ as the fraction of sources on a given sample with spectroscopic redshift. The 20k sample has $f_{\rm spec} = 1$, while the COSMOS photometric catalogue has $f_{\rm spec} = 0.34$ for $i^+ \leq 22.5$ galaxies. In this section we only use the $N = 10542$ sources at $0.2 \leq z < 0.9$ with a high reliable spectroscopic redshift from the 20k sample.

To test our method at intermediate $f_{\rm spec}$, we created synthetic catalogues by assigning their measured $z_{\rm phot}$ and $\sigma_{z_{\rm phot}}$ to $N(1-f_{\rm spec}$) random sources of the 20k sample (we denote this case as $S = 1$ in the following). To explore different values of $\Delta_{z}$, we assigned to the previous random sources a redshift as drawn for a Gaussian distribution with median $z_{\rm phot}$ and $\sigma^{2} = (S^2-1)\,\sigma_{z_{\rm phot}}^2$, where $S > 1$ is the factor by which we increase the initial $\Delta_{z}$ of the sample. In this case, the redshift error of the source is set to $S\sigma_{z_{\rm phot}}$. Then, we measured
\begin{equation}
\delta f_{\rm m} \equiv \frac{f_{\rm m}^{\rm syn}}{f_{\rm m}^{\rm 20k}} - 1,
\end{equation}
where $f_{\rm m}^{\rm 20k}$ is the measured merger fraction in the 20k spectroscopic sample at $0.2 \leq z < 0.9$ without imposing any mass or luminosity difference and $f_{\rm m}^{\rm syn}$ is the merger fraction from the synthetic samples in the same redshift range. When $S > 1$, we repeated the process ten times and averaged the results. 

We explored several cases with our synthetic catalogues. For example, we assumed that all sources in the synthetic principal catalogue (subindex 1) and in the companion one (subindex 2) have a photometric redshift, $f_{\rm spec,1} = f_{\rm spec,2} = 0$, and that $\Delta_{\rm z,1} = \Delta_{\rm z,2} = 0.007\ (S_1 = S_2 = 1)$. We also considered more realistic cases, as $f_{\rm spec,1} = 0.3$ and $\Delta_{\rm z,1} = 0.007\ (S_1 = 1)$ for principals, and $f_{\rm spec,2} = 0$ and $\Delta_{\rm z,2} = 0.042\ (S_2 = 6)$ for companions. We found that $\delta f_{\rm m}$ is higher than 10\% for $r_{\rm p}^{\rm max} = 30h^{-1}$ kpc close pairs for $\Delta_{\rm z,2} \gtrsim 0.05\ (S_2 \gtrsim 7)$ and realistic values of $\Delta_{\rm z,1}$. We checked that $|\,\delta f_{\rm m}\,| \lesssim 10$\% for $\Delta_{\rm z,2} \leq 0.04$ and $r_{\rm p}^{\rm max} = 30h^{-1}$ kpc, justifying the upper limit $\Delta_{\rm z} = 0.04$ imposed in Sect.~\ref{deltazphot}. For higher $r_{\rm p}^{\rm max}$ the method overestimates the merger fraction by about 50\% in the $\Delta_{\rm z,2} = 0.04$ case. Because we are interested on faint companions, we set $r_{\rm p}^{\rm max} = 30h^{-1}$ kpc in the following to ensure reliable merger fractions. 

On the other hand, we found that the $\sigma_{\rm stat}$ of the $f_{\rm m}^{\rm syn}$ is $\sim5$\% of the measured value, i.e., two times lower than the estimated $|\,\delta f_{\rm m}\,| \sim 10$\%. Because of this, and to ensure reliable uncertainties in the merger fractions, we impose a minimum error in $f_{\rm m}$ of 10\%, and we take as final merger fraction error $\sigma_{\rm f_{\rm m}} = {\rm max}(0.1f_{\rm m},\sigma_{\rm stat})$.

In the next section we test further our methodology by comparing the merger fraction from a spectroscopic survey ($f_{\rm spec}~=~1$) against that in COSMOS from our photometric catalogue.

\begin{figure}[t!]
\resizebox{\hsize}{!}{\includegraphics{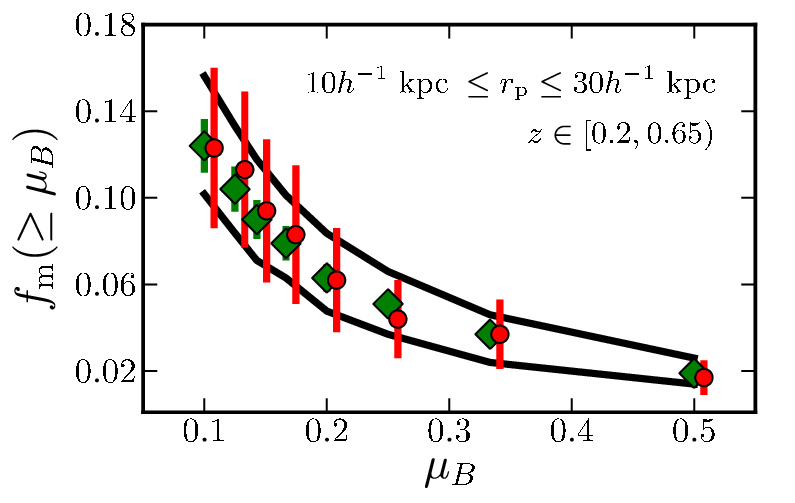}}
\resizebox{\hsize}{!}{\includegraphics{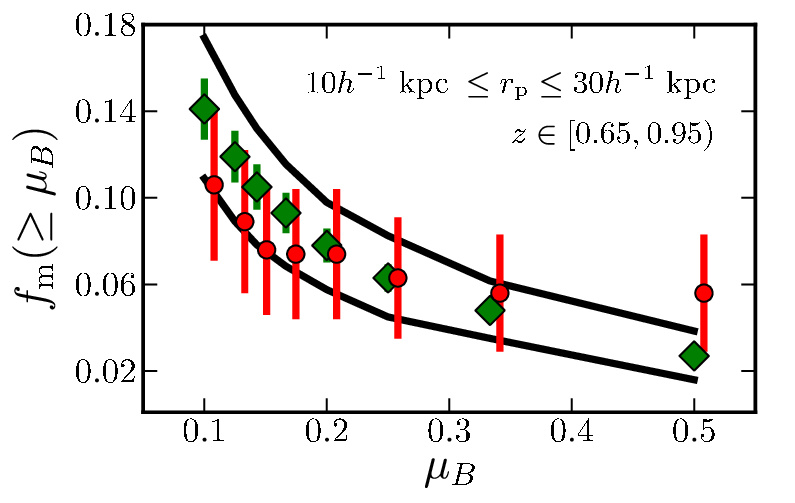}}
\caption{Merger fraction of $M_{B}^{\rm e} \leq -20$ galaxies as a function of luminosity difference in the $B-$band, $\mu_{B}$, at $z \in [0.2,0.65)$ (top) and $z \in [0.65,0.95)$ (bottom) for $10h^{-1}\ {\rm kpc} \leq r_{\rm p} \leq 30h^{-1}$ kpc close pairs. Diamonds are from present work in COSMOS (photometric catalogue) while dots are from VVDS-Deep (LS11, spectroscopic catalogue). The black solid lines in both panels show the maximum and minimum merger fractions, including $1\sigma_{\rm f_{\rm m}}$ errors, when we split the COSMOS field in VVDS-Deep size subfields ($\sim$0.5 deg$^2$).}
\label{ffvvdsfig}
\end{figure}

\subsection{Comparison with merger fractions in VVDS-Deep: cosmic variance effect}\label{cosvar}
In a previous work in VVDS-Deep, LS11 measured the merger fraction of $M_{B}^{\rm e} \leq -20$ galaxies with spectroscopic redshifts, where $M_{B}^{\rm e} = M_{B} + Qz$ and $Q = 1.1$ accounts for the evolution of the luminosity function with redshift, as a function of luminosity difference in the $B-$band, $\mu_{B} = L_{B,2}/L_{B,1}$. As an additional test of our methodology, in this section we compare the merger fraction in the COSMOS photometric catalogue with that measured by LS11 down to $\mu_{B} = 1/10$, reaching the minor merger regime in which we are interested on. To minimise the systematic biases, we used the same redshift ranges, $z_{\rm r, 1} = [0.2,0.65)$ and $z_{\rm r, 2} = [0.65,0.95)$, close pair definition ($r_{\rm p}^{\rm max} = 30h^{-1}$ kpc), principal sample ($M_{B}^{\rm e} \leq -20$), and companion sample ($M_{B}^{\rm e} \leq -17.5$) than LS11. We checked that the photometric redshift errors are $\Delta_{\rm z} \lesssim 0.04$ up to $z \sim 0.95$ for faint companion galaxies (see Sect.~\ref{20ktest}). Note that LS11 use $r_{\rm p}^{\rm min} = 5h^{-1}$ kpc, while we take $r_{\rm p}^{\rm min} = 10h^{-1}$ kpc. Hence, we recomputed the merger fractions in VVDS-Deep for $r_{\rm p}^{\rm min} = 10h^{-1}$ kpc. We show the merger fractions from COSMOS and VVDS-Deep for different values of $\mu_{B}$ in Fig.~\ref{ffvvdsfig}.

We find that VVDS-Deep and COSMOS merger fractions are in excellent agreement in the first redshift range, while in the second redshift range some discrepancies exist, with the merger fraction in COSMOS being higher than in VVDS-Deep at $\mu \lesssim 1/5$. However, both studies are compatible within error bars. Note that merger fraction uncertainties in COSMOS are $\sim3$ times lower than in VVDS-Deep because of the higher number of principals in COSMOS. We checked the effect of comic variance in this comparison. For that, we split the zCOSMOS area in several VVDS-Deep size ($\sim$0.5 deg$^{2}$) subfields and measured the merger fraction in these subfields. The maximum and minimum values of $f_{\rm m}$ in these subfields, including $1\sigma_{\rm f_{\rm m}}$ errors, are marked in Fig.~\ref{ffvvdsfig} with solid lines. We find that, within $1\sigma_{\rm f_{\rm m}}$, there is a zCOSMOS subfield with merger properties similar to the VVDS-Deep field. Because the zCOSMOS subfields are contiguous, this exercise provides a lower limit to the actual cosmic variance in the COSMOS field \citep[e.g.,][]{moster11}. Hence, we conclude that our methodology is able to recover reliable minor merger fractions from photometric samples in the COSMOS field.

\section{The merger fraction of massive ETGs in the COSMOS field}\label{ffcosmos}
The final goal of the present paper is to estimate the role of mergers (minor and major) in the mass assembly and size evolution of massive ETGs. To facilitate future comparison, we present first the merger properties of the global massive population in Sect.~\ref{ffms}. Then, we focus in the ETGs population in Sect.~\ref{ffmor}. 

The evolution of the merger fraction with redshift up to $z \sim 1.5$ is well parametrised by a power-law function \citep[e.g.,][]{lefevre00,clsj09ffgoods,deravel09},
\begin{equation}
f_{\rm m}\,(z) = f_{\rm m,0}\,(1+z)^{m}, \label{ffzeq}
\end{equation}
so we take this parametrisation in the following.

\subsection{The merger fraction of the global massive population}\label{ffms}
We summarise the minor, major and total merger fractions for $M_{\star} \geq 10^{11}\ M_{\odot}$ galaxies in the COSMOS field in Table~\ref{ffmstab} and we show them in Fig.~\ref{ffmsfig}. We defined five redshift bins between $z_{\rm down} = 0.2$ and $z_{\rm up} = 0.9$ both for minor and major mergers. The ranges $0.3 < z < 0.375$, $0.7 < z < 0.75$ and $0.825 < z < 0.85$ are dominated by Large Scale Structures (LSS, \citealt{kovac10}), so we use these LSS as natural boundaries in our study. This minimises the impact of LSS in our measurements, since the merger fraction depends on environment \citep{lin10,deravel11,pawel12}. We identify a total of 56.2 major mergers and 71.1 minor ones at $0.2 \leq z < 0.9$. Note that the number of mergers can take non integer values because of the weighting scheme used in our methodology (Sect.~\ref{method}). We compare the previous number of mergers (measured as $\Sigma_{k} N_{\rm p}^{k}$, Eq.~[\ref{ncphot2}]) with the total number of close pair systems ($N_{\rm p}$), obtaining that the fraction of real close pairs over the total number of systems is $\sim65$\%. We find that

\begin{itemize}
\item The minor merger fraction is nearly constant with redshift, $f_{\rm mm} \sim 0.051$ . The least-squares fit to the minor merger fraction data is
	\begin{equation}	
		f_{\rm mm} = (0.052 \pm 0.009) (1+z)^{-0.1 \pm 0.3}.
	\end{equation}
	The negative value of the power-law index implies that the minor merger fraction decreases slightly with redshift, but it is consistent with a null evolution ($m_{\rm mm} = 0$). This confirms the trend found by LS11 for bright galaxies, and by \citet{jogee09} and \citet{lotz11} for less massive ($M_{\star} \gtrsim 10^{10}\ M_{\odot}$) galaxies, and extend it to the high mass regime.

\begin{figure}[t!]
\resizebox{\hsize}{!}{\includegraphics{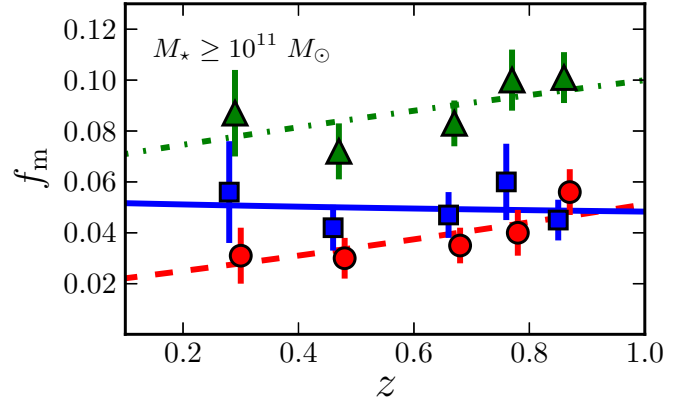}}
\caption{Major (dots), minor (squares) and total (major + minor, triangles) merger fraction of $M_{\star} \geq 10^{11}\ M_{\odot}$ galaxies as a function of redshift in the COSMOS field. Dashed, solid and dott-dashed curves are the least-squares best fit of a power-law function, $f_{\rm m} \propto (1+z)^{m}$, to the major ($m_{\rm MM} = 1.4$), minor ($m_{\rm mm} = -0.1$) and total ($m_{\rm m} = 0.6$) merger fraction data, respectively.}
\label{ffmsfig}
\end{figure}

\item The major merger fraction of massive galaxies increases with redshift as
	\begin{equation}	
		f_{\rm MM} = (0.019 \pm 0.003) (1+z)^{1.4 \pm 0.3}.\label{ffMMeq}
	\end{equation}
	This increase with $z$ contrasts with the nearly constant minor merger fraction. In Fig.~\ref{ffMMfig} we compare our measurements with those from the literature for massive galaxies and for $r_{\rm p}^{\rm max} \sim 30h^{-1}$ kpc close pairs. \citet{deravel11} measure the major merger fraction by $r_{\rm p} \leq 30h^{-1}$ kpc spectroscopic close pairs in the 10k zCOSMOS sample, so their sample is included in ours. Because they assume a different inner radius than us, we apply a factor 2/3 to their original values (see Sect.~\ref{mrcosmos}, for details). Both merger fractions are in good agreement, supporting our methodology. Note that our uncertainties are lower by a factor of three than those in \citet{deravel11} because our principal sample is a factor of four larger than theirs. \citet{xu12} measure the merger fraction from photometric close pairs also in the COSMOS field. They provide the fraction of galaxies in close pairs with $\mu \geq 1/2.5$, so we apply a factor 0.7 to obtain the number of close pairs (this is the fraction of principal galaxies in their massive sample) and a factor 1.6 to estimate the number of $\mu \geq 1/4$ systems (the merger fraction depends on $\mu$ as $f_{\rm m} \propto \mu^{s}$, as shown by LS11, and $s = -0.95$ for massive galaxies in COSMOS, Sect.~\ref{massgrowth}). On the other hand, \citet{bundy09} and \citet{bluck09} measure the major ($\mu \geq 1/4$) merger fraction in GOODS\footnote{http://www.stsci.edu/science/goods/} (Great Observatories Origins Deep Survey, \citealt{giavalisco04}) and Palomar/DEEP2 \citep{powir} surveys, respectively. These studies are also in good agreement with our values, with the point at $z = 0.8$ from \cite{bluck09} being the only discrepancy. The least-squeres fit to all the close pair studies in Fig.~\ref{ffMMfig} yields similar parameters to those from our COSMOS data alone, Eq.~(\ref{ffMMeq}).

\begin{figure}[t!]
\resizebox{\hsize}{!}{\includegraphics{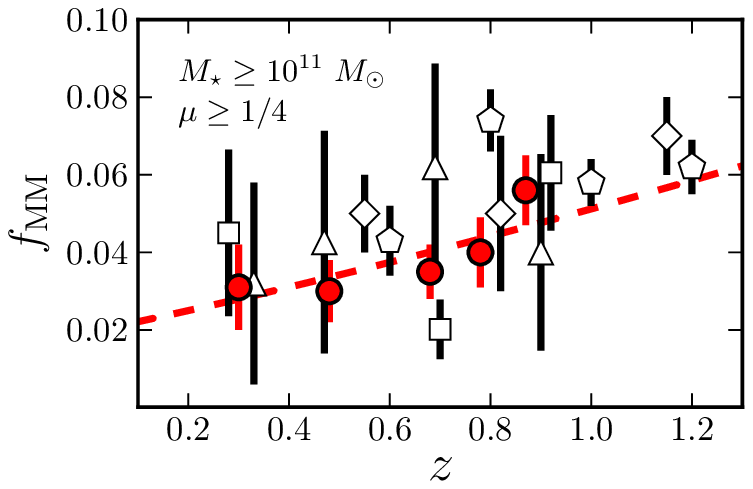}}
\caption{Major ($\mu \geq 1/4$) merger fraction for $M_{\star} \geq 10^{11}\ M_{\odot}$ galaxies from $r_{\rm p}^{\rm max} \sim 30h^{-1}$ kpc close pairs. The dots are from present work, triangles are form \citet{deravel11} in the zCOSMOS 10k sample, squares from \citet{xu12} in the COSMOS field, pentagons from \citet{bluck09} in the Palomar/DEEP2 survey, and diamonds from \citet{bundy09} in the GOODS fields. Some points are slightly shifted when needed to avoid overlap. The dashed line is the least-squares best fit of a power-law function, $f_{\rm MM} \propto (1+z)^{1.4}$, to the major merger fraction data in the present work.}
\label{ffMMfig}
\end{figure}

\begin{figure}[t!]
\resizebox{\hsize}{!}{\includegraphics{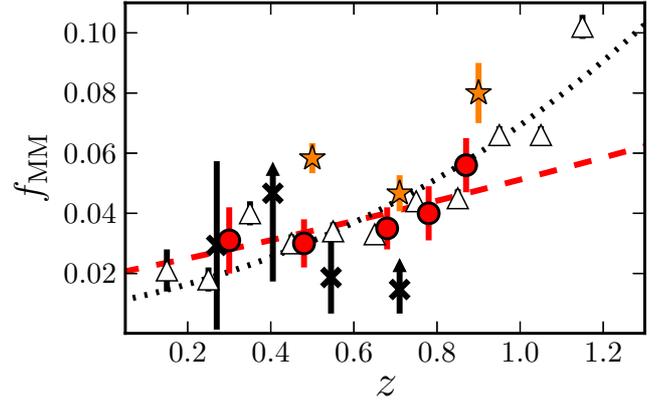}}
\caption{Major merger fraction as a function of redshift. The dots are from present work for $M_{\star} \geq 10^{11}\ M_{\odot}$ galaxies from $10 h^{-1}$ kpc $\leq r_{\rm p}^{\rm max} \leq 30h^{-1}$ kpc close pairs. The triangles are from \citet{kar07} in the COSMOS field for $M_{V} \leq -19.8$ galaxies from $5 h^{-1}$ kpc $\leq r_{\rm p}^{\rm max} \leq 20h^{-1}$ kpc close pairs. The stars are from \citet{bridge10} in the CFHTLS by morphological criteria for $M_{\star} \gtrsim 5 \times 10^{10}\ M_{\odot}$ galaxies, and crosses are from \citet{jogee09} for $M_{\star} \geq 2.5 \times 10^{10}\ M_{\odot}$ galaxies by morphological criteria in GEMS (upward arrows mark those points that are lower limits). The dashed line is the least-squares best fit of a power-law function, $f_{\rm MM} \propto (1+z)^{1.4}$, to the major merger fraction data in the present work. The dotted line is the evolution from \citet{kar07}, $f_{\rm MM} \propto (1+z)^{2.8}$.}
\label{ffMMfig2}
\end{figure}

For completeness, if Fig.~\ref{ffMMfig2} we compare our major merger fractions with other works that are either based on morphological criteria or come from luminosity-selected samples. Regarding morphological studies, \citet{bridge10} provide the major merger fraction of $M_{\star} \gtrsim 5 \times 10^{10}\ M_{\odot}$ galaxies in two CFHTLS\footnote{http://cfht.hawaii.edu/Science/CFHLS/} (Canada-France-Hawaii Telescope Legacy Survey, \citealt{cfhtls}) Deep fields, including the COSMOS field. They perform a visual classification of the sources, finding 286 merging systems of that mass. In their work, \citet{jogee09} estimate a lower limit of the major merger fraction of $M_{\star} \geq 2.5 \times 10^{10}\ M_{\odot}$ galaxies in the GEMS\footnote{http://www.mpia-hd.mpg.de/GEMS/gems.htm} (Galaxy Evolution From Morphology And SEDs, \citealt{rix04}) survey. We cannot compare directly the merger fractions from these two morphological studies with ours because of the different methodologies \citep[e.g.,][]{bridge10,lotz11}. Thus, we translate their merger rates into the expected close pair fraction following the prescriptions in Sect.~\ref{mrcosmos}. Giving the uncertainties in the merger time scales of both methods and the difficulties to assign a precise mass ratio $\mu$ to the merger candidates in morphological studies, the merger fractions from \citet{bridge10} and \citet{jogee09} are in nice agreement with our results.

\citet{kar07} estimate the merger fraction of luminous galaxies ($L_{V} \leq -19.8$) in the COSMOS field. They take these luminous galaxies to define the principal and the companion sample, i.e., they are incomplete for low luminosity major companions near the selection boundary. We find that both studies in the COSMOS field are compatible in the common redshift range ($0.2 < z < 0.9$). The different evolution of the major merger fraction in both works, $m = 2.8$ in \citet{kar07} vs $m = 1.4$ in our study, is due to the $z > 0.9$ data. We conclude that both studies are consistent, even if a direct quantitative comparison is not possible because of the different sample selection and companion definition.

\item The fit to the total merger fraction is
	\begin{equation}	
		f_{\rm m} = (0.067 \pm 0.008) (1+z)^{0.6 \pm 0.3}.
	\end{equation}
This evolution is slower than the major merger one, reflecting the different properties of minor and major mergers. We compare our total merger fractions with others in the literature in Fig.~\ref{fftotalfig}. \citet{marmol12} study the total merger fraction of massive galaxies by $r_{\rm p}^{\rm max} = 70h^{-1}$ kpc close companions. The merger fraction depends on the search radius as $f_{\rm m} \propto r_{\rm p}^{-0.95}$ (LS11), so we translate the merger fractions provided by \citet{marmol12} to our search radius. On the other hand, \citet{newman12} measure the merger fraction of $M_{\star} \geq 5 \times 10^{10}\ M_{\odot}$ galaxies from $r_{\rm p}^{\rm max} = 30h^{-1}$ kpc close pairs. The values from both close pair studies are consistent with ours. Also the results of \citet{williams11} suggest a slow/null evolution in the total ($\mu \geq 1/10$) merger faction of massive galaxies up to $z \sim 2$.

\begin{figure}[t!]
\resizebox{\hsize}{!}{\includegraphics{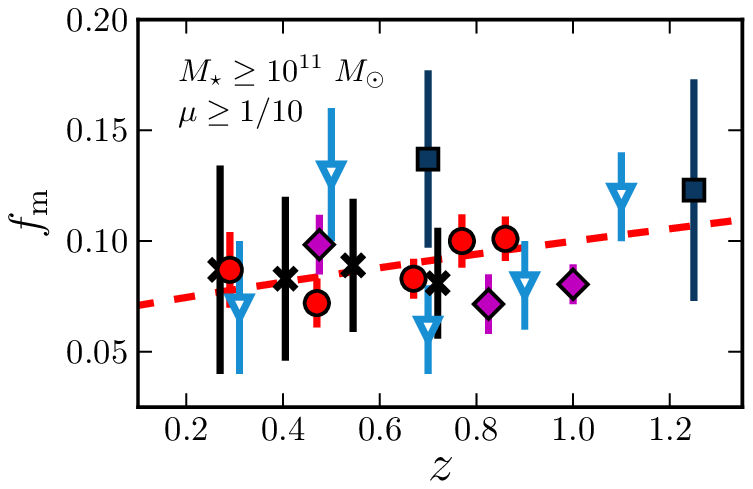}}
\caption{Total (major + minor, $\mu \geq 1/10$) merger fraction as a function of redshift. Dots are from the present work in the COSMOS field for $M_{\star} \geq 10^{11}\ M_{\odot}$ galaxies, diamonds are from \citet{marmol12} for massive galaxies, squares are from \citet{newman12} for $M_{\star} \geq 5 \times 10^{10}\ M_{\odot}$ galaxies, crosses are from \citet{jogee09} for $M_{\star} \geq 2.5 \times 10^{10}\ M_{\odot}$ galaxies by morphological criteria, and inverted triangles are from \citet{lotz11} for $M_{\star} \geq 10^{10}\ M_{\odot}$ galaxies by morphological criteria. The dashed line is the least-squares best fit of a power-law function, $f_{\rm m} \propto (1+z)^{0.6}$, to the total merger fraction data in the present work.}
\label{fftotalfig}
\end{figure}

Regarding morphological studies, \citet{jogee09} estimate the total ($\mu \geq 1/10$) merger fraction of $M_{\star} \geq 2.5 \times 10^{10}\ M_{\odot}$ galaxies in the GEMS survey. Their values, $f_{\rm m} \sim 0.08$, are consistent with ours. We also show the merger fraction from \citet{lotz11} for $M_{\star} \geq 10^{10}\ M_{\odot}$ galaxies in the AEGIS\footnote{http://aegis.ucolick.org/} (All-Wavelength Extended Groth Strip International Survey, \citealt{davis07}) survey. The different methodologies between these works and ours, and the different stellar mass regimes probed, make direct comparisons difficult \citep[see][for a review of this topic]{bridge10,lotz11}. In summary, previous work is compatible with a mild evolution of the total merger fraction, as we observe.
\end{itemize}

\begin{table*}
\caption{Minor, major and total merger fraction of $M_{\star} \geq 10^{11}\ M_{\odot}$ galaxies}
\label{ffmstab}
\begin{center}
\begin{tabular}{lccccc}
\hline\hline\noalign{\smallskip}
Merger fraction  & $z = 0.29$ & $z = 0.46$ & $z = 0.65$ & $z = 0.77$ & $z = 0.86$\\
\noalign{\smallskip}
 & $0.2 \leq z < 0.36$ & $0.36 \leq z < 0.57$ & $0.57 \leq z < 0.73$ & $0.73 \leq z < 0.83$ & $0.83 \leq z < 0.9$\\
\noalign{\smallskip}
\hline
\noalign{\smallskip}
$f_{\rm MM}$ 	& $0.031  \pm 0.011$ & $0.030 \pm 0.008$  & $0.035 \pm 0.007$ & $0.040 \pm 0.009$ & $0.056 \pm 0.009$ \\
$f_{\rm mm}$  	& $0.056  \pm 0.015$ & $0.042 \pm 0.009$  & $0.047 \pm 0.008$ & $0.060 \pm 0.010$ & $0.045 \pm 0.008$ \\
$f_{\rm m}$  	& $0.087  \pm 0.017$ & $0.072 \pm 0.011$  & $0.083 \pm 0.009$ & $0.100 \pm 0.012$ & $0.101 \pm 0.010$ \\
\noalign{\smallskip}
\hline
\end{tabular}
\end{center}
\end{table*}

\subsection{The merger fraction of ETGs}\label{ffmor}
\begin{figure}[t!]
\resizebox{\hsize}{!}{\includegraphics{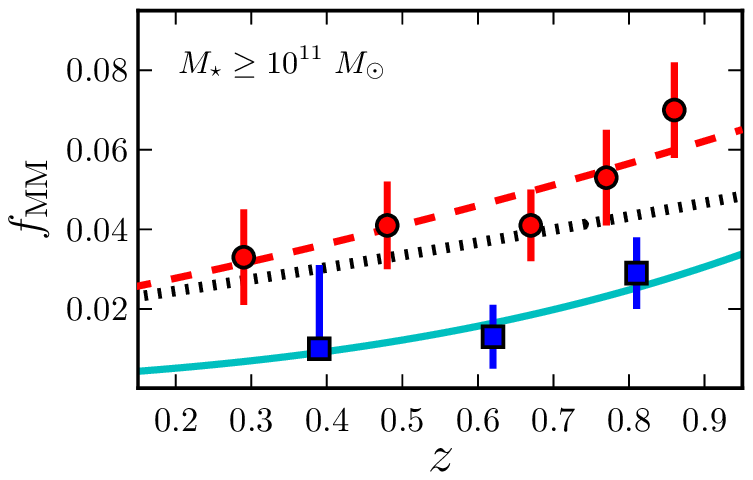}}
\resizebox{\hsize}{!}{\includegraphics{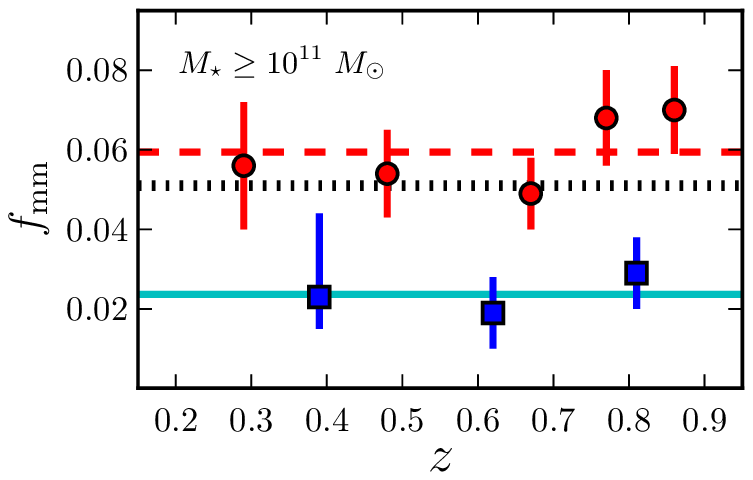}}
\caption{Major (upper penel) and minor (lower panel) merger fractions of $M_{\star} \geq 10^{11}\ M_{\odot}$ galaxies as a function of redshift and morphology. Dots are for ETGs, while squares are for LTGs. Dashed (solid) lines are the best fit to the ETGs (LTGs) data, while dotted lines are the fits for the global population.}
\label{ffmorfig}
\end{figure}

We summarise the minor and major merger fractions for both massive ($M_{\star} \geq 10^{11}\ M_{\odot}$) ETGs and LTGs in the COSMOS field in Tables~\ref{ffettab} and \ref{fflttab}, respectively, while we show them in Fig.~\ref{ffmorfig}. We defined five redshift bins between $z_{\rm down} = 0.2$ and $z_{\rm up} = 0.9$ for ETGs, as for the global population, but only three in the case of LTGs because of the lower number of principal sources. We do not split the companion sample by neither morphology or colour in this section, and we study the properties of the companion galaxies in Sect.~\ref{colsec}.

We assume $m_{\rm mm} = 0$ in the following for the minor merger fraction, as for the global population (Sect.~\ref{ffms}). The mean minor merger fraction of ETGs is $f_{\rm mm}^{\rm ET} = 0.060$, while $f_{\rm mm}^{\rm LT} = 0.023$ for LTGs. There is therefore a factor of three difference between the merger fractions of early type and late type populations. LS11 also find a similar result when comparing the minor merger fraction of red and blue bright galaxies.

On the other hand, the major merger fraction of ETGs is also higher than that of LTGs by a factor of two. The fit to the major merger data yields 
\begin{eqnarray}
f_{\rm MM}^{\rm ET} = (0.020 \pm 0.003) (1+z)^{1.8 \pm 0.3},\\
f_{\rm MM}^{\rm LT} \sim 0.003 (1+z)^{4}.
\end{eqnarray}
Because we only have three data points for LTGs and of the high uncertainty in the first redshift bin, the reported value of $m_{\rm MM}$ for massive LTGs is only tentative. Nevertheless, that the major merger fraction of LTGs evolves faster than that of ETGs is in agreement with previous studies which compare early-types/red and late-types/blue galaxies (e.g., \citealt{lin08,deravel09,bundy09,chou10}; LS11). 

As shown by \citet{lotz11}, the merger rate evolution depends on the selection of the sample, with samples selected to prove a constant number density population over cosmic time showing a faster evolution ($m \sim 3$) than those with a constant mass selection ($m \sim 1.5$). To check the impact of the selection in the merger fraction of ETGs, we computed the major and minor merger fraction of ETGs with $\log\,(M_{\star}/M_{\odot}) \geq 11.15 - 0.15z$ ($n$-selected sample, in the following). As shown by \citet{vandokkum10}, this provides a nearly constant number-density selection for massive galaxies. We find that the major and minor merger fractions from the $n$-selected sample are compatible with those from the mass-selected sample. Regarding their evolution, the major merger fraction evolves faster in the $n$-selected sample, $m = 2.5 \pm 0.4$, that in the mass-selected sample, $m = 1.8 \pm 0.3$, as expected. The minor merger fraction remains the same, $f_{\rm mm} = 0.064 \pm 0.006$ ($n$-selected sample) vs $f_{\rm mm} = 0.060 \pm 0.008$ (mass-selected sample). In addition, we checked that the results presented in Sect.~\ref{discussion} remain the same when we use the merger fractions from the $n-$selected sample instead of those from the mass-selected one. Therefore, we conclude that the selection of the massive ETGs sample has limited impact in our results.

In summary, the merger fraction of massive ($M_{\star} \geq 10^{11}\ M_{\odot}$) ETGs, both major and minor, is higher by a factor of 2-3 than that of massive LTGs \citep[see also][for a similar result]{marmol12}. We estimate the merger rate of ETGs in Sect.~\ref{mrcosmos}.

\begin{table*}
\caption{Minor and major merger fraction of ETGs with $M_{\star} \geq 10^{11}\ M_{\odot}$}
\label{ffettab}
\begin{center}
\begin{tabular}{lccccc}
\hline\hline\noalign{\smallskip}
Merger fraction  & $z = 0.29$ & $z = 0.48$ & $z = 0.67$ & $z = 0.77$ & $z = 0.86$\\
\noalign{\smallskip}
 & $0.2 \leq z < 0.36$ & $0.36 \leq z < 0.57$ & $0.57 \leq z < 0.73$ & $0.73 \leq z < 0.83$ & $0.83 \leq z < 0.9$\\
\noalign{\smallskip}
\hline
\noalign{\smallskip}
$f_{\rm MM}^{\rm ET}$ 	& $0.033  \pm 0.012$ & $0.041 \pm 0.011$  & $0.041 \pm 0.009$ & $0.053 \pm 0.012$ & $0.070 \pm 0.012$ \\
$f_{\rm mm}^{\rm ET}$  & $0.056  \pm 0.016$ & $0.054 \pm 0.011$  & $0.049 \pm 0.009$ & $0.068 \pm 0.013$ & $0.070 \pm 0.011$ \\
\noalign{\smallskip}
\hline
\end{tabular}
\end{center}
\end{table*}

\begin{table*}
\caption{Minor and major merger fraction of LTGs with $M_{\star} \geq 10^{11}\ M_{\odot}$}
\label{fflttab}
\begin{center}
\begin{tabular}{lccc}
\hline\hline\noalign{\smallskip}
Merger fraction  & $z = 0.39$ & $z = 0.62$ & $z = 0.81$\\
\noalign{\smallskip}
 & $0.2 \leq z < 0.5$ & $0.5 \leq z < 0.7$ & $0.7 \leq z < 0.9$\\
\noalign{\smallskip}
\hline
\noalign{\smallskip}
$f_{\rm MM}^{\rm LT}$ 		& $0.010^{+0.021}_{-0.003}$ & $0.013 \pm 0.008$	& $0.029 \pm 0.009$ \\
$f_{\rm mm}^{\rm LT}$  		& $0.023^{+0.021}_{-0.008}$ & $0.019 \pm 0.009$  & $0.029 \pm 0.009$ \\
\noalign{\smallskip}
\hline
\end{tabular}
\end{center}
\end{table*}

\subsection{Colour properties of companion galaxies}\label{colsec}
In this section we attempt to identify the types of galaxies in the companion population. As the morphological classification is not reliable for all companions because they are faint, we instead use a colour selection. We took as red (quiescent) companions those with SED (rest-frame, dust reddening corrected) colour $NUV-r^{+} \geq 3.5$, while as blue (star-forming) those with $NUV-r^{+} < 3.5$ \citep[see][for details]{ilbert10}, and we measured the fraction of red companions ($f_{\rm red}$) of massive galaxies at $0.2 \leq z < 0.9$. 

We find that $62$\% of the companions of the whole principal sample are red, while $\sim38$\% are blue. Furthermore, the red fraction remains nearly the same for minor ($f_{\rm red} = 60$\%) and major ($f_{\rm red} = 64$\%) companions. When we repeated the previous study focusing on massive ETGs as principals, we find $f_{\rm red} \sim 65$\%, both for minor and major companions. Because $\sim95$\% of our massive ETGs are also red, most of the ETG close pairs are ''dry'' (i.e., red - red).

\section{The merger rate of massive ETGs in the COSMOS field}\label{mrcosmos}
In this section we estimate the minor ($R_{\rm mm}$) and major ($R_{\rm MM}$) merger rate, defined as the number of mergers per galaxy and Gyr, of massive ETGs. We recall here the steps to compute the merger rate from the merger fraction, focusing first on the major merger rate.

Following \citet{deravel09}, we define the major merger rate as
\begin{equation}
R_{\rm MM} = f_{\rm MM}\,C_{\rm p}\, C_{\rm m}\, T_{\rm MM}^{-1},\label{mrparMM}
\end{equation}
where the factor $C_{\rm p}$ takes into account the lost companions in the inner $10h^{-1}$ kpc \citep{bell06} and 
the factor $C_{\rm m}$ is the fraction of the observed close pairs that finally merge in a typical time scale $T_{\rm MM}$. 
We take $C_{\rm p} = 3/2$. The typical merger time scale depends 
on $r_{\rm p}^{\rm max}$ and can be estimated by cosmological and $N$-body simulations. In our case, we compute 
the major merger time scale from the cosmological simulations of \citet{kit08}, based on the Millennium simulation \citep{springel05}. This major merger time scale refers to major mergers ($\mu > 1/4$ in stellar mass), and depends mainly on $r_{\rm p}^{\rm max}$ and on the stellar mass of the principal galaxy, 
with a weak dependence on redshift in our range of interest (see \citealt{deravel09}, for details). 
Taking $\log\, (M_{\star}/M_{\odot}) = 11.2$ as the average stellar mass of our principal galaxies with a close companion, we obtain $T_{\rm MM} = 1.0\pm0.2$ Gyr for $r_{\rm p}^{\rm max} = 30h^{-1}$ kpc and $\Delta v^{\rm max} = 500$ km s$^{-1}$. We assumed an uncertainty of 0.2 dex in the average mass of the principal galaxies to estimate the error in $T_{\rm MM}$. This time scale already includes the factor $C_{\rm m}$ \citep[see][LS11]{patton08,bundy09,lin10}, so we take $C_{\rm m} = 1$ in the following. In addition, LS11 show that time scales from \citet{kit08} are equivalent to those from the $N-$body/hydrodynamical simulations by \citet{lotz10t}, and that they account properly for the observed increase of the merger fraction with $r_{\rm p}^{\rm max}$ \citep[see also][]{deravel09}. We stress that these merger time scales have an additional factor of two uncertainty in their normalisation \citep[e.g.,][]{hopkins10mer,lotz11}.

The minor merger rate is
\begin{equation}
R_{\rm mm} = f_{\rm mm}\,C_{\rm p}\, C_{\rm m}\, T_{\rm mm}^{-1},\label{mrparmm}
\end{equation}
where $T_{\rm mm} = \Upsilon \times T_{\rm MM}$. Following LS11, we take $\Upsilon = 1.5\pm0.1$ from the $N-$body/hydrodynamical simulations of major and minor mergers performed by \citet[][see also \citealt{lotz11}]{lotz10t,lotz10gas}. As for major mergers, we assume $C_{\rm p} = 3/2$ and $C_{\rm m} = 1$.

\begin{figure}[t!]
\resizebox{\hsize}{!}{\includegraphics{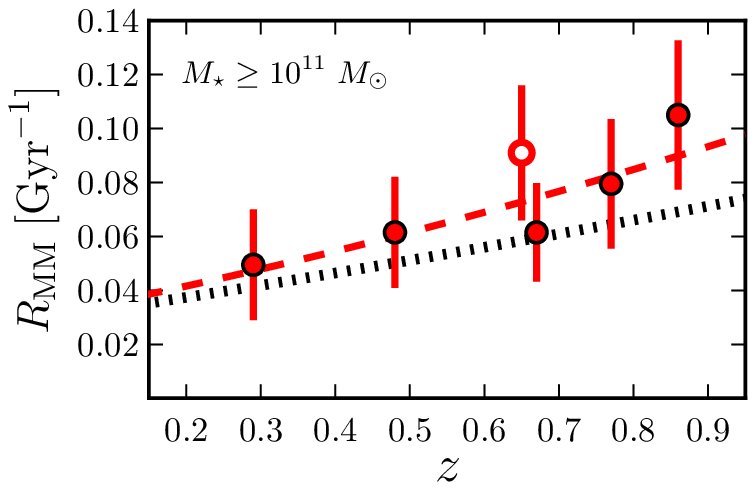}}
\resizebox{\hsize}{!}{\includegraphics{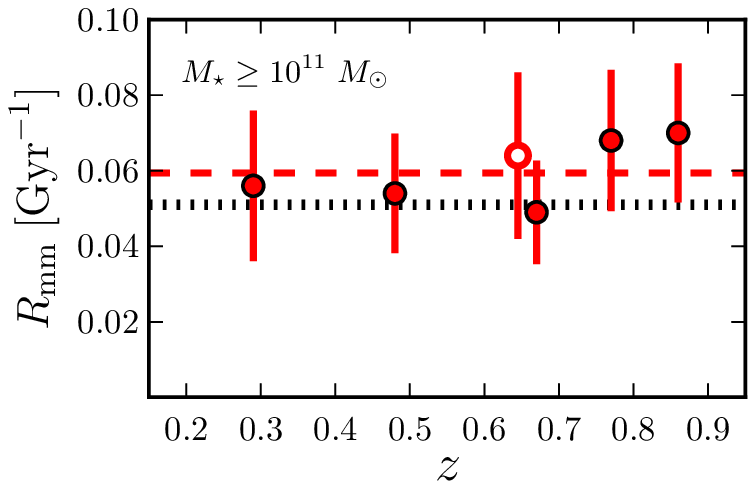}}
\caption{Major (upper panel) and minor (lower panel) merger rate of $M_{\star} \geq 10^{11}\ M_{\odot}$ ETGs as a function of redshift. Filled symbols are from the present work, while open ones are from LS11 in VVDS-Deep for red galaxies. Dashed lines are the best fit to the ETGs data, while dotted lines are the fits for the global population.}
\label{mrmorfig}
\end{figure}

We summarise the major and minor merger rates of massive ETGs in Table~\ref{mrettab}, and show them in Fig.~\ref{mrmorfig}. We parametrise their redshift evolution as
\begin{equation}
R_{\rm m}\,(z) = R_{\rm m,0}\,(1+z)^{n}. \label{mrzeq}
\end{equation}

\begin{table*}
\caption{Minor and major merger rate of ETGs with $M_{\star} \geq 10^{11}\ M_{\odot}$}
\label{mrettab}
\begin{center}
\begin{tabular}{lccccc}
\hline\hline\noalign{\smallskip}
Merger rate  & $z = 0.29$ & $z = 0.48$ & $z = 0.67$ & $z = 0.77$ & $z = 0.86$\\
\noalign{\smallskip}
(Gyr$^{-1}$)  & $0.2 \leq z < 0.36$ & $0.36 \leq z < 0.57$ & $0.57 \leq z < 0.73$ & $0.73 \leq z < 0.83$ & $0.83 \leq z < 0.9$\\
\noalign{\smallskip}
\hline
\noalign{\smallskip}
$R_{\rm MM}^{\rm ET}$ 	& $0.049  \pm 0.020$ & $0.061 \pm 0.021$  & $0.062 \pm 0.018$ & $0.080 \pm 0.024$ & $0.105 \pm 0.028$ \\
$R_{\rm mm}^{\rm ET}$  & $0.056  \pm 0.020$ & $0.054 \pm 0.016$  & $0.049 \pm 0.014$ & $0.068 \pm 0.019$ & $0.070 \pm 0.018$ \\
\noalign{\smallskip}
\hline
\end{tabular}
\end{center}
\end{table*}

Assuming $n_{\rm mm} = 0$ for minor mergers, as for the merger fraction (Sect.~\ref{ffmor}), we find $R_{\rm mm}^{\rm ET} = 0.060 \pm 0.008\ {\rm Gyr}^{-1}$. The fit to the major merger rate of massive ETGs is
\begin{equation}
R_{\rm MM}^{\rm ET} = (0.030 \pm 0.006)\,(1+z)^{1.8 \pm 0.3}\ {\rm Gyr}^{-1}.\label{rMMet}\\
\end{equation}
Our results imply that the minor merger rate is higher than the major merger one at $z \lesssim 0.5$. In addition, the minor and major merger rates of massive ETGs are $\sim20$\% higher than for the global population.

In Fig.~\ref{mrmorfig} we also show the minor and major merger rates of red bright galaxies measured by LS11. We find that red galaxies have similar merger rates, both minor and major, than our massive ETGs. This suggests that massive red sequence galaxies have similar merger properties: nearly 95\% of our ETGs are red, while the mean mass of the red galaxies in LS11 is $\overline{M_{\star}}_{\rm ,red} \sim 10^{10.8}\ M_{\odot}$, a factor of two less massive than our ETGs, $\overline{M_{\star}}_{\rm ,ET} \sim 10^{11.2}\ M_{\odot}$. The study of the merger properties of the red sequence galaxies as a function of stellar mass is beyond the scope of this paper and we explore this issue in a future work.

\section{The role of mergers in the evolution of massive ETGs since $z = 1$}\label{discussion}
In this section we use the previous merger rates to estimate the number of minor and major mergers per massive ($M_{\star} \geq 10^{11}\ M_{\odot}$) ETG since $z = 1$ (Sect.~\ref{nmerger}) and the impact of mergers in the mass growth (Sect.~\ref{massgrowth}) and size evolution (Sect.~\ref{sizegrowth}) of ETGs in the last $\sim$8 Gyr.

\subsection{Number of minor mergers since $z = 1$}\label{nmerger}
We can obtain the average number of minor mergers per ETG between $z_2$ and $z_1 < z_2$ as
\begin{equation}
N_{\rm mm}^{\rm ET}(z_1,z_2) = \int_{z_1}^{z_2} \frac{R_{\rm mm}^{\rm ET}\,{\rm d}z}{(1+z)H_0E(z)},\label{numm}
\end{equation}
where $E(z) = \sqrt{\Omega_{\Lambda} + \Omega_{m}(1+z)^3}$ in a flat universe. 
The definition of $N_{\rm MM}^{\rm ET}$ for major mergers is analogous. Using the merger rates in previous section, we obtain $N_{\rm m}^{\rm ET} = 0.89\pm0.14$, with $N_{\rm MM}^{\rm ET} = 0.43\pm0.13$ and $N_{\rm mm}^{\rm ET} = 0.46\pm0.06$ between $z = 1$ and $z = 0$. The number of minor mergers per massive ETGs since $z = 1$ is therefore similar to the number of major ones. Note that these values and those reported in the following have an additional factor of two uncertainty due to the uncertainty on the merger time scales derived from simulations (Sect.~\ref{mrcosmos}).

The number of major mergers per red bright galaxy measured by LS11 is $N_{\rm MM}^{\rm red} = 0.7 \pm 0.2$, higher than our measurement, while the number of minor mergers is similar, $N_{\rm mm}^{\rm red} = 0.5 \pm 0.2$. The discrepancy in the major merger case can be explained by the evolution of the merger rate in both studies, since LS11 assumed $n_{\rm MM}^{\rm red} = 0$ and we measure $n_{\rm MM}^{\rm ET} = 1.8$.

On the other hand, LTGs have a significantly lower number of mergers, $N_{\rm m}^{\rm LT} \sim 0.35$, with $N_{\rm MM}^{\rm LT} \sim 0.15$ and $N_{\rm mm}^{\rm LT} \sim 0.20$. We refer the reader to LS11 for the discussion about the role of major and minor mergers in the evolution of LTGs. In their work, \citet{pozzetti10} find that almost all the evolution in the stellar mass function since $z \sim 1$ is a consequence of the observed star formation \citep[see also][]{vergani08}, and estimate that $N_{\rm m} \sim 0.7$ mergers since $z \sim 1$ per $\log\,(M_{\star}/M_{\odot}) \sim 10.6$ galaxy are needed to explain the remaining evolution. Their result is similar to our direct estimation for the global massive population (ETGs + LTGs), $N_{\rm m} = 0.75 \pm 0.14$, but they infer $N_{\rm MM} < 0.2$. This value is half of ours, $N_{\rm MM} = 0.36 \pm 0.13$, pointing out that close pair studies are needed to understand accurately the role of major/minor mergers in galaxy evolution.

\subsection{Mass assembled through mergers since $z = 1$}\label{massgrowth}
Following LS11, we estimate the mass assembled due to mergers by weighting the number of mergers in the previous section with the average major ($\overline{\mu}_{\rm MM}$) and minor merger ($\overline{\mu}_{\rm mm}$) mass ratio, 
\begin{equation}
\delta M_{\star}(z) \equiv \frac{M_{\star}(0)}{M_{\star}(z)} - 1 = \overline{\mu}_{\rm MM} N^{\rm ET}_{\rm MM}(0,z) + \overline{\mu}_{\rm mm} N^{\rm ET}_{\rm mm}(0,z).\label{deltamass}
\end{equation}
To obtain the average mass ratios we measured the merger fraction of massive ETGs at $0.2 \leq z < 0.9$ for different values of $\mu$, from $\mu = 1/2$ to $1/10$. Then, we fitted to the data a power-law, $f_{\rm m}\,(\geq \mu) \propto \mu^{s}$, and used the prescription in LS11 to estimate the average merger mass ratio from the value of the power-law index $s$. Following those steps we find $s = -0.95$ for massive ETGs in COSMOS, while the average merger mass ratios are $\overline{\mu}_{\rm MM} = 0.48$ and $\overline{\mu}_{\rm mm} = 0.15$, similar to those values reported by LS11. With all previous results we obtain that {\it mergers with $\mu \geq 1/10$ increase the stellar mass of massive ETGs by $\delta M_{\star} = 28 \pm 8$\% since $z = 1$}. LS11 find $\delta M_{\star}(1) = 40 \pm 10$\% for red bright galaxies in VVDS-Deep, consistent with our measurement within errors. We note that they use $B-$band luminosity as a proxy of stellar mass, so their value is an upper limit due to the lower mass-to-light ratio of blue companions. \citet{bluck12} study the major and minor ($\mu \geq 1/100$) merger fraction of massive galaxies at $1.7 < z < 3$ in GNS\footnote{http://www.nottingham.ac.uk/astronomy/gns/} (GOODS NICMOS Survey, \citealt{conselice11}). They extrapolate their results to lower redshifts, estimating $\delta M_{\star}(1) = 30 \pm 25$\% for $\mu \geq 1/10$ mergers. Their value is in good agreement with our measurement, but its large uncertainty prevents a quantitative comparison.

The relative contribution of major/minor mergers to our inferred mass growth is 75\%/25\% because the average major merger is three times more massive than the average minor one, as already pointed out by LS11. In their cosmological model, \citet{hopkins10fusbul} predict that the relative contribution of major and minor mergers in the spheroids assembly of $\log\,(M_{\star}/M_{\odot}) \sim 11.2$ galaxies is $\sim80\%$/20\%, in good agreement with our observational result.

On the other hand, several authors have studied luminosity functions and clustering to constrain the evolution of luminous red galaxies (LRGs) with redshift, finding that LRGs have increased their mass $\delta M_{\star} \sim30$\%$-50$\% by merging since $z = 1$ \citep{brown07,brown08,cool08}. Their results are similar to our direct estimation, but we must take this agreement with caution. \citet{tal12} show that LRGs have a lack of major companions, excluding major mergers as an important growth channel \citep[see also][]{depropris10}. Typically LRGs have $L \gtrsim 3L^*$, and a low impact of major mergers in this systems is indeed expected by cosmological models, where the contribution of major mergers in galaxy mass assembly peaks at $\sim M_{\star}^{*}$ \citep{khochfar09,hopkins10fusbul,cattaneo11}. Thus, even if the values of $\delta M_{\star}$ are similar for LRGs and our massive galaxies, they could have a different origin. A better approach to estimate indirectly the impact of mergers in mass growth is to study the evolution of massive red galaxies at a fixed number density: because they are red (i.e., they have low star formation), their mass is expected to grow only by merging. Following this approach, \citet{vandokkum10} and \citet{brammer11} estimate $\delta M_{\star}(1) \sim 40$\% for massive galaxies in the NEWFIRM Medium-Band Survey\footnote{http://www.astro.yale.edu/nmbs/Overview.html} \citep{vandokkum09}. Their result represents the integral over all possible $\mu$ values, so in combination with our $\delta M_{\star}(1) \sim 30$\% for $\mu \geq 1/10$, this would imply that (i) $\mu \geq 1/10$ mergers dominates the mass assembly of massive galaxies since $z = 1$ and (ii) there is room for an extra $\delta M_{\star} \sim 10$\% growth due to very minor mergers ($\mu < 1/10$).

\subsection{Size growth due to mergers since $z = 1$}\label{sizegrowth}
\begin{figure}[t!]
\resizebox{\hsize}{!}{\includegraphics{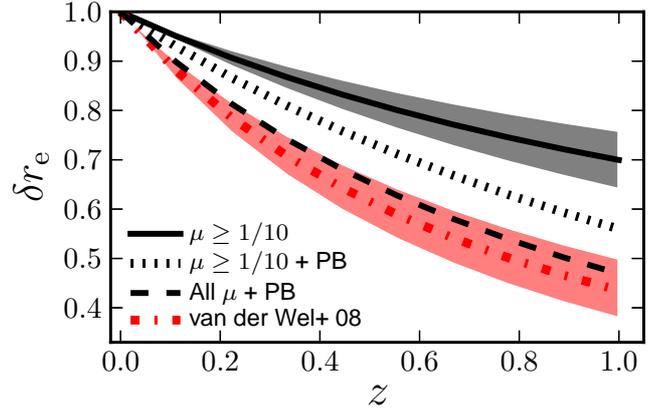}}
\caption{Effective radius normalised to its local value, $\delta r_{\rm e}$, as a function of redshift. The dott-dashed line is the observational evolution from \citet{vanderwel08esize}, $\delta r_{\rm e} = (1+z)^{-1.2}$. The solid line is the evolution due to major and minor mergers ($\mu \geq 1/10$) expected from our results. The shaded areas in both cases mark the 68\% confidence interval. The dotted line is the expected evolution when the progenitor bias (PB) is taken into account. The dashed line is the expected evolution when PB and very minor mergers ($\mu < 1/10$) are included (see text for details).}
\label{sizemodelfig}
\end{figure}

Since the first results of \citet{daddi05} and \citet{trujillo06}, several authors have studied in details the size evolution of massive ETGs with cosmic time. It is now well established that ETGs were smaller, on average, than their local counterparts of a given stellar mass by a factor of two at $z = 1$ and of four at $z = 2$ (Sect.~\ref{intro}). The size evolution is usually parametrised as
\begin{equation}
\delta r_{\rm e}\,(z) \equiv \frac{r_{\rm e}\,(z)}{r_{\rm e}(0)} = (1 + z)^{-\alpha},
\end{equation}
where $r_{\rm e}$ is the effective radius of the galaxy. Despite of all observational efforts, the value of $\alpha$ is still in debate, spanning the range $\alpha = 0.9 - 1.5$ (see references in Sect.~\ref{intro}), as well as its dependency on stellar mass (massive galaxies evolve faster, \citealt{williams10}, or not, \citealt{damjanov11}). In the following we assume as fiducial $\alpha$ value the value reported by \citet{vanderwel08esize} from a combination of several analysis, $\alpha = 1.2$ ($\delta r_{\rm e} = 0.43$ at $z = 1$), with an uncertainty of 0.2 (dott-dashed line in Fig.~\ref{sizemodelfig}).

Two main effects could explain the size evolution of ETGs: the progenitor bias and genuine size growth. The number density of massive (red) galaxies at $z = 2$ is $\sim 15-30$\% of that in the local universe \citep[e.g.,][]{arnouts07,pgon08,williams10,ilbert10}, and those ETGs that have reached the red sequence at later times are systematically more extended than those which did it at high redshift. This effect is called the {\it progenitor bias} and mimic a size growth \citep[see][for further datails]{vanderwel09, vale10a, vale10b, cassata11}. Both \citet{vanderwel09} and \citet{saglia10} estimate that the progenitor bias of massive ETGs accounts for a factor 1.25 ($\delta r_{\rm e} = 0.8$) of the size evolution since $z = 1$, and we assume this value in the following.

Regarding size growth, several authors have suggested that compact galaxies at $z \sim 2$ are the cores of present day massive ellipticals, and that they increase their size by adding stellar mass in the outskirts of the compact high redshift galaxy \citep{bezanson09,hopkins09core,vandokkum10,weinzirl11}. The fact that the more compact galaxies at $z \sim 0.1$ \citep{trujillo09} and $z \sim 1$ \citep{martinez11} have similar young ages ($\sim 1-2$ Gyr), combined with their paucity in the local universe \citep{trujillo09,taylor10,cassata11} also support the size evolution of these systems along cosmic time. Mergers, specially the minor ones, have been proposed to explain this evolution \citep[e.g.,][]{naab09,bezanson09,hopkins10size,weinzirl11}. Adiabatic expansion due to AGN activity \citep{fan10} or stellar evolution \citep{damjanov09} could also play a role. Thanks to our direct measurements of the minor and major merger rate of massive ETGs, we are able to explore the contribution of mergers to the size growth of these galaxies in the last $\sim 8$ Gyr.

Theory and simulations show that equal-mass mergers between two spheroidal galaxies are less effective in increasing the size of ETGs than a major/minor merger with a less dense galaxy, both spiral and spheroidal. In the first case the increase in size is proportional to the accreted mass, $r_{\rm e} \propto M_{\star}^{\beta}$, with $\beta = 1$, while in the second case the index $\beta$ is higher and spans a wide range, $\beta \sim 1.5-2.5$ \citep[e.g.,][]{bezanson09,hopkins10size}. In our case, we estimate $\beta$ for a given $\mu$ from the relation between the initial ($r_{\rm e,i}$) and the final effective radius ($r_{\rm e,f}$) of an ETG in a merger process derived by \citet{fan10},
\begin{equation}
\frac{r_{\rm e,f}}{r_{\rm e,i}} = \frac{(1 + \mu)^2}{1 + \mu^{2 - \epsilon}} = (1 + \mu)^{\beta},\label{req}
\end{equation}
where $\epsilon$ is the slope of the stellar mass vs size relation. In their work, \citet{damjanov11} find $\epsilon = 0.51$ for early-type galaxies in the range $0.2 < z < 2$ \citep[see also][]{williams10,newman12}, similar to the $\epsilon = 0.56$ from \citet{shen03} in SDSS\footnote{http://www.sdss.org/} (Sloan Digital Sky Survey, \citealt{sdssdr7}) or the $\epsilon \sim 0.5$ expected from the Faber-Jackson relation \citep{faber76}. However, not all the observed mergers are between two early-type galaxies. Using colour as a proxy for the morphology of our companion galaxies, we find that 65\% of the mergers are "dry" (red - red), while 35\% are "mixed" (red - blue), for both major and minor mergers (Sect.~\ref{colsec}). In the mixed case we use $\epsilon = 0.27$, a value estimated from the data of \citet{shen03} for late-type galaxies in our mass range of interest. Finally, we obtain the $\beta$ for a given $\mu$ as $0.65\beta_{\rm dry} + 0.35\beta_{\rm mixed}$. Using the average mass ratios $\overline{\mu}_{\rm MM}$ and $\overline{\mu}_{\rm mm}$ in Sect.~\ref{massgrowth}, we find $\beta_{\rm MM} = 1.30$ for major mergers and $\beta_{\rm mm} = 1.65$ for minor ones.

Following Eq.~(\ref{deltamass}), we trace the mass growth of massive ETGs with redshift for both minor, $\delta M_{\star,{\rm mm}}(z)$, and major mergers, $\delta M_{\star,{\rm MM}}(z)$. Then, we translate these mass growths to a size growth with the previous values of $\beta$,
\begin{equation}
\delta r_{\rm e}^{\rm m}\,(z) = [1 + \delta M_{\star,{\rm MM}}(z)]^{-\beta_{\rm MM}} \times [1 + \delta M_{\star,{\rm mm}}(z)]^{-\beta_{\rm mm}}. 
\end{equation}
Finally, we estimate the contribution of mergers to the total size evolution since $z = 1$ as
\begin{equation}
\Delta r_{\rm e} = \frac{1 - \delta r_{\rm e}^{\rm m}(1)}{1 - \delta r_{\rm e}(1)}.
\end{equation}

This model yields a size evolution due to mergers of $\delta r_{\rm e}^{\rm m}(1) = 0.70$ ($\alpha = 0.52 \pm 0.12$, solid line in Fig.~\ref{sizemodelfig}). This implies that {\it observed major and minor mergers can explain $\Delta r_{\rm e}\sim55$\% of the size evolution in massive ETGs since $z \sim 1$}. In the following, all quoted $\Delta r_{\rm e}$ have a typical $\sim15$\% uncertainty due to the errors in the merger rates and in the observed size evolution. 

We take into account the progenitor bias by applying a linear function to the previous size growth due to mergers (dotted line in Fig.~\ref{sizemodelfig}), 
\begin{equation}
\delta r_{\rm e}^{\rm PB}\,(z) = (1 - 0.2z) \times \delta r_{\rm e}^{\rm m}\,(z). 
\end{equation}
We obtain $\delta r_{\rm e}^{\rm PB}(1) = 0.56$ ($\alpha = 0.84 \pm 0.12$), thus explaining $\Delta r_{\rm e}\sim75$\% of the size evolution with our current observations. We note that this value is similar to the $\delta r_{\rm e}(1) = 0.63$ estimated by the simple model of \citet{vanderwel09}, which only includes the progenitor bias and a merger prescription from cosmological simulations. The remaining $\Delta r_{\rm e}\sim25$\% of the evolution should be explained by other physical processes (e.g., very minor mergers with $\mu < 1/10$ or adiabatic expansion) or by systematic errors in the measurements (e.g., lower merger time scales or an overestimation of the size evolution). We explore these processes/systematics in the following.

\begin{itemize}
\item {\bf Very minor mergers} ($\mu < 1/10$). Cosmological simulations find that $\mu \geq 1/10$ mergers are not the more common ones, with the merger history of massive galaxies being dominated by $\mu < 1/10$ mergers \citep{shankar10,jimenez11,oser12}. However, in this simulations the mass accretion is dominated by $\mu \geq 10$ events due to the low mass of the very minor companions. As we show in Sect.~\ref{massgrowth}, a mass growth of $\delta M_{\star} \sim 10$\% due to very minor mergers since $z = 1$ is compatible with the observed mass assembly of massive red galaxies \citep{vandokkum10,brammer11}. This translates to $N_{\rm vm}\sim4$ very minor mergers per massive ETG since $z \sim 1$ (we assumed that very minor mergers have $1/100 \leq \mu < 1/10$ and estimated that $\overline{\mu}_{\rm vm} \sim 0.025 = 1/40$ following the prescriptions in Sect.~\ref{massgrowth}). Note that we can increase arbitrarily the number of very minor mergers by lowering $\overline{\mu}_{\rm vm}$, but not their contribution to the mass growth, which is fixed. We checked that the conclusions in this section are independent of $\overline{\mu}_{\rm vm}$.

We estimate $\beta_{\rm vm} = 1.85$ for very minor mergers, thus obtaining an extra size growth of $\Delta r_{\rm e}\sim20$\% due to mergers, $\delta r_{\rm e}^{\rm m}(1) = 0.58$ and $\alpha = 0.78 \pm 0.12$ when all $\mu$ values are taken into account. Hence, mergers since $z \sim 1$ may explain $\Delta r_{\rm e}\sim 75$\% of the observed size evolution, while $\Delta r_{\rm e}\sim 95$\%, with $\delta r_{\rm e}^{\rm PB}(1) = 0.47$ and $\alpha = 1.1$, when the progenitor bias is taken into account (dashed line in Fig.~\ref{sizemodelfig}). In this picture, nearly half of the evolution due to mergers is related to minor ($\mu < 1/4$) events. This result reinforces our conclusion that mergers are the main contributors to the size evolution of massive ETGs since $z = 1$, but observational estimations of the very minor merger rate ($\mu < 1/10$) are needed to constraint their role. As a first attempt, \citet{marmol12} find that the merger fraction of massive galaxies at $z \lesssim 1$ for $\mu \geq 1/100$ satellites is two times that of $\mu \geq 1/10$ satellites. That suggests $N_{\rm vm}\sim1$, and an additional contribution for even smaller satellites ($\mu < 1/100$) could be possible.

\item {\bf Adiabatic expansion}. This will occur in a relaxed system that is losing mass. As mass is lost the potential becomes shallower, so the system expands into a new stable equilibrium. The amount that a system expands depends on both the ejected mass ($M_{\star,{\rm eject}}$) and on the time scale of the process ($T_{\rm eject}$). \citet{fan08,fan10} suggest adiabatic expansion due to quasar activity and/or supernova winds as an alternative process to explain the size growth of massive early-types, specially at $z \gtrsim 1$. These processes occur on very short time scales after the formation of the spheroid ($T_{\rm eject} \lesssim 0.5$ Gyr, \citealt{ragone11}), so we expect those galaxies with stellar populations older than $\sim 1$ Gyr to be already located in the local stellar mass-size relation. This is not supported by observations, in which galaxies older than $\gtrsim 3$ Gyr at $z \sim 1$ are still smaller than the local ones \citep[see][for details]{trujillo11}. Interestingly, minor mergers with gas-rich satellites ($\sim 35$\% of our observed mergers) could trigger recent star formation and AGN activity in massive early types \citep[e.g.,][]{kaviraj09,onti11}, therefore favouring some degree of adiabatic expansion and adding an extra size growth to the merging process. Devoted N-body simulations are needed to explore this topic in details.

It is also to be noted that the mass loss due to stellar winds from the passive evolution of stellar populations in a galaxy may lead to adiabatic expansion \citep{damjanov09}. \citet{ragone11} show that a typical massive galaxy is able to eject enough mass due to galactic winds to increase its size by a factor of 1.2 in $\sim 8$ Gyr. This result assumes that the potential of the galaxy is not able to retain any of the ejected mass, so this could indicate that at most $\Delta r_{\rm e}\sim 20$\% of the size evolution since $z = 1$ could be explained by stellar winds.

\item {\bf Overestimation of the size evolution}. Results from \citet{martinez11} suggest that the photometric stellar masses of \citet{trujillo07} are an order of magnitude higher than those estimated from velocity dispersion measurements. This does not erase the size evolution, but makes it smaller (massive galaxies are more extended than less massive ones at a given redshift, e.g., \citealt{damjanov11}). Taking dynamical masses ($M_{\rm dym}$) as a reference instead of photometric ones, \citet{vanderwel08esize} find $\alpha = 0.98 \pm 0.11$, smaller than the $\alpha = 1.20$ found by the same authors from photometric studies. The same trend is found by \citet{saglia10} from the ESO Distant Cluster Survey\footnote{http://www.mpa-garching.mpg.de/galform/ediscs/index.shtml} \citep[EDisCS;][]{white05} galaxies: $\alpha \sim 0.65$ from dynamical masses vs $\alpha \sim 0.85$ from stellar masses after the progenitor bias is accounted for. Finally, \citet{newman10} find $\alpha \sim 0.75$ for $M_{\rm dym} \geq 10^{11}\ M_{\odot}$ galaxies. Assuming these smaller $\alpha$ values from dynamical masses, major and minor mergers account for $\Delta r_{\rm e}\sim65$\% of the size evolution, and all the evolution is explained when the progenitor bias and very minor mergers are taken into account.

It is also possible that the extended, low-surface brightness envelopes of high-z galaxies were missed and their $r_{\rm e}$ were correspondingly underestimated. However, deep observations in the near infrared (optical rest-frame) from space \citep{szomoru10,cassata10,weinzirl11} and from ground-based facilities with adaptive optics \citep{carrasco10} confirm the compactness of $1 \lesssim z \lesssim 3$ massive galaxies.

On the other hand, also higher values of $\alpha$ than our fiducial value $\alpha = 1.2 \pm 0.2$ are present in the literature. For example, \citet{buitrago08} find $\alpha = 1.51\pm0.04$ at $z < 2$, while \citet{damjanov11} find $\alpha = 1.62\pm0.34$. Assuming $\alpha = 1.5$, $\mu \geq 1/10$ mergers would explain $\Delta r_{\rm e}\sim 45$\% of the size evolution, while the addition of very minor mergers would increase the role of mergers up to $\Delta r_{\rm e}\sim 65$\%. In that case, the contribution of other processes would increases to $\Delta r_{\rm e}\sim 20$\%. Thus, even if the size evolution is faster than our fiducial $\alpha$ value, mergers would be still the dominant mechanism. 

\item {\bf Merger time scale}. The main uncertainty in our merger rates is the assumed merger time scale, which typically has a factor of two uncertainty in their normalisation \citep[e.g.,][]{hopkins10mer}. The $T_{\rm MM}$ from \citet{kit08} are typically longer than others in the literature \citep[e.g.,][]{patton08,lin10} or similar to those from N-body/hydrodynamical simulations \citep{lotz10t,lotz10gas}. Thus, we expect, if anything, a shorter $T_{\rm MM}$, which implies a larger role of mergers in size evolution (i.e., higher merger rates and number of mergers since $z \sim 1$). In fact, a shorter $T_{\rm MM}$ by a factor of 1.5 is enough to explain the observed mass growth and size evolution without the contribution of very minor mergers.

\item {\bf Uncertainties in $\beta$}. Equation~(\ref{req}), which we used to derive the values of $\beta$ in our model, assumes parabolic orbits and dissipationless (gas-free) mergers. About the first assumption, \cite{khochfar06orbit} and \citet{wetzel10} show that most dark matter halos in cosmological simulations merge on parabolic orbits. On the other hand, we find that $\sim65$\% of our mergers are dry, but the other $\sim35$\% are mixed and an extra dissipative component is present (Sect~\ref{colsec}). In these cases simulations suggest that $\beta$ should be higher than derived from Eq.~(\ref{req}), even reaching $\beta \sim 2.5$ \citep[][]{hopkins10size}. This does not change our conclusions because it translates to a higher size evolution due to mergers. In addition, \citet{oser12} show that the size growth expected from Eq.~(\ref{req}) is in nice agreement with the growth measured in hydrodynamical simulations settled in a cosmological contest.
\end{itemize}

In summary, our results suggest that merging is the main contributor to the size evolution of massive ETGs, accounting for $\Delta r_{\rm e}\sim 50$\%$-75$\% of the observed evolution since $z \sim 1$. Nearly half of the evolution due to mergers is related to minor ($\mu < 1/4$) events.

\subsubsection{Additional constraints from velocity dispersion evolution}
\begin{figure}[t!]
\resizebox{\hsize}{!}{\includegraphics{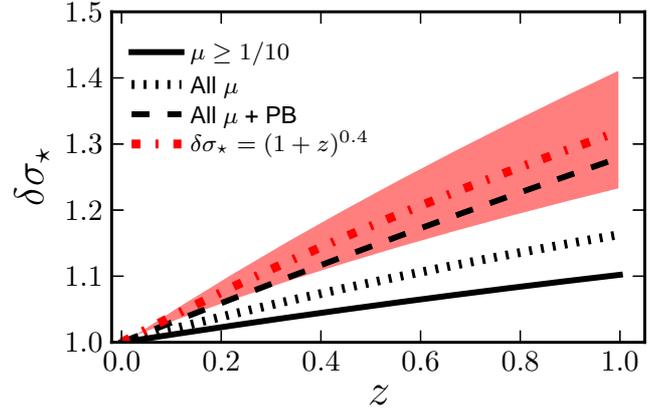}}
\caption{Velocity dispersion normalised to its local value, $\delta \sigma_{\star}$, as a function of redshift. The dott-dashed line is the observed evolution, $\delta \sigma_{\star} = (1+z)^{0.4}$. The shaded area marks the 68\% confidence interval. The solid line is the evolution due to major and minor mergers ($\mu \geq 1/10$) expected from our results. The dotted line is the expected evolution when very minor mergers ($\mu < 1/10$) are taken into account. The dashed line is the expected evolution when very minor mergers and the progenitor bias (PB) are included (see text for details).}
\label{sigmodelfig}
\end{figure}

In addition to their mass and size, the velocity dispersion ($\sigma_{\star}$) of ETGs evolves with redshift as $\delta \sigma_{\star} (z) = \sigma_{\star} (z)/\sigma_{\star} (0) = (1+z)^{a}$. We assume $a = 0.4 \pm 0.1$ in the following \citep[$\delta \sigma_{\star} = 1.32 $ at $z = 1$,][]{cenarro09,cappellari09,saglia10,vandesande11}. When we apply our simple model using the prescriptions of \citet{fan10} for the evolution of $\sigma_{\star}$ in merger events, $\mu \geq 1/10$ mergers are only able to explain $15$\% of the observed evolution, $\delta \sigma_{\star}^{\rm m}(1) = 1.05$. \citet{hopkins09scale} propose another prescription to trace the evolution in $\sigma_{\star}$ from the evolution in size that takes into account the dark matter component of the galaxy,
\begin{equation}
\delta \sigma_{\star}^{\rm m}(z) = \sqrt{\frac{\gamma + 1/\delta r_{\rm e}^{\rm m}(z)}{1 + \gamma}},
\end{equation}
where $\gamma \sim 1$ for $M_{\star} \sim 10^{11}\ M_{\odot}$ galaxies. Using this prescription, the evolution of $\sigma_{\star}$ is faster, but we still explain only $\sim35$\% of the observed evolution, $\delta \sigma_{\star}^{\rm m}(1) = 1.10$ (solid line in Fig.~\ref{sigmodelfig}). The addition of very minor mergers increase the contribution to $\sim50$\%, $\delta \sigma_{\star}^{\rm m}(1) = 1.16$ (dotted line in Fig.~\ref{sigmodelfig}). However, a small change of $\sigma_{\star}$ due to mergers is consistent with the picture from \citet{bernardi11}. They study in details the colour$-M_{\star}$ and colour$-\sigma_{\star}$ relation of ETGs in SDSS, finding that $M_{\star} \sim 2 \times 10^{11}\ M_{\odot}$ is a transition mass ($M_{\rm tran}$) for which the curvature of the colour$-M_{\star}$ relation change, while no deviation is present in the colour$-\sigma_{\star}$ relation. These authors claim that (dry) mergers are the main process in the evolution of those ETGs with $M_{\star} \gtrsim M_{\rm tran} \sim M_{\star}^*$ \citep[see also][for a similar conclusion]{vanderwel09ba,clsj10megoods,oesch10,eliche10I,jairo12}, as our results also suggest.

One missing ingredient in the model described in this section is the progenitor bias: new early-types which appeared since $z \sim 1$ are not only more extended that previous ones, but also have a lower velocity dispersion \citep{vanderwel09}. Thus, the progenitor bias also mimic a decrease of $\sigma_{\star}$ with cosmic time. The results of \citet{saglia10} suggest that a factor of 1.1 in the $\sigma_{\star}$ evolution is due to the progenitor bias. Applying this extra evolution as a factor $1 + 0.1z$ to that from mergers (very minor ones included), we are able to explain $90$\% of the increase in velocity dispersion, $\delta \sigma_{\star}^{\rm PB}(1) = 1.28$ (dashed line in Fig.~\ref{sigmodelfig}). Including this, our model is compatible with the observed evolution and suggests that mergers and the progenitor bias have a similar contribution to $\sigma_{\star}$ evolution, somewhat different from the dominant role of mergers in size evolution. 

\subsubsection{Additional constraints from S\'ersic index evolution}
Other structural parameters of ETGs, as the S\'ersic index $n_{\rm s}$ \citep{sersic68}, also evolve with redshift. Depending on the value of $n_{\rm s}$, galaxies can be described as disc-like with a $n_{\rm s} = 1$ exponential profile or bulge-like with higher $n_{\rm s}$ values, where ellipticals are expected to have $n_{\rm s} = 4$ profiles. We refer to the change in the S\'ersic index with redshift as
\begin{equation}
\Delta n_{\rm s}(z) = n_{\rm s}(0) - n_{\rm s}(z),
\end{equation}
where $n_{\rm s}(z)$ is the S\'ersic index at redshift $z$. Several studies find that the S\'ersic index of the global massive population decreases with redshift \citep[e.g.,][]{vandokkum10,weinzirl11,buitrago12}, evolving from $n_{\rm s}(1) \sim 3$ to $n_{\rm s}(0) \sim 6$, $\Delta n_{\rm s}(1) \sim 3$. Focusing on massive ETGs, the study of \citet{buitrago12} find an evolution in $n_{\rm s}$ consistent with $\Delta n_{\rm s}(z) = 1.4z$ up to $z \sim 2.5$, and we take this evolution as a reference.

We used the theoretical results in \citet{hopkins10size} to estimate the change in the S\'ersic index of ETGs since $z \sim 1$ due to mergers. We took their results for minor mergers/late accretion as representative, so we can roughly estimate $\Delta n_{\rm s}(z)$ for a given increase in effective radius (their Fig.~3). From the values of $\delta r_{\rm e}^{\rm m}$ estimated in Sect.~\ref{sizegrowth}, we expect $\Delta n_{\rm s}(1) \sim 1-2$. This evolution is in agreement with the $\Delta n_{\rm s}(1) = 1.4$ measured by \citet{buitrago12} for massive ETGs. We conclude that the observed merging activity is also consistent with the observed evolution in the S\'ersic index of massive ETGs, supporting the dominant role of mergers in the evolution of these systems since $z \sim 1$.

\subsubsection{Additional constraints from scaling relations}
\citet{nipoti09} point out that the tightness of the local scaling laws of ETGs posses an important limit to the growth of these systems by (dry) merging (see also~\citealt{ciotti07,nair11}). Using these local scaling laws, they conclude that typical present-day massive ETGs could not have assembled more than $\sim45$\% of their present stellar mass and grew more than a factor $\sim$1.9 in size via merging. Even if uncertain, we can extrapolate our observed trends up to $z \sim 2$ and compare the inferred mass and size growths with these upper limits provided by \citet{nipoti09}. We obtain a mass growth by merging (including very minor mergers) of $\delta M_{\star}\sim 60$\% since $z = 2$, which implies that $\delta M_{\star}(2)M_{\star}(2)/M_{\star}(0) \sim 40$\% of the total mass at $z = 0$ was assembled by merging since $z = 2$. The size grows by a factor of $\sim 2$ due to this merging in the same cosmic time lapse. Therefore, merging seems compatible with the upper limits in mass assembly and size growth imposed by the tightness of the local scaling laws, although a more complex model is needed to fully explore how these laws evolve due to our observed merger history.

\subsubsection{Comparison with previous studies}
In a previous work, \citet{trujillo11} use a similar model than ours to estimate the number of mergers needed since $z \sim 1$ to explain size evolution if merging is the only process involved. They conclude that $N_{\rm m} = 5.0 \pm 1.64$ mergers with $\mu = 1/3$ are needed. This number of mergers is higher than our direct measurement by a factor of five, $N_{\rm m} = 0.89 \pm 0.14$ (our average merger with $\mu \geq 1/10$ has $\overline{\mu} \sim 1/3$). If we take into account our estimated very minor mergers, our numbers are $N_{\rm m} \sim 5$ and $\overline{\mu} \sim 1/10$. For this value of $\mu$ they infer $N_{\rm m} = 11.20 \pm 3.66$, still higher than our estimation. The model of \citet{trujillo11} also estimates the mass growth due to mergers since $z \sim 1$, which is a factor of $3-5$, also higher than any observational estimation or constraint (a factor of $\sim1.4$, Sect.~\ref{massgrowth}). 

\citet{newman12} study the size evolution of red galaxies in the CANDELS\footnote{http://candels.ucolick.org/About.html} (Cosmic Assembly Near-infrared Deep Extragalactic Legacy Survey, \citealt{candels,candels2}) survey and the role of mergers with $\mu \geq 1/10$ in this size growth at $0.4 < z < 2$. Applying a similar model than ours to translate their observed total merger fraction to a size growth, they conclude that merging can reasonably account for the size evolution observed at $z \lesssim 1$ after the progenitor bias is taking into account, while at $z \gtrsim 1$ mergers are not common enough. Despite the fact that they only have one merger fraction data point at $0.4 < z < 1$ (Fig.~\ref{fftotalfig}), their conclusion is consistent with our more detailed study at $z \lesssim 1$.

\subsubsection{Expectations from cosmological models}\label{models}
Several theoretical efforts have been conducted to explain the size evolution of ETGs. In this section we compare the predicted size evolution from cosmological models with our best model, which suggests that $\Delta r_{\rm e}\sim75$\% of the evolution in size is due to mergers, $\Delta r_{\rm e}\sim20$\% to the progenitor bias and $\Delta r_{\rm e}\sim5$\% to other processes (e.g., adiabatic expansion).

The model of \citet{hopkins10size} predicts that, since $z = 2$, un-equal mass mergers explain $\Delta r_{\rm e}\sim60$\% of the observed size evolution, in agreement with our result. However, these authors only track the evolution of compact galaxies since $z = 2$ and do not take into account the possible contribution of the progenitor bias, but argue that it should impact their predictions. In fact, they predict that $\sim 45$\% of the size evolution since $z = 1$ is due to un-equal mass mergers, another $\sim 45$\% is accounted for by systematics in size measurements and the extra $\sim10$\% is due to adiabatic expansion, probably reflecting their biased population.

The model of \citet{shankar11} predicts $\delta r_{\rm e} \sim 0.7$ for massive galaxies, in agreement with our observational derivation due only to mergers \citep[see also][]{khochfar06}. Interestingly, the evolution increases to $\delta r_{\rm e} \sim 0.5$ when individual galaxies are tracked along their evolution without any stellar mass selection. They predict that $\sim40$\% of the mass accreted by merging in massive galaxies is due to major mergers with $\mu \geq 1/3$. Our best model implies that $\sim47$\% of mass and size growth is due to major mergers with $\mu \geq 1/3$. The qualitative agreement between both works is remarkable.

On the other hand, \citet{oser12} find $\alpha = 1.12$ ($\alpha = 1.44$ for passive galaxies) and $a \sim 0.4$ by re-simulating with high resolution a set of 40 galaxies with $M_{\star} \geq 6.3 \times 10^{10}\ M_{\odot}$ in a cosmological context. They find that the number-averaged merger has $\overline{\mu} = 1/16$, while the mass-averaged merger has $\overline{\mu}_{\star} = 1/5$. From our model we estimate $\overline{\mu}_{\star} = 1/3$ and $\overline{\mu} = 1/13$. We check that $\overline{\mu}_{\star}$ is independent of the assumed number of very minor mergers $N_{\rm vm}$, while we can vary $\overline{\mu}$ arbitrarily by changing $N_{\rm vm}$. Thus, only the comparison with $\overline{\mu}_{\star}$ is representative. The predicted value is lower than our measurement, but they find a higher role of major mergers at $M_{\star} \sim 10^{11.1}\ M_{\odot}$ \citep[see also][]{khochfar09, hopkins10fusbul, cattaneo11}, with $\overline{\mu}_{\star} \sim 1/3$ and a big dispersion due to the low statistics (see their Fig. 6). Future simulations with higher number of galaxies are needed to explore in more details this issue.

In summary, our result that merging is the main process involved in size evolution mostly agrees with simulations, but more observational and theoretical studies are needed to understand the remaining discrepancies.

\section{Conclusions}\label{conclusion}
We have measured the minor and major merger fraction and rate of massive ($M_{\star} \geq 10^{11}\ M_{\odot}$) galaxies from close pairs in the COSMOS field, and explored the role of mergers in the mass growth and size evolution of massive ETGs since $z \sim 1$.

We find that the merger fraction and rate of massive galaxies evolves as a power-law $(1+z)^{n}$, with no or only small evolution of the minor merger rate, $n_{\rm mm} \sim 0$, in contrast with the increase of the major merger rate, $n_{\rm MM} = 1.4$. The total (major + minor) merger rate evolves slower than the major one, with $n_{\rm m} = 0.6$. When splitting galaxies according to their HST/ACS morphology, the minor merger fraction for ETGs is higher by a factor of three than that for LTGs, and both are nearly constant with redshift. The fraction of major mergers for LTGs evolves faster ($n_{\rm MM}^{\rm LT} \sim 4$) than for ETGs ($n_{\rm MM}^{\rm ET} = 1.8$). We also find that when we repeat our study with a constant number-density sample, comprising ETGs with $\log\,(M_{\star}/M_{\odot}) \geq 11.15 - 0.15z$, the evolution of the major merger fraction is faster ($n_{\rm MM}^{\rm ET} = 2.5$), whilst the minor merger fraction and other derived quantities remain the same. Therefore, we conclude that the selection of the massive ETGs sample has limited impact in our results below.

Our results imply that massive ETGs have undergone 0.89 mergers (0.43 major and 0.46 minor) since $z \sim 1$, leading to a mass growth of $\sim30$\% (75\%/25\% due to major/minor mergers). We use a simple model to translate the estimated mass growth due to mergers into an effective radius growth. With this model we find that $\mu \geq 1/10$ mergers can explain $\sim 55$\% of the observed size evolution since $z \sim 1$. We infer that another $\sim20$\% is due to the progenitor bias (the new ETGs appeared since $z = 1$ are more extended than their high-z counterparts) and we estimate that very minor mergers ($\mu < 1/10$) could contribute with an additional $\sim20$\%. The remaining $\sim5$\% could come from adiabatic expansion due to stellar winds or from observational effects. In addition, our picture also reproduces the mass growth and the velocity dispersion evolution of these massive ETGs galaxies since $z \sim 1$.

We conclude from these results, and after exploring all the possible uncertainties in our model, that merging is the main contributor to the size evolution of massive ETGs at $z \lesssim 1$, accounting for $\sim 50-75$\% of that evolution in the last 8 Gyr. Nearly half of the evolution due to mergers is related to minor ($\mu < 1/4$) events.

Studies in larger sky areas are needed to improve the statistics, especially at lower redshifts when the cosmological volume probed is still the main source of uncertainty. We point out that a local measurement of the minor merger fraction and rate is needed to better constrain its evolution with redshift. Understanding the dependency of the minor merger rate on stellar mass, as well as extending observations to the very minor merger regime ($\mu \leq 1/10$) will be important to further improve this picture. In addition, extending the observational work at $z > 1$, when the massive red sequence seems to emerge, will be necessary to probe the early epochs of mass assembly.

\begin{acknowledgements}
We dedicate this paper to the memory of our six IAC colleagues and friends who
met with a fatal accident in Piedra de los Cochinos, Tenerife, in February 2007,
with a special thanks to Maurizio Panniello, whose teachings of \texttt{python}
were so important for this paper.

We thank the comments and suggestions of the anonymous referee. We also thank 
Ignacio Trujillo, Fernando Buitrago, Carmen Eliche-Moral, Paolo Cassata, 
Jairo M\'endez-Abreu, David Patton, Sara Ellison, Trevor Mendel, and Jorge Moreno 
for useful discussions. 

This work is supported by funding from ANR-07-BLAN-0228 and ERC-2010-AdG-268107-EARLY.

\end{acknowledgements}

\bibliography{biblio}

\begin{thebibliography}{153}
\expandafter\ifx\csname natexlab\endcsname\relax\def\natexlab#1{#1}\fi

\bibitem[{{Abazajian} {et~al.}(2009){Abazajian}, {Adelman-McCarthy},
  {Ag{\"u}eros}, {Allam}, {Allende Prieto}, {An}, {Anderson}, {Anderson},
  {Annis}, {Bahcall}, \& et~al.}]{sdssdr7}
{Abazajian}, K.~N., {Adelman-McCarthy}, J.~K., {Ag{\"u}eros}, M.~A., {et~al.}
  2009, \apjs, 182, 543

\bibitem[{{Arnouts} {et~al.}(2007){Arnouts}, {Walcher}, {Le F{\`e}vre},
  {Zamorani}, {Ilbert}, {Le Brun}, {Pozzetti}, {Bardelli}, {Tresse}, {Zucca},
  {Charlot}, {Lamareille}, {McCracken}, {Bolzonella}, {Iovino}, {Lonsdale},
  {Polletta}, {Surace}, {Bottini}, {Garilli}, {Maccagni}, {Picat},
  {Scaramella}, {Scodeggio}, {Vettolani}, {Zanichelli}, {Adami}, {Cappi},
  {Ciliegi}, {Contini}, {de La Torre}, {Foucaud}, {Franzetti}, {Gavignaud},
  {Guzzo}, {Marano}, {Marinoni}, {Mazure}, {Meneux}, {Merighi}, {Paltani},
  {Pell{\`o}}, {Pollo}, {Radovich}, {Temporin}, \& {Vergani}}]{arnouts07}
{Arnouts}, S., {Walcher}, C.~J., {Le F{\`e}vre}, O., {et~al.} 2007, \aap, 476,
  137

\bibitem[{{Baldry} {et~al.}(2004){Baldry}, {Glazebrook}, {Brinkmann},
  {Ivezi{\'c}}, {Lupton}, {Nichol}, \& {Szalay}}]{baldry04}
{Baldry}, I.~K., {Glazebrook}, K., {Brinkmann}, J., {et~al.} 2004, \apj, 600,
  681

\bibitem[{{Barro} {et~al.}(2011){Barro}, {P{\'e}rez-Gonz{\'a}lez}, {Gallego},
  {Ashby}, {Kajisawa}, {Miyazaki}, {Villar}, {Yamada}, \&
  {Zamorano}}]{barro11mass}
{Barro}, G., {P{\'e}rez-Gonz{\'a}lez}, P.~G., {Gallego}, J., {et~al.} 2011,
  \apjs, 193, 30

\bibitem[{{Bell} {et~al.}(2003){Bell}, {McIntosh}, {Katz}, \&
  {Weinberg}}]{bell03}
{Bell}, E.~F., {McIntosh}, D.~H., {Katz}, N., \& {Weinberg}, M.~D. 2003, \apjs,
  149, 289

\bibitem[{{Bell} {et~al.}(2006){Bell}, {Phleps}, {Somerville}, {Wolf}, {Borch},
  \& {Meisenheimer}}]{bell06}
{Bell}, E.~F., {Phleps}, S., {Somerville}, R.~S., {et~al.} 2006, \apj, 652, 270

\bibitem[{{Bernardi} {et~al.}(2011){Bernardi}, {Roche}, {Shankar}, \&
  {Sheth}}]{bernardi11}
{Bernardi}, M., {Roche}, N., {Shankar}, F., \& {Sheth}, R.~K. 2011, \mnras,
  412, 684

\bibitem[{{Bezanson} {et~al.}(2009){Bezanson}, {van Dokkum}, {Tal},
  {Marchesini}, {Kriek}, {Franx}, \& {Coppi}}]{bezanson09}
{Bezanson}, R., {van Dokkum}, P.~G., {Tal}, T., {et~al.} 2009, \apj, 697, 1290

\bibitem[{{Bluck} {et~al.}(2009){Bluck}, {Conselice}, {Bouwens}, {Daddi},
  {Dickinson}, {Papovich}, \& {Yan}}]{bluck09}
{Bluck}, A.~F.~L., {Conselice}, C.~J., {Bouwens}, R.~J., {et~al.} 2009, \mnras,
  394, L51

\bibitem[{{Bluck} {et~al.}(2012){Bluck}, {Conselice}, {Buitrago},
  {Gr{\"u}tzbauch}, {Hoyos}, {Mortlock}, \& {Bauer}}]{bluck12}
{Bluck}, A.~F.~L., {Conselice}, C.~J., {Buitrago}, F., {et~al.} 2012, \apj,
  747, 34

\bibitem[{{Brammer} {et~al.}(2011){Brammer}, {Whitaker}, {van Dokkum},
  {Marchesini}, {Franx}, {Kriek}, {Labb{\'e}}, {Lee}, {Muzzin}, {Quadri},
  {Rudnick}, \& {Williams}}]{brammer11}
{Brammer}, G.~B., {Whitaker}, K.~E., {van Dokkum}, P.~G., {et~al.} 2011, \apj,
  739, 24

\bibitem[{{Bridge} {et~al.}(2010){Bridge}, {Carlberg}, \&
  {Sullivan}}]{bridge10}
{Bridge}, C.~R., {Carlberg}, R.~G., \& {Sullivan}, M. 2010, \apj, 709, 1067

\bibitem[{{Brown} {et~al.}(2007){Brown}, {Dey}, {Jannuzi}, {Brand}, {Benson},
  {Brodwin}, {Croton}, \& {Eisenhardt}}]{brown07}
{Brown}, M.~J.~I., {Dey}, A., {Jannuzi}, B.~T., {et~al.} 2007, \apj, 654, 858

\bibitem[{{Brown} {et~al.}(2008){Brown}, {Zheng}, {White}, {Dey}, {Jannuzi},
  {Benson}, {Brand}, {Brodwin}, \& {Croton}}]{brown08}
{Brown}, M.~J.~I., {Zheng}, Z., {White}, M., {et~al.} 2008, \apj, 682, 937

\bibitem[{{Bruzual} \& {Charlot}(2003)}]{bc03}
{Bruzual}, G. \& {Charlot}, S. 2003, \mnras, 344, 1000

\bibitem[{{Buitrago} {et~al.}(2008){Buitrago}, {Trujillo}, {Conselice},
  {Bouwens}, {Dickinson}, \& {Yan}}]{buitrago08}
{Buitrago}, F., {Trujillo}, I., {Conselice}, C.~J., {et~al.} 2008, \apjl, 687,
  L61

\bibitem[{{Buitrago} {et~al.}(2011){Buitrago}, {Trujillo}, {Conselice}, \&
  {Haeussler}}]{buitrago12}
{Buitrago}, F., {Trujillo}, I., {Conselice}, C.~J., \& {Haeussler}, B. 2011,
  \mnras, submitted [arXiv: 1111.6993]

\bibitem[{{Bundy} {et~al.}(2006){Bundy}, {Ellis}, {Conselice}, {Taylor},
  {Cooper}, {Willmer}, {Weiner}, {Coil}, {Noeske}, \& {Eisenhardt}}]{bundy06}
{Bundy}, K., {Ellis}, R.~S., {Conselice}, C.~J., {et~al.} 2006, \apj, 651, 120

\bibitem[{{Bundy} {et~al.}(2009){Bundy}, {Fukugita}, {Ellis}, {Targett},
  {Belli}, \& {Kodama}}]{bundy09}
{Bundy}, K., {Fukugita}, M., {Ellis}, R.~S., {et~al.} 2009, \apj, 697, 1369

\bibitem[{{Calzetti} {et~al.}(2000){Calzetti}, {Armus}, {Bohlin}, {Kinney},
  {Koornneef}, \& {Storchi-Bergmann}}]{calzetti00}
{Calzetti}, D., {Armus}, L., {Bohlin}, R.~C., {et~al.} 2000, \apj, 533, 682

\bibitem[{{Cameron}(2011)}]{cameron11}
{Cameron}, E. 2011, \pasa, 28, 128

\bibitem[{{Capak} {et~al.}(2007){Capak}, {Aussel}, {Ajiki}, {McCracken},
  {Mobasher}, {Scoville}, {Shopbell}, {Taniguchi}, {Thompson}, {Tribiano},
  {Sasaki}, {Blain}, {Brusa}, {Carilli}, {Comastri}, {Carollo}, {Cassata},
  {Colbert}, {Ellis}, {Elvis}, {Giavalisco}, {Green}, {Guzzo}, {Hasinger},
  {Ilbert}, {Impey}, {Jahnke}, {Kartaltepe}, {Kneib}, {Koda}, {Koekemoer},
  {Komiyama}, {Leauthaud}, {Le Fevre}, {Lilly}, {Liu}, {Massey}, {Miyazaki},
  {Murayama}, {Nagao}, {Peacock}, {Pickles}, {Porciani}, {Renzini}, {Rhodes},
  {Rich}, {Salvato}, {Sanders}, {Scarlata}, {Schiminovich}, {Schinnerer},
  {Scodeggio}, {Sheth}, {Shioya}, {Tasca}, {Taylor}, {Yan}, \&
  {Zamorani}}]{capak07}
{Capak}, P., {Aussel}, H., {Ajiki}, M., {et~al.} 2007, \apjs, 172, 99

\bibitem[{{Cappellari} {et~al.}(2009){Cappellari}, {di Serego Alighieri},
  {Cimatti}, {Daddi}, {Renzini}, {Kurk}, {Cassata}, {Dickinson},
  {Franceschini}, {Mignoli}, {Pozzetti}, {Rodighiero}, {Rosati}, \&
  {Zamorani}}]{cappellari09}
{Cappellari}, M., {di Serego Alighieri}, S., {Cimatti}, A., {et~al.} 2009,
  \apjl, 704, L34

\bibitem[{{Carrasco} {et~al.}(2010){Carrasco}, {Conselice}, \&
  {Trujillo}}]{carrasco10}
{Carrasco}, E.~R., {Conselice}, C.~J., \& {Trujillo}, I. 2010, \mnras, 405,
  2253

\bibitem[{{Cassata} {et~al.}(2010){Cassata}, {Giavalisco}, {Guo}, {Ferguson},
  {Koekemoer}, {Renzini}, {Fontana}, {Salimbeni}, {Dickinson}, {Casertano},
  {Conselice}, {Grogin}, {Lotz}, {Papovich}, {Lucas}, {Straughn}, {Gardner}, \&
  {Moustakas}}]{cassata10}
{Cassata}, P., {Giavalisco}, M., {Guo}, Y., {et~al.} 2010, \apjl, 714, L79

\bibitem[{{Cassata} {et~al.}(2011){Cassata}, {Giavalisco}, {Guo}, {Renzini},
  {Ferguson}, {Koekemoer}, {Salimbeni}, {Scarlata}, {Grogin}, {Conselice},
  {Dahlen}, {Lotz}, {Dickinson}, \& {Lin}}]{cassata11}
{Cassata}, P., {Giavalisco}, M., {Guo}, Y., {et~al.} 2011, \apj, 743, 96

\bibitem[{{Cattaneo} {et~al.}(2011){Cattaneo}, {Mamon}, {Warnick}, \&
  {Knebe}}]{cattaneo11}
{Cattaneo}, A., {Mamon}, G.~A., {Warnick}, K., \& {Knebe}, A. 2011, \aap, 533,
  A5

\bibitem[{{Cenarro} \& {Trujillo}(2009)}]{cenarro09}
{Cenarro}, A.~J. \& {Trujillo}, I. 2009, \apjl, 696, L43

\bibitem[{{Chabrier}(2003)}]{chabrier03}
{Chabrier}, G. 2003, \pasp, 115, 763

\bibitem[{{Chou} {et~al.}(2011){Chou}, {Bridge}, \& {Abraham}}]{chou10}
{Chou}, R.~C.~Y., {Bridge}, C.~R., \& {Abraham}, R.~G. 2011, \aj, 141, 87

\bibitem[{{Ciotti} {et~al.}(2007){Ciotti}, {Lanzoni}, \&
  {Volonteri}}]{ciotti07}
{Ciotti}, L., {Lanzoni}, B., \& {Volonteri}, M. 2007, \apj, 658, 65

\bibitem[{{Conselice} {et~al.}(2011){Conselice}, {Bluck}, {Ravindranath},
  {Mortlock}, {Koekemoer}, {Buitrago}, {Gr{\"u}tzbauch}, \&
  {Penny}}]{conselice11}
{Conselice}, C.~J., {Bluck}, A.~F.~L., {Ravindranath}, S., {et~al.} 2011,
  \mnras, 417, 2770

\bibitem[{{Conselice} {et~al.}(2007){Conselice}, {Bundy}, {Trujillo}, {Coil},
  {Eisenhardt}, {Ellis}, {Georgakakis}, {Huang}, {Lotz}, {Nandra}, {Newman},
  {Papovich}, {Weiner}, \& {Willmer}}]{powir}
{Conselice}, C.~J., {Bundy}, K., {Trujillo}, I., {et~al.} 2007, \mnras, 381,
  962

\bibitem[{{Cool} {et~al.}(2008){Cool}, {Eisenstein}, {Fan}, {Fukugita},
  {Jiang}, {Maraston}, {Meiksin}, {Schneider}, \& {Wake}}]{cool08}
{Cool}, R.~J., {Eisenstein}, D.~J., {Fan}, X., {et~al.} 2008, \apj, 682, 919

\bibitem[{{Coupon} {et~al.}(2009){Coupon}, {Ilbert}, {Kilbinger}, {McCracken},
  {Mellier}, {Arnouts}, {Bertin}, {Hudelot}, {Schultheis}, {Le F{\`e}vre}, {Le
  Brun}, {Guzzo}, {Bardelli}, {Zucca}, {Bolzonella}, {Garilli}, {Zamorani},
  {Zanichelli}, {Tresse}, \& {Aussel}}]{cfhtls}
{Coupon}, J., {Ilbert}, O., {Kilbinger}, M., {et~al.} 2009, \aap, 500, 981

\bibitem[{{Daddi} {et~al.}(2005){Daddi}, {Renzini}, {Pirzkal}, {Cimatti},
  {Malhotra}, {Stiavelli}, {Xu}, {Pasquali}, {Rhoads}, {Brusa}, {di Serego
  Alighieri}, {Ferguson}, {Koekemoer}, {Moustakas}, {Panagia}, \&
  {Windhorst}}]{daddi05}
{Daddi}, E., {Renzini}, A., {Pirzkal}, N., {et~al.} 2005, \apj, 626, 680

\bibitem[{{Damjanov} {et~al.}(2011){Damjanov}, {Abraham}, {Glazebrook},
  {McCarthy}, {Caris}, {Carlberg}, {Chen}, {Crampton}, {Green}, {J{\o}rgensen},
  {Juneau}, {Le Borgne}, {Marzke}, {Mentuch}, {Murowinski}, {Roth}, {Savaglio},
  \& {Yan}}]{damjanov11}
{Damjanov}, I., {Abraham}, R.~G., {Glazebrook}, K., {et~al.} 2011, \apjl, 739,
  L44

\bibitem[{{Damjanov} {et~al.}(2009){Damjanov}, {McCarthy}, {Abraham},
  {Glazebrook}, {Yan}, {Mentuch}, {Le Borgne}, {Savaglio}, {Crampton},
  {Murowinski}, {Juneau}, {Carlberg}, {J{\o}rgensen}, {Roth}, {Chen}, \&
  {Marzke}}]{damjanov09}
{Damjanov}, I., {McCarthy}, P.~J., {Abraham}, R.~G., {et~al.} 2009, \apj, 695,
  101

\bibitem[{{Davis} {et~al.}(2007){Davis}, {Guhathakurta}, {Konidaris}, {Newman},
  {Ashby}, {Biggs}, {Barmby}, {Bundy}, {Chapman}, {Coil}, {Conselice},
  {Cooper}, {Croton}, {Eisenhardt}, {Ellis}, {Faber}, {Fang}, {Fazio},
  {Georgakakis}, {Gerke}, {Goss}, {Gwyn}, {Harker}, {Hopkins}, {Huang},
  {Ivison}, {Kassin}, {Kirby}, {Koekemoer}, {Koo}, {Laird}, {Le Floc'h}, {Lin},
  {Lotz}, {Marshall}, {Martin}, {Metevier}, {Moustakas}, {Nandra}, {Noeske},
  {Papovich}, {Phillips}, {Rich}, {Rieke}, {Rigopoulou}, {Salim},
  {Schiminovich}, {Simard}, {Smail}, {Small}, {Weiner}, {Willmer}, {Willner},
  {Wilson}, {Wright}, \& {Yan}}]{davis07}
{Davis}, M., {Guhathakurta}, P., {Konidaris}, N.~P., {et~al.} 2007, \apjl, 660,
  L1

\bibitem[{{De Propris} {et~al.}(2010){De Propris}, {Driver}, {Colless},
  {Drinkwater}, {Loveday}, {Ross}, {Bland-Hawthorn}, {York}, \&
  {Pimbblet}}]{depropris10}
{De Propris}, R., {Driver}, S.~P., {Colless}, M., {et~al.} 2010, \aj, 139, 794

\bibitem[{{de Ravel} {et~al.}(2011){de Ravel}, {Kampczyk}, {Le F{\`e}vre},
  {Lilly}, {Tasca}, {Tresse}, {Lopez-Sanjuan}, {Bolzonella}, {Kovac}, {Abbas},
  {Bardelli}, {Bongiorno}, {Caputi}, {Contini}, {Coppa}, {Cucciati}, {de la
  Torre}, {Dunlop}, {Franzetti}, {Garilli}, {Iovino}, {Kneib}, {Koekemoer},
  {Knobel}, {Lamareille}, {Le Borgne}, {Le Brun}, {Leauthaud}, {Maier},
  {Mainieri}, {Mignoli}, {Pello}, {Peng}, {Perez Montero}, {Ricciardelli},
  {Scodeggio}, {Silverman}, {Tanaka}, {Vergani}, {Zamorani}, {Zucca},
  {Bottini}, {Cappi}, {Carollo}, {Cassata}, {Cimatti}, {Fumana}, {Guzzo},
  {Maccagni}, {Marinoni}, {McCracken}, {Memeo}, {Meneux}, {Oesch}, {Porciani},
  {Pozzetti}, {Renzini}, {Scaramella}, \& {Scarlata}}]{deravel11}
{de Ravel}, L., {Kampczyk}, P., {Le F{\`e}vre}, O., {et~al.} 2011, \aap,
  submitted [ArXiv: 1104.5470]

\bibitem[{{de Ravel} {et~al.}(2009){de Ravel}, {Le F{\`e}vre}, {Tresse},
  {Bottini}, {Garilli}, {Le Brun}, {Maccagni}, {Scaramella}, {Scodeggio},
  {Vettolani}, {Zanichelli}, {Adami}, {Arnouts}, {Bardelli}, {Bolzonella},
  {Cappi}, {Charlot}, {Ciliegi}, {Contini}, {Foucaud}, {Franzetti},
  {Gavignaud}, {Guzzo}, {Ilbert}, {Iovino}, {Lamareille}, {McCracken},
  {Marano}, {Marinoni}, {Mazure}, {Meneux}, {Merighi}, {Paltani}, {Pell{\`o}},
  {Pollo}, {Pozzetti}, {Radovich}, {Vergani}, {Zamorani}, {Zucca}, {Bondi},
  {Bongiorno}, {Brinchmann}, {Cucciati}, {de La Torre}, {Gregorini}, {Memeo},
  {Perez-Montero}, {Mellier}, {Merluzzi}, \& {Temporin}}]{deravel09}
{de Ravel}, L., {Le F{\`e}vre}, O., {Tresse}, L., {et~al.} 2009, \aap, 498, 379

\bibitem[{{Desai} {et~al.}(2011){Desai}, {Dey}, {Cohen}, {Le Floc'h}, \&
  {Soifer}}]{desai11}
{Desai}, V., {Dey}, A., {Cohen}, E., {Le Floc'h}, E., \& {Soifer}, B.~T. 2011,
  \apj, 730, 130

\bibitem[{{Drory} {et~al.}(2009){Drory}, {Bundy}, {Leauthaud}, {Scoville},
  {Capak}, {Ilbert}, {Kartaltepe}, {Kneib}, {McCracken}, {Salvato}, {Sanders},
  {Thompson}, \& {Willott}}]{drory09}
{Drory}, N., {Bundy}, K., {Leauthaud}, A., {et~al.} 2009, \apj, 707, 1595

\bibitem[{{Efron}(1982)}]{efron82}
{Efron}, B. 1982

\bibitem[{{Eliche-Moral} {et~al.}(2010){Eliche-Moral}, {Prieto}, {Gallego},
  {Barro}, {Zamorano}, {Lopez-Sanjuan}, {Balcells}, {Guzman}, \&
  {Munoz-Mateos}}]{eliche10I}
{Eliche-Moral}, M.~C., {Prieto}, M., {Gallego}, J., {et~al.} 2010, \aap, 519,
  A55

\bibitem[{{Faber} \& {Jackson}(1976)}]{faber76}
{Faber}, S.~M. \& {Jackson}, R.~E. 1976, \apj, 204, 668

\bibitem[{{Fan} {et~al.}(2010){Fan}, {Lapi}, {Bressan}, {Bernardi}, {De Zotti},
  \& {Danese}}]{fan10}
{Fan}, L., {Lapi}, A., {Bressan}, A., {et~al.} 2010, \apj, 718, 1460

\bibitem[{{Fan} {et~al.}(2008){Fan}, {Lapi}, {De Zotti}, \& {Danese}}]{fan08}
{Fan}, L., {Lapi}, A., {De Zotti}, G., \& {Danese}, L. 2008, \apjl, 689, L101

\bibitem[{{Feldmann} {et~al.}(2010){Feldmann}, {Carollo}, {Mayer}, {Renzini},
  {Lake}, {Quinn}, {Stinson}, \& {Yepes}}]{feldmann10}
{Feldmann}, R., {Carollo}, C.~M., {Mayer}, L., {et~al.} 2010, \apj, 709, 218

\bibitem[{{Fern{\'a}ndez-Ontiveros} {et~al.}(2011){Fern{\'a}ndez-Ontiveros},
  {L{\'o}pez-Sanjuan}, {Montes}, {Prieto}, \& {Acosta-Pulido}}]{onti11}
{Fern{\'a}ndez-Ontiveros}, J.~A., {L{\'o}pez-Sanjuan}, C., {Montes}, M.,
  {Prieto}, M.~A., \& {Acosta-Pulido}, J.~A. 2011, \mnras, 411, L21

\bibitem[{{Fontana} {et~al.}(2004){Fontana}, {Pozzetti}, {Donnarumma},
  {Renzini}, {Cimatti}, {Zamorani}, {Menci}, {Daddi}, {Giallongo}, {Mignoli},
  {Perna}, {Salimbeni}, {Saracco}, {Broadhurst}, {Cristiani}, {D'Odorico}, \&
  {Gilmozzi}}]{fontana04}
{Fontana}, A., {Pozzetti}, L., {Donnarumma}, I., {et~al.} 2004, \aap, 424, 23

\bibitem[{{Fontana} {et~al.}(2006){Fontana}, {Salimbeni}, {Grazian},
  {Giallongo}, {Pentericci}, {Nonino}, {Fontanot}, {Menci}, {Monaco},
  {Cristiani}, {Vanzella}, {de Santis}, \& {Gallozzi}}]{fontana06}
{Fontana}, A., {Salimbeni}, S., {Grazian}, A., {et~al.} 2006, \aap, 459, 745

\bibitem[{{George} {et~al.}(2011){George}, {Leauthaud}, {Bundy}, {Finoguenov},
  {Tinker}, {Lin}, {Mei}, {Kneib}, {Aussel}, {Behroozi}, {Busha}, {Capak},
  {Coccato}, {Covone}, {Faure}, {Fiorenza}, {Ilbert}, {Le Floc'h}, {Koekemoer},
  {Tanaka}, {Wechsler}, \& {Wolk}}]{george11}
{George}, M.~R., {Leauthaud}, A., {Bundy}, K., {et~al.} 2011, \apj, 742, 125

\bibitem[{{Giavalisco} {et~al.}(2004){Giavalisco}, {Ferguson}, {Koekemoer},
  {Dickinson}, {Alexander}, {Bauer}, {Bergeron}, {Biagetti}, {Brandt},
  {Casertano}, {Cesarsky}, {Chatzichristou}, {Conselice}, {Cristiani}, {Da
  Costa}, {Dahlen}, {de Mello}, {Eisenhardt}, {Erben}, {Fall}, {Fassnacht},
  {Fosbury}, {Fruchter}, {Gardner}, {Grogin}, {Hook}, {Hornschemeier}, {Idzi},
  {Jogee}, {Kretchmer}, {Laidler}, {Lee}, {Livio}, {Lucas}, {Madau},
  {Mobasher}, {Moustakas}, {Nonino}, {Padovani}, {Papovich}, {Park},
  {Ravindranath}, {Renzini}, {Richardson}, {Riess}, {Rosati}, {Schirmer},
  {Schreier}, {Somerville}, {Spinrad}, {Stern}, {Stiavelli}, {Strolger},
  {Urry}, {Vandame}, {Williams}, \& {Wolf}}]{giavalisco04}
{Giavalisco}, M., {Ferguson}, H.~C., {Koekemoer}, A.~M., {et~al.} 2004, \apjl,
  600, L93

\bibitem[{{Grogin} {et~al.}(2011){Grogin}, {Kocevski}, {Faber}, {Ferguson},
  {Koekemoer}, {Riess}, {Acquaviva}, {Alexander}, {Almaini}, {Ashby}, {Barden},
  {Bell}, {Bournaud}, {Brown}, {Caputi}, {Casertano}, {Cassata}, {Castellano},
  {Challis}, {Chary}, {Cheung}, {Cirasuolo}, {Conselice}, {Roshan Cooray},
  {Croton}, {Daddi}, {Dahlen}, {Dav{\'e}}, {de Mello}, {Dekel}, {Dickinson},
  {Dolch}, {Donley}, {Dunlop}, {Dutton}, {Elbaz}, {Fazio}, {Filippenko},
  {Finkelstein}, {Fontana}, {Gardner}, {Garnavich}, {Gawiser}, {Giavalisco},
  {Grazian}, {Guo}, {Hathi}, {H{\"a}ussler}, {Hopkins}, {Huang}, {Huang},
  {Jha}, {Kartaltepe}, {Kirshner}, {Koo}, {Lai}, {Lee}, {Li}, {Lotz}, {Lucas},
  {Madau}, {McCarthy}, {McGrath}, {McIntosh}, {McLure}, {Mobasher},
  {Moustakas}, {Mozena}, {Nandra}, {Newman}, {Niemi}, {Noeske}, {Papovich},
  {Pentericci}, {Pope}, {Primack}, {Rajan}, {Ravindranath}, {Reddy}, {Renzini},
  {Rix}, {Robaina}, {Rodney}, {Rosario}, {Rosati}, {Salimbeni}, {Scarlata},
  {Siana}, {Simard}, {Smidt}, {Somerville}, {Spinrad}, {Straughn}, {Strolger},
  {Telford}, {Teplitz}, {Trump}, {van der Wel}, {Villforth}, {Wechsler},
  {Weiner}, {Wiklind}, {Wild}, {Wilson}, {Wuyts}, {Yan}, \& {Yun}}]{candels}
{Grogin}, N.~A., {Kocevski}, D.~D., {Faber}, S.~M., {et~al.} 2011, \apjs, 197,
  35

\bibitem[{{Hern{\'a}ndez-Toledo} {et~al.}(2005){Hern{\'a}ndez-Toledo},
  {Avila-Reese}, {Conselice}, \& {Puerari}}]{htoledo05}
{Hern{\'a}ndez-Toledo}, H.~M., {Avila-Reese}, V., {Conselice}, C.~J., \&
  {Puerari}, I. 2005, \aj, 129, 682

\bibitem[{{Hern{\'a}ndez-Toledo} {et~al.}(2006){Hern{\'a}ndez-Toledo},
  {Avila-Reese}, {Salazar-Contreras}, \& {Conselice}}]{htoledo06}
{Hern{\'a}ndez-Toledo}, H.~M., {Avila-Reese}, V., {Salazar-Contreras}, J.~R.,
  \& {Conselice}, C.~J. 2006, \aj, 132, 71

\bibitem[{{Hopkins} {et~al.}(2010{\natexlab{a}}){Hopkins}, {Bundy}, {Croton},
  {Hernquist}, {Keres}, {Khochfar}, {Stewart}, {Wetzel}, \&
  {Younger}}]{hopkins10fusbul}
{Hopkins}, P.~F., {Bundy}, K., {Croton}, D., {et~al.} 2010{\natexlab{a}}, \apj,
  715, 202

\bibitem[{{Hopkins} {et~al.}(2010{\natexlab{b}}){Hopkins}, {Bundy},
  {Hernquist}, {Wuyts}, \& {Cox}}]{hopkins10size}
{Hopkins}, P.~F., {Bundy}, K., {Hernquist}, L., {Wuyts}, S., \& {Cox}, T.~J.
  2010{\natexlab{b}}, \mnras, 401, 1099

\bibitem[{{Hopkins} {et~al.}(2009{\natexlab{a}}){Hopkins}, {Bundy}, {Murray},
  {Quataert}, {Lauer}, \& {Ma}}]{hopkins09core}
{Hopkins}, P.~F., {Bundy}, K., {Murray}, N., {et~al.} 2009{\natexlab{a}},
  \mnras, 398, 898

\bibitem[{{Hopkins} {et~al.}(2010{\natexlab{c}}){Hopkins}, {Croton}, {Bundy},
  {Khochfar}, {van den Bosch}, {Somerville}, {Wetzel}, {Keres}, {Hernquist},
  {Stewart}, {Younger}, {Genel}, \& {Ma}}]{hopkins10mer}
{Hopkins}, P.~F., {Croton}, D., {Bundy}, K., {et~al.} 2010{\natexlab{c}}, \apj,
  724, 915

\bibitem[{{Hopkins} {et~al.}(2009{\natexlab{b}}){Hopkins}, {Hernquist}, {Cox},
  {Keres}, \& {Wuyts}}]{hopkins09scale}
{Hopkins}, P.~F., {Hernquist}, L., {Cox}, T.~J., {Keres}, D., \& {Wuyts}, S.
  2009{\natexlab{b}}, \apj, 691, 1424

\bibitem[{{Ilbert} {et~al.}(2009){Ilbert}, {Capak}, {Salvato}, {Aussel},
  {McCracken}, {Sanders}, {Scoville}, {Kartaltepe}, {Arnouts}, {Le Floc'h},
  {Mobasher}, {Taniguchi}, {Lamareille}, {Leauthaud}, {Sasaki}, {Thompson},
  {Zamojski}, {Zamorani}, {Bardelli}, {Bolzonella}, {Bongiorno}, {Brusa},
  {Caputi}, {Carollo}, {Contini}, {Cook}, {Coppa}, {Cucciati}, {de la Torre},
  {de Ravel}, {Franzetti}, {Garilli}, {Hasinger}, {Iovino}, {Kampczyk},
  {Kneib}, {Knobel}, {Kovac}, {Le Borgne}, {Le Brun}, {F{\`e}vre}, {Lilly},
  {Looper}, {Maier}, {Mainieri}, {Mellier}, {Mignoli}, {Murayama}, {Pell{\`o}},
  {Peng}, {P{\'e}rez-Montero}, {Renzini}, {Ricciardelli}, {Schiminovich},
  {Scodeggio}, {Shioya}, {Silverman}, {Surace}, {Tanaka}, {Tasca}, {Tresse},
  {Vergani}, \& {Zucca}}]{ilbert09}
{Ilbert}, O., {Capak}, P., {Salvato}, M., {et~al.} 2009, \apj, 690, 1236

\bibitem[{{Ilbert} {et~al.}(2010){Ilbert}, {Salvato}, {Le Floc'h}, {Aussel},
  {Capak}, {McCracken}, {Mobasher}, {Kartaltepe}, {Scoville}, {Sanders},
  {Arnouts}, {Bundy}, {Cassata}, {Kneib}, {Koekemoer}, {Le F{\`e}vre}, {Lilly},
  {Surace}, {Taniguchi}, {Tasca}, {Thompson}, {Tresse}, {Zamojski}, {Zamorani},
  \& {Zucca}}]{ilbert10}
{Ilbert}, O., {Salvato}, M., {Le Floc'h}, E., {et~al.} 2010, \apj, 709, 644

\bibitem[{{Jim{\'e}nez} {et~al.}(2011){Jim{\'e}nez}, {Cora}, {Bassino},
  {Tecce}, \& {Smith Castelli}}]{jimenez11}
{Jim{\'e}nez}, N., {Cora}, S.~A., {Bassino}, L.~P., {Tecce}, T.~E., \& {Smith
  Castelli}, A.~V. 2011, \mnras, 417, 785

\bibitem[{{Jogee} {et~al.}(2009){Jogee}, {Miller}, {Penner}, {Skelton},
  {Conselice}, {Somerville}, {Bell}, {Zheng}, {Rix}, {Robaina}, {Barazza},
  {Barden}, {Borch}, {Beckwith}, {Caldwell}, {Peng}, {Heymans}, {McIntosh},
  {H{\"a}u{\ss}ler}, {Jahnke}, {Meisenheimer}, {Sanchez}, {Wisotzki}, {Wolf},
  \& {Papovich}}]{jogee09}
{Jogee}, S., {Miller}, S.~H., {Penner}, K., {et~al.} 2009, \apj, 697, 1971

\bibitem[{{Kampczyk} {et~al.}(2011){Kampczyk}, {Lilly}, {de Ravel}, {Le
  F{\`e}vre}, {Bolzonella}, {Carollo}, {Diener}, {Knobel}, {Kovac}, {Maier},
  {Renzini}, {Sargent}, {Vergani}, {Abbas}, {Bardelli}, {Bongiorno},
  {Bordoloi}, {Caputi}, {Contini}, {Coppa}, {Cucciati}, {de la Torre},
  {Franzetti}, {Garilli}, {Iovino}, {Kneib}, {Koekemoer}, {Lamareille}, {Le
  Borgne}, {Le Brun}, {Leauthaud}, {Mainieri}, {Mignoli}, {Pello}, {Peng},
  {Perez Montero}, {Ricciardelli}, {Scodeggio}, {Silverman}, {Tanaka}, {Tasca},
  {Tresse}, {Zamorani}, {Zucca}, {Bottini}, {Cappi}, {Cassata}, {Cimatti},
  {Fumana}, {Guzzo}, {Kartaltepe}, {Marinoni}, {McCracken}, {Memeo}, {Meneux},
  {Oesch}, {Porciani}, {Pozzetti}, \& {Scaramella}}]{pawel12}
{Kampczyk}, P., {Lilly}, S.~J., {de Ravel}, L., {et~al.} 2011, \apj, submitted
  [ArXiv:1112.4842]

\bibitem[{{Kartaltepe} {et~al.}(2007){Kartaltepe}, {Sanders}, {Scoville},
  {Calzetti}, {Capak}, {Koekemoer}, {Mobasher}, {Murayama}, {Salvato},
  {Sasaki}, \& {Taniguchi}}]{kar07}
{Kartaltepe}, J.~S., {Sanders}, D.~B., {Scoville}, N.~Z., {et~al.} 2007, \apjs,
  172, 320

\bibitem[{{Kaviraj} {et~al.}(2008){Kaviraj}, {Khochfar}, {Schawinski}, {Yi},
  {Gawiser}, {Silk}, {Virani}, {Cardamone}, {van Dokkum}, \&
  {Urry}}]{kaviraj08}
{Kaviraj}, S., {Khochfar}, S., {Schawinski}, K., {et~al.} 2008, \mnras, 388, 67

\bibitem[{{Kaviraj} {et~al.}(2009){Kaviraj}, {Peirani}, {Khochfar}, {Silk}, \&
  {Kay}}]{kaviraj09}
{Kaviraj}, S., {Peirani}, S., {Khochfar}, S., {Silk}, J., \& {Kay}, S. 2009,
  \mnras, 394, 1713

\bibitem[{{Kaviraj} {et~al.}(2007){Kaviraj}, {Schawinski}, {Devriendt},
  {Ferreras}, {Khochfar}, {Yoon}, {Yi}, {Deharveng}, {Boselli}, {Barlow},
  {Conrow}, {Forster}, {Friedman}, {Martin}, {Morrissey}, {Neff},
  {Schiminovich}, {Seibert}, {Small}, {Wyder}, {Bianchi}, {Donas}, {Heckman},
  {Lee}, {Madore}, {Milliard}, {Rich}, \& {Szalay}}]{kaviraj07}
{Kaviraj}, S., {Schawinski}, K., {Devriendt}, J.~E.~G., {et~al.} 2007, \apjs,
  173, 619

\bibitem[{{Kaviraj} {et~al.}(2011){Kaviraj}, {Tan}, {Ellis}, \&
  {Silk}}]{kaviraj11}
{Kaviraj}, S., {Tan}, K.-M., {Ellis}, R.~S., \& {Silk}, J. 2011, \mnras, 411,
  2148

\bibitem[{{Khochfar} \& {Burkert}(2006)}]{khochfar06orbit}
{Khochfar}, S. \& {Burkert}, A. 2006, \aap, 445, 403

\bibitem[{{Khochfar} \& {Silk}(2006)}]{khochfar06}
{Khochfar}, S. \& {Silk}, J. 2006, \apjl, 648, L21

\bibitem[{{Khochfar} \& {Silk}(2009)}]{khochfar09}
{Khochfar}, S. \& {Silk}, J. 2009, \mnras, 397, 506

\bibitem[{{Kitzbichler} \& {White}(2008)}]{kit08}
{Kitzbichler}, M.~G. \& {White}, S.~D.~M. 2008, \mnras, 1300

\bibitem[{{Koekemoer} {et~al.}(2007){Koekemoer}, {Aussel}, {Calzetti}, {Capak},
  {Giavalisco}, {Kneib}, {Leauthaud}, {Le F{\`e}vre}, {McCracken}, {Massey},
  {Mobasher}, {Rhodes}, {Scoville}, \& {Shopbell}}]{koekemoer07}
{Koekemoer}, A.~M., {Aussel}, H., {Calzetti}, D., {et~al.} 2007, \apjs, 172,
  196

\bibitem[{{Koekemoer} {et~al.}(2011){Koekemoer}, {Faber}, {Ferguson}, {Grogin},
  {Kocevski}, {Koo}, {Lai}, {Lotz}, {Lucas}, {McGrath}, {Ogaz}, {Rajan},
  {Riess}, {Rodney}, {Strolger}, {Casertano}, {Castellano}, {Dahlen},
  {Dickinson}, {Dolch}, {Fontana}, {Giavalisco}, {Grazian}, {Guo}, {Hathi},
  {Huang}, {van der Wel}, {Yan}, {Acquaviva}, {Alexander}, {Almaini}, {Ashby},
  {Barden}, {Bell}, {Bournaud}, {Brown}, {Caputi}, {Cassata}, {Challis},
  {Chary}, {Cheung}, {Cirasuolo}, {Conselice}, {Roshan Cooray}, {Croton},
  {Daddi}, {Dav{\'e}}, {de Mello}, {de Ravel}, {Dekel}, {Donley}, {Dunlop},
  {Dutton}, {Elbaz}, {Fazio}, {Filippenko}, {Finkelstein}, {Frazer}, {Gardner},
  {Garnavich}, {Gawiser}, {Gruetzbauch}, {Hartley}, {H{\"a}ussler},
  {Herrington}, {Hopkins}, {Huang}, {Jha}, {Johnson}, {Kartaltepe},
  {Khostovan}, {Kirshner}, {Lani}, {Lee}, {Li}, {Madau}, {McCarthy},
  {McIntosh}, {McLure}, {McPartland}, {Mobasher}, {Moreira}, {Mortlock},
  {Moustakas}, {Mozena}, {Nandra}, {Newman}, {Nielsen}, {Niemi}, {Noeske},
  {Papovich}, {Pentericci}, {Pope}, {Primack}, {Ravindranath}, {Reddy},
  {Renzini}, {Rix}, {Robaina}, {Rosario}, {Rosati}, {Salimbeni}, {Scarlata},
  {Siana}, {Simard}, {Smidt}, {Snyder}, {Somerville}, {Spinrad}, {Straughn},
  {Telford}, {Teplitz}, {Trump}, {Vargas}, {Villforth}, {Wagner}, {Wandro},
  {Wechsler}, {Weiner}, {Wiklind}, {Wild}, {Wilson}, {Wuyts}, \&
  {Yun}}]{candels2}
{Koekemoer}, A.~M., {Faber}, S.~M., {Ferguson}, H.~C., {et~al.} 2011, \apjs,
  197, 36

\bibitem[{{Kova{\v c}} {et~al.}(2010){Kova{\v c}}, {Lilly}, {Cucciati},
  {Porciani}, {Iovino}, {Zamorani}, {Oesch}, {Bolzonella}, {Knobel},
  {Finoguenov}, {Peng}, {Carollo}, {Pozzetti}, {Caputi}, {Silverman}, {Tasca},
  {Scodeggio}, {Vergani}, {Scoville}, {Capak}, {Contini}, {Kneib}, {Le
  F{\`e}vre}, {Mainieri}, {Renzini}, {Bardelli}, {Bongiorno}, {Coppa}, {de la
  Torre}, {de Ravel}, {Franzetti}, {Garilli}, {Guzzo}, {Kampczyk},
  {Lamareille}, {Le Borgne}, {Le Brun}, {Maier}, {Mignoli}, {Pello}, {Perez
  Montero}, {Ricciardelli}, {Tanaka}, {Tresse}, {Zucca}, {Abbas}, {Bottini},
  {Cappi}, {Cassata}, {Cimatti}, {Fumana}, {Koekemoer}, {Maccagni}, {Marinoni},
  {McCracken}, {Memeo}, {Meneux}, \& {Scaramella}}]{kovac10}
{Kova{\v c}}, K., {Lilly}, S.~J., {Cucciati}, O., {et~al.} 2010, \apj, 708, 505

\bibitem[{{Le F{\`e}vre} {et~al.}(2000){Le F{\`e}vre}, {Abraham}, {Lilly},
  {Ellis}, {Brinchmann}, {Schade}, {Tresse}, {Colless}, {Crampton},
  {Glazebrook}, {Hammer}, \& {Broadhurst}}]{lefevre00}
{Le F{\`e}vre}, O., {Abraham}, R., {Lilly}, S.~J., {et~al.} 2000, \mnras, 311,
  565

\bibitem[{{Le F{\`e}vre} {et~al.}(2003){Le F{\`e}vre}, {Saisse}, {Mancini},
  {Brau-Nogue}, {Caputi}, {Castinel}, {D'Odorico}, {Garilli}, {Kissler-Patig},
  {Lucuix}, {Mancini}, {Pauget}, {Sciarretta}, {Scodeggio}, {Tresse}, \&
  {Vettolani}}]{lefevre03}
{Le F{\`e}vre}, O., {Saisse}, M., {Mancini}, D., {et~al.} 2003, in Society of
  Photo-Optical Instrumentation Engineers (SPIE) Conference Series, Vol. 4841,
  Society of Photo-Optical Instrumentation Engineers (SPIE) Conference Series,
  ed. {M.~Iye \& A.~F.~M.~Moorwood}, 1670--1681

\bibitem[{{Le F{\`e}vre} {et~al.}(2005){Le F{\`e}vre}, {Vettolani}, {Garilli},
  {Tresse}, {Bottini}, {Le Brun}, {Maccagni}, {Picat}, {Scaramella},
  {Scodeggio}, {Zanichelli}, {Adami}, {Arnaboldi}, {Arnouts}, {Bardelli},
  {Bolzonella}, {Cappi}, {Charlot}, {Ciliegi}, {Contini}, {Foucaud},
  {Franzetti}, {Gavignaud}, {Guzzo}, {Ilbert}, {Iovino}, {McCracken}, {Marano},
  {Marinoni}, {Mathez}, {Mazure}, {Meneux}, {Merighi}, {Paltani}, {Pell{\`o}},
  {Pollo}, {Pozzetti}, {Radovich}, {Zamorani}, {Zucca}, {Bondi}, {Bongiorno},
  {Busarello}, {Lamareille}, {Mellier}, {Merluzzi}, {Ripepi}, \&
  {Rizzo}}]{lefevre05}
{Le F{\`e}vre}, O., {Vettolani}, G., {Garilli}, B., {et~al.} 2005, \aap, 439,
  845

\bibitem[{{Lilly} {et~al.}(2009){Lilly}, {Le Brun}, {Maier}, {Mainieri},
  {Mignoli}, {Scodeggio}, {Zamorani}, {Carollo}, {Contini}, {Kneib}, {Le
  F{\`e}vre}, {Renzini}, {Bardelli}, {Bolzonella}, {Bongiorno}, {Caputi},
  {Coppa}, {Cucciati}, {de la Torre}, {de Ravel}, {Franzetti}, {Garilli},
  {Iovino}, {Kampczyk}, {Kovac}, {Knobel}, {Lamareille}, {Le Borgne}, {Pello},
  {Peng}, {P{\'e}rez-Montero}, {Ricciardelli}, {Silverman}, {Tanaka}, {Tasca},
  {Tresse}, {Vergani}, {Zucca}, {Ilbert}, {Salvato}, {Oesch}, {Abbas},
  {Bottini}, {Capak}, {Cappi}, {Cassata}, {Cimatti}, {Elvis}, {Fumana},
  {Guzzo}, {Hasinger}, {Koekemoer}, {Leauthaud}, {Maccagni}, {Marinoni},
  {McCracken}, {Memeo}, {Meneux}, {Porciani}, {Pozzetti}, {Sanders},
  {Scaramella}, {Scarlata}, {Scoville}, {Shopbell}, \&
  {Taniguchi}}]{zcosmos10k}
{Lilly}, S.~J., {Le Brun}, V., {Maier}, C., {et~al.} 2009, \apjs, 184, 218

\bibitem[{{Lilly} {et~al.}(2007){Lilly}, {Le F{\`e}vre}, {Renzini}, {Zamorani},
  {Scodeggio}, {Contini}, {Carollo}, {Hasinger}, {Kneib}, {Iovino}, {Le Brun},
  {Maier}, {Mainieri}, {Mignoli}, {Silverman}, {Tasca}, {Bolzonella},
  {Bongiorno}, {Bottini}, {Capak}, {Caputi}, {Cimatti}, {Cucciati}, {Daddi},
  {Feldmann}, {Franzetti}, {Garilli}, {Guzzo}, {Ilbert}, {Kampczyk}, {Kovac},
  {Lamareille}, {Leauthaud}, {Borgne}, {McCracken}, {Marinoni}, {Pello},
  {Ricciardelli}, {Scarlata}, {Vergani}, {Sanders}, {Schinnerer}, {Scoville},
  {Taniguchi}, {Arnouts}, {Aussel}, {Bardelli}, {Brusa}, {Cappi}, {Ciliegi},
  {Finoguenov}, {Foucaud}, {Franceschini}, {Halliday}, {Impey}, {Knobel},
  {Koekemoer}, {Kurk}, {Maccagni}, {Maddox}, {Marano}, {Marconi}, {Meneux},
  {Mobasher}, {Moreau}, {Peacock}, {Porciani}, {Pozzetti}, {Scaramella},
  {Schiminovich}, {Shopbell}, {Smail}, {Thompson}, {Tresse}, {Vettolani},
  {Zanichelli}, \& {Zucca}}]{lilly07}
{Lilly}, S.~J., {Le F{\`e}vre}, O., {Renzini}, A., {et~al.} 2007, \apjs, 172,
  70

\bibitem[{{Lin} {et~al.}(2010){Lin}, {Cooper}, {Jian}, {Koo}, {Patton}, {Yan},
  {Willmer}, {Coil}, {Chiueh}, {Croton}, {Gerke}, {Lotz}, {Guhathakurta}, \&
  {Newman}}]{lin10}
{Lin}, L., {Cooper}, M.~C., {Jian}, H., {et~al.} 2010, \apj, 718, 1158

\bibitem[{{Lin} {et~al.}(2004){Lin}, {Koo}, {Willmer}, {Patton}, {Conselice},
  {Yan}, {Coil}, {Cooper}, {Davis}, {Faber}, {Gerke}, {Guhathakurta}, \&
  {Newman}}]{lin04}
{Lin}, L., {Koo}, D.~C., {Willmer}, C.~N.~A., {et~al.} 2004, \apjl, 617, L9

\bibitem[{{Lin} {et~al.}(2008){Lin}, {Patton}, {Koo}, {Casteels}, {Conselice},
  {Faber}, {Lotz}, {Willmer}, {Hsieh}, {Chiueh}, {Newman}, {Novak}, {Weiner},
  \& {Cooper}}]{lin08}
{Lin}, L., {Patton}, D.~R., {Koo}, D.~C., {et~al.} 2008, \apj, 681, 232

\bibitem[{{L{\'o}pez-Sanjuan} {et~al.}(2010{\natexlab{a}}){L{\'o}pez-Sanjuan},
  {Balcells}, {P{\'e}rez-Gonz{\'a}lez}, {Barro}, {Gallego}, \&
  {Zamorano}}]{clsj10pargoods}
{L{\'o}pez-Sanjuan}, C., {Balcells}, M., {P{\'e}rez-Gonz{\'a}lez}, P.~G.,
  {et~al.} 2010{\natexlab{a}}, \aap, 518, A20

\bibitem[{{L{\'o}pez-Sanjuan} {et~al.}(2009){L{\'o}pez-Sanjuan}, {Balcells},
  {P{\'e}rez-Gonz{\'a}lez}, {Barro}, {Garc{\'{\i}}a-Dab{\'o}}, {Gallego}, \&
  {Zamorano}}]{clsj09ffgoods}
{L{\'o}pez-Sanjuan}, C., {Balcells}, M., {P{\'e}rez-Gonz{\'a}lez}, P.~G.,
  {et~al.} 2009, \aap, 501, 505

\bibitem[{{L{\'o}pez-Sanjuan} {et~al.}(2010{\natexlab{b}}){L{\'o}pez-Sanjuan},
  {Balcells}, {P{\'e}rez-Gonz{\'a}lez}, {Barro}, {Garc{\'{\i}}a-Dab{\'o}},
  {Gallego}, \& {Zamorano}}]{clsj10megoods}
{L{\'o}pez-Sanjuan}, C., {Balcells}, M., {P{\'e}rez-Gonz{\'a}lez}, P.~G.,
  {et~al.} 2010{\natexlab{b}}, \apj, 710, 1170

\bibitem[{{L{\'o}pez-Sanjuan} {et~al.}(2011){L{\'o}pez-Sanjuan}, {Le
  F{\`e}vre}, {de Ravel}, {Cucciati}, {Ilbert}, {Tresse}, {Bardelli},
  {Bolzonella}, {Contini}, {Garilli}, {Guzzo}, {Maccagni}, {McCracken},
  {Mellier}, {Pollo}, {Vergani}, \& {Zucca}}]{clsj11mmvvds}
{L{\'o}pez-Sanjuan}, C., {Le F{\`e}vre}, O., {de Ravel}, L., {et~al.} 2011,
  \aap, 530, A20

\bibitem[{{Lotz} {et~al.}(2011){Lotz}, {Jonsson}, {Cox}, {Croton}, {Primack},
  {Somerville}, \& {Stewart}}]{lotz11}
{Lotz}, J.~M., {Jonsson}, P., {Cox}, T.~J., {et~al.} 2011, \apj, 742, 103

\bibitem[{{Lotz} {et~al.}(2010{\natexlab{a}}){Lotz}, {Jonsson}, {Cox}, \&
  {Primack}}]{lotz10gas}
{Lotz}, J.~M., {Jonsson}, P., {Cox}, T.~J., \& {Primack}, J.~R.
  2010{\natexlab{a}}, \mnras, 404, 590

\bibitem[{{Lotz} {et~al.}(2010{\natexlab{b}}){Lotz}, {Jonsson}, {Cox}, \&
  {Primack}}]{lotz10t}
{Lotz}, J.~M., {Jonsson}, P., {Cox}, T.~J., \& {Primack}, J.~R.
  2010{\natexlab{b}}, \mnras, 404, 575

\bibitem[{{M{\'a}rmol-Queralt{\'o}} {et~al.}(2012){M{\'a}rmol-Queralt{\'o}},
  {Trujillo}, {P{\'e}rez-Gonz{\'a}lez}, {Varela}, \& {Barro}}]{marmol12}
{M{\'a}rmol-Queralt{\'o}}, E., {Trujillo}, I., {P{\'e}rez-Gonz{\'a}lez}, P.~G.,
  {Varela}, J., \& {Barro}, G. 2012, \mnras, 422, 2187

\bibitem[{{Martinez-Manso} {et~al.}(2011){Martinez-Manso}, {Guzman}, {Barro},
  {Cenarro}, {Perez-Gonzalez}, {Sanchez-Blazquez}, {Trujillo}, {Balcells},
  {Cardiel}, {Gallego}, {Hempel}, \& {Prieto}}]{martinez11}
{Martinez-Manso}, J., {Guzman}, R., {Barro}, G., {et~al.} 2011, \apjl, 738, L22

\bibitem[{{M{\'e}ndez-Abreu} {et~al.}(2012){M{\'e}ndez-Abreu}, {Aguerri},
  {Barrena}, {S{\'a}nchez-Janssen}, {Boschin}, {Castro-Rodriguez}, {Corsini},
  {Del Burgo}, {D'Onghia}, {Girardi}, {Iglesias-P{\'a}ramo}, {Napolitano},
  {Vilchez}, \& {Zarattini}}]{jairo12}
{M{\'e}ndez-Abreu}, J., {Aguerri}, J.~A.~L., {Barrena}, R., {et~al.} 2012,
  \aap, 537, A25

\bibitem[{{Moster} {et~al.}(2011){Moster}, {Somerville}, {Newman}, \&
  {Rix}}]{moster11}
{Moster}, B.~P., {Somerville}, R.~S., {Newman}, J.~A., \& {Rix}, H.-W. 2011,
  \apj, 731, 113

\bibitem[{{Naab} {et~al.}(2009){Naab}, {Johansson}, \& {Ostriker}}]{naab09}
{Naab}, T., {Johansson}, P.~H., \& {Ostriker}, J.~P. 2009, \apjl, 699, L178

\bibitem[{{Nair} {et~al.}(2011){Nair}, {van den Bergh}, \& {Abraham}}]{nair11}
{Nair}, P., {van den Bergh}, S., \& {Abraham}, R.~G. 2011, \apjl, 734, L31+

\bibitem[{{Newman} {et~al.}(2012){Newman}, {Ellis}, {Bundy}, \&
  {Treu}}]{newman12}
{Newman}, A.~B., {Ellis}, R.~S., {Bundy}, K., \& {Treu}, T. 2012, \apj, 746,
  162

\bibitem[{{Newman} {et~al.}(2010){Newman}, {Ellis}, {Treu}, \&
  {Bundy}}]{newman10}
{Newman}, A.~B., {Ellis}, R.~S., {Treu}, T., \& {Bundy}, K. 2010, \apjl, 717,
  L103

\bibitem[{{Nipoti} {et~al.}(2009){Nipoti}, {Treu}, {Auger}, \&
  {Bolton}}]{nipoti09}
{Nipoti}, C., {Treu}, T., {Auger}, M.~W., \& {Bolton}, A.~S. 2009, \apjl, 706,
  L86

\bibitem[{{Oesch} {et~al.}(2010){Oesch}, {Carollo}, {Feldmann}, {Hahn},
  {Lilly}, {Sargent}, {Scarlata}, {Aller}, {Aussel}, {Bolzonella}, {Bschorr},
  {Bundy}, {Capak}, {Ilbert}, {Kneib}, {Koekemoer}, {Kova{\v c}}, {Leauthaud},
  {Le Floc'h}, {Massey}, {McCracken}, {Pozzetti}, {Renzini}, {Rhodes},
  {Salvato}, {Sanders}, {Scoville}, {Sheth}, {Taniguchi}, \&
  {Thompson}}]{oesch10}
{Oesch}, P.~A., {Carollo}, C.~M., {Feldmann}, R., {et~al.} 2010, \apjl, 714,
  L47

\bibitem[{{Oser} {et~al.}(2012){Oser}, {Naab}, {Ostriker}, \&
  {Johansson}}]{oser12}
{Oser}, L., {Naab}, T., {Ostriker}, J.~P., \& {Johansson}, P.~H. 2012, \apj,
  744, 63

\bibitem[{{Patton} \& {Atfield}(2008)}]{patton08}
{Patton}, D.~R. \& {Atfield}, J.~E. 2008, \apj, 685, 235

\bibitem[{{Patton} {et~al.}(2000){Patton}, {Carlberg}, {Marzke}, {Pritchet},
  {da Costa}, \& {Pellegrini}}]{patton00}
{Patton}, D.~R., {Carlberg}, R.~G., {Marzke}, R.~O., {et~al.} 2000, \apj, 536,
  153

\bibitem[{{Peng} {et~al.}(2010){Peng}, {Lilly}, {Kova{\v c}}, {Bolzonella},
  {Pozzetti}, {Renzini}, {Zamorani}, {Ilbert}, {Knobel}, {Iovino}, {Maier},
  {Cucciati}, {Tasca}, {Carollo}, {Silverman}, {Kampczyk}, {de Ravel},
  {Sanders}, {Scoville}, {Contini}, {Mainieri}, {Scodeggio}, {Kneib}, {Le
  F{\`e}vre}, {Bardelli}, {Bongiorno}, {Caputi}, {Coppa}, {de la Torre},
  {Franzetti}, {Garilli}, {Lamareille}, {Le Borgne}, {Le Brun}, {Mignoli},
  {Perez Montero}, {Pello}, {Ricciardelli}, {Tanaka}, {Tresse}, {Vergani},
  {Welikala}, {Zucca}, {Oesch}, {Abbas}, {Barnes}, {Bordoloi}, {Bottini},
  {Cappi}, {Cassata}, {Cimatti}, {Fumana}, {Hasinger}, {Koekemoer},
  {Leauthaud}, {Maccagni}, {Marinoni}, {McCracken}, {Memeo}, {Meneux}, {Nair},
  {Porciani}, {Presotto}, \& {Scaramella}}]{peng10}
{Peng}, Y., {Lilly}, S.~J., {Kova{\v c}}, K., {et~al.} 2010, \apj, 721, 193

\bibitem[{{P{\'e}rez-Gonz{\'a}lez} {et~al.}(2008){P{\'e}rez-Gonz{\'a}lez},
  {Rieke}, {Villar}, {Barro}, {Blaylock}, {Egami}, {Gallego}, {Gil de Paz},
  {Pascual}, {Zamorano}, \& {Donley}}]{pgon08}
{P{\'e}rez-Gonz{\'a}lez}, P.~G., {Rieke}, G.~H., {Villar}, V., {et~al.} 2008,
  \apj, 675, 234

\bibitem[{{Pozzetti} {et~al.}(2007){Pozzetti}, {Bolzonella}, {Lamareille},
  {Zamorani}, {Franzetti}, {Le F{\`e}vre}, {Iovino}, {Temporin}, {Ilbert},
  {Arnouts}, {Charlot}, {Brinchmann}, {Zucca}, {Tresse}, {Scodeggio}, {Guzzo},
  {Bottini}, {Garilli}, {Le Brun}, {Maccagni}, {Picat}, {Scaramella},
  {Vettolani}, {Zanichelli}, {Adami}, {Bardelli}, {Cappi}, {Ciliegi},
  {Contini}, {Foucaud}, {Gavignaud}, {McCracken}, {Marano}, {Marinoni},
  {Mazure}, {Meneux}, {Merighi}, {Paltani}, {Pell{\`o}}, {Pollo}, {Radovich},
  {Bondi}, {Bongiorno}, {Cucciati}, {de la Torre}, {Gregorini}, {Mellier},
  {Merluzzi}, {Vergani}, \& {Walcher}}]{pozzetti07}
{Pozzetti}, L., {Bolzonella}, M., {Lamareille}, F., {et~al.} 2007, \aap, 474,
  443

\bibitem[{{Pozzetti} {et~al.}(2010){Pozzetti}, {Bolzonella}, {Zucca},
  {Zamorani}, {Lilly}, {Renzini}, {Moresco}, {Mignoli}, {Cassata}, {Tasca},
  {Lamareille}, {Maier}, {Meneux}, {Halliday}, {Oesch}, {Vergani}, {Caputi},
  {Kova{\v c}}, {Cimatti}, {Cucciati}, {Iovino}, {Peng}, {Carollo}, {Contini},
  {Kneib}, {Le F{\'e}vre}, {Mainieri}, {Scodeggio}, {Bardelli}, {Bongiorno},
  {Coppa}, {de la Torre}, {de Ravel}, {Franzetti}, {Garilli}, {Kampczyk},
  {Knobel}, {Le Borgne}, {Le Brun}, {Pell{\`o}}, {Perez Montero},
  {Ricciardelli}, {Silverman}, {Tanaka}, {Tresse}, {Abbas}, {Bottini}, {Cappi},
  {Guzzo}, {Koekemoer}, {Leauthaud}, {Maccagni}, {Marinoni}, {McCracken},
  {Memeo}, {Porciani}, {Scaramella}, {Scarlata}, \& {Scoville}}]{pozzetti10}
{Pozzetti}, L., {Bolzonella}, M., {Zucca}, E., {et~al.} 2010, \aap, 523, A13

\bibitem[{{Ragone-Figueroa} \& {Granato}(2011)}]{ragone11}
{Ragone-Figueroa}, C. \& {Granato}, G.~L. 2011, \mnras, 414, 3690

\bibitem[{{Rix} {et~al.}(2004){Rix}, {Barden}, {Beckwith}, {Bell}, {Borch},
  {Caldwell}, {H{\"a}ussler}, {Jahnke}, {Jogee}, {McIntosh}, {Meisenheimer},
  {Peng}, {Sanchez}, {Somerville}, {Wisotzki}, \& {Wolf}}]{rix04}
{Rix}, H.-W., {Barden}, M., {Beckwith}, S.~V.~W., {et~al.} 2004, \apjs, 152,
  163

\bibitem[{{Robaina} {et~al.}(2010){Robaina}, {Bell}, {van der Wel},
  {Somerville}, {Skelton}, {McIntosh}, {Meisenheimer}, \& {Wolf}}]{robaina10}
{Robaina}, A.~R., {Bell}, E.~F., {van der Wel}, A., {et~al.} 2010, \apj, 719,
  844

\bibitem[{{Saglia} {et~al.}(2010){Saglia}, {S{\'a}nchez-Bl{\'a}zquez},
  {Bender}, {Simard}, {Desai}, {Arag{\'o}n-Salamanca}, {Milvang-Jensen},
  {Halliday}, {Jablonka}, {Noll}, {Poggianti}, {Clowe}, {De Lucia},
  {Pell{\'o}}, {Rudnick}, {Valentinuzzi}, {White}, \& {Zaritsky}}]{saglia10}
{Saglia}, R.~P., {S{\'a}nchez-Bl{\'a}zquez}, P., {Bender}, R., {et~al.} 2010,
  \aap, 524, A6+

\bibitem[{{Saracco} {et~al.}(2010){Saracco}, {Longhetti}, \&
  {Gargiulo}}]{saracco10}
{Saracco}, P., {Longhetti}, M., \& {Gargiulo}, A. 2010, \mnras, 408, L21

\bibitem[{{Scarlata} {et~al.}(2007){Scarlata}, {Carollo}, {Lilly}, {Feldmann},
  {Kampczyk}, {Renzini}, {Cimatti}, {Halliday}, {Daddi}, {Sargent},
  {Koekemoer}, {Scoville}, {Kneib}, {Leauthaud}, {Massey}, {Rhodes}, {Tasca},
  {Capak}, {McCracken}, {Mobasher}, {Taniguchi}, {Thompson}, {Ajiki}, {Aussel},
  {Murayama}, {Sanders}, {Sasaki}, {Shioya}, \& {Takahashi}}]{scarlata07ee}
{Scarlata}, C., {Carollo}, C.~M., {Lilly}, S.~J., {et~al.} 2007, \apjs, 172,
  494

\bibitem[{{Scodeggio} {et~al.}(2005){Scodeggio}, {Franzetti}, {Garilli},
  {Zanichelli}, {Paltani}, {Maccagni}, {Bottini}, {Le Brun}, {Contini},
  {Scaramella}, {Adami}, {Bardelli}, {Zucca}, {Tresse}, {Ilbert}, {Foucaud},
  {Iovino}, {Merighi}, {Zamorani}, {Gavignaud}, {Rizzo}, {McCracken}, {Le
  F{\`e}vre}, {Picat}, {Vettolani}, {Arnaboldi}, {Arnouts}, {Bolzonella},
  {Cappi}, {Charlot}, {Ciliegi}, {Guzzo}, {Marano}, {Marinoni}, {Mathez},
  {Mazure}, {Meneux}, {Pell{\`o}}, {Pollo}, {Pozzetti}, \&
  {Radovich}}]{scodeggio05}
{Scodeggio}, M., {Franzetti}, P., {Garilli}, B., {et~al.} 2005, \pasp, 117,
  1284

\bibitem[{{Scoville} {et~al.}(2007){Scoville}, {Aussel}, {Brusa}, {Capak},
  {Carollo}, {Elvis}, {Giavalisco}, {Guzzo}, {Hasinger}, {Impey}, {Kneib},
  {LeFevre}, {Lilly}, {Mobasher}, {Renzini}, {Rich}, {Sanders}, {Schinnerer},
  {Schminovich}, {Shopbell}, {Taniguchi}, \& {Tyson}}]{cosmos}
{Scoville}, N., {Aussel}, H., {Brusa}, M., {et~al.} 2007, \apjs, 172, 1

\bibitem[{{S\'ersic}(1968)}]{sersic68}
{S\'ersic}, J.~L. 1968, {Atlas de galaxias australes} (Cordoba, Argentina:
  Observatorio Astronomico, 1968)

\bibitem[{{Shankar} {et~al.}(2010){Shankar}, {Marulli}, {Bernardi}, {Dai},
  {Hyde}, \& {Sheth}}]{shankar10}
{Shankar}, F., {Marulli}, F., {Bernardi}, M., {et~al.} 2010, \mnras, 403, 117

\bibitem[{{Shankar} {et~al.}(2011){Shankar}, {Marulli}, {Bernardi}, {Mei},
  {Meert}, \& {Vikram}}]{shankar11}
{Shankar}, F., {Marulli}, F., {Bernardi}, M., {et~al.} 2011, \mnras, in press
  [ArXiv: 1105.6043]

\bibitem[{{Shen} {et~al.}(2003){Shen}, {Mo}, {White}, {Blanton}, {Kauffmann},
  {Voges}, {Brinkmann}, \& {Csabai}}]{shen03}
{Shen}, S., {Mo}, H.~J., {White}, S.~D.~M., {et~al.} 2003, \mnras, 343, 978

\bibitem[{{Springel} {et~al.}(2005){Springel}, {White}, {Jenkins}, {Frenk},
  {Yoshida}, {Gao}, {Navarro}, {Thacker}, {Croton}, {Helly}, {Peacock}, {Cole},
  {Thomas}, {Couchman}, {Evrard}, {Colberg}, \& {Pearce}}]{springel05}
{Springel}, V., {White}, S.~D.~M., {Jenkins}, A., {et~al.} 2005, \nat, 435, 629

\bibitem[{{Strateva} {et~al.}(2001){Strateva}, {Ivezi{\'c}}, {Knapp},
  {Narayanan}, {Strauss}, {Gunn}, {Lupton}, {Schlegel}, {Bahcall}, {Brinkmann},
  {Brunner}, {Budav{\'a}ri}, {Csabai}, {Castander}, {Doi}, {Fukugita}, {Gy{\H
  o}ry}, {Hamabe}, {Hennessy}, {Ichikawa}, {Kunszt}, {Lamb}, {McKay},
  {Okamura}, {Racusin}, {Sekiguchi}, {Schneider}, {Shimasaku}, \&
  {York}}]{strateva01}
{Strateva}, I., {Ivezi{\'c}}, {\v Z}., {Knapp}, G.~R., {et~al.} 2001, \aj, 122,
  1861

\bibitem[{{Szomoru} {et~al.}(2010){Szomoru}, {Franx}, {van Dokkum}, {Trenti},
  {Illingworth}, {Labb{\'e}}, {Bouwens}, {Oesch}, \& {Carollo}}]{szomoru10}
{Szomoru}, D., {Franx}, M., {van Dokkum}, P.~G., {et~al.} 2010, \apjl, 714,
  L244

\bibitem[{{Tal} {et~al.}(2009){Tal}, {van Dokkum}, {Nelan}, \&
  {Bezanson}}]{tal09}
{Tal}, T., {van Dokkum}, P.~G., {Nelan}, J., \& {Bezanson}, R. 2009, \aj, 138,
  1417

\bibitem[{{Tal} {et~al.}(2012){Tal}, {Wake}, {van Dokkum}, {van den Bosch},
  {Schneider}, {Brinkmann}, \& {Weaver}}]{tal12}
{Tal}, T., {Wake}, D.~A., {van Dokkum}, P.~G., {et~al.} 2012, \apj, 746, 138

\bibitem[{{Tasca} {et~al.}(2009){Tasca}, {Kneib}, {Iovino}, {Le F{\`e}vre},
  {Kova{\v c}}, {Bolzonella}, {Lilly}, {Abraham}, {Cassata}, {Cucciati},
  {Guzzo}, {Tresse}, {Zamorani}, {Capak}, {Garilli}, {Scodeggio}, {Sheth},
  {Zucca}, {Carollo}, {Contini}, {Mainieri}, {Renzini}, {Bardelli},
  {Bongiorno}, {Caputi}, {Coppa}, {de La Torre}, {de Ravel}, {Franzetti},
  {Kampczyk}, {Knobel}, {Koekemoer}, {Lamareille}, {Le Borgne}, {Le Brun},
  {Maier}, {Mignoli}, {Pello}, {Peng}, {Perez Montero}, {Ricciardelli},
  {Silverman}, {Vergani}, {Tanaka}, {Abbas}, {Bottini}, {Cappi}, {Cimatti},
  {Ilbert}, {Leauthaud}, {Maccagni}, {Marinoni}, {McCracken}, {Memeo},
  {Meneux}, {Oesch}, {Porciani}, {Pozzetti}, {Scaramella}, \&
  {Scarlata}}]{tasca09}
{Tasca}, L.~A.~M., {Kneib}, J.-P., {Iovino}, A., {et~al.} 2009, \aap, 503, 379

\bibitem[{{Taylor} {et~al.}(2010){Taylor}, {Franx}, {Glazebrook}, {Brinchmann},
  {van der Wel}, \& {van Dokkum}}]{taylor10}
{Taylor}, E.~N., {Franx}, M., {Glazebrook}, K., {et~al.} 2010, \apj, 720, 723

\bibitem[{{Toft} {et~al.}(2009){Toft}, {Franx}, {van Dokkum}, {F{\"o}rster
  Schreiber}, {Labbe}, {Wuyts}, \& {Marchesini}}]{toft09}
{Toft}, S., {Franx}, M., {van Dokkum}, P., {et~al.} 2009, \apj, 705, 255

\bibitem[{{Trujillo} {et~al.}(2009){Trujillo}, {Cenarro}, {de
  Lorenzo-C{\'a}ceres}, {Vazdekis}, {de la Rosa}, \& {Cava}}]{trujillo09}
{Trujillo}, I., {Cenarro}, A.~J., {de Lorenzo-C{\'a}ceres}, A., {et~al.} 2009,
  \apjl, 692, L118

\bibitem[{{Trujillo} {et~al.}(2007){Trujillo}, {Conselice}, {Bundy}, {Cooper},
  {Eisenhardt}, \& {Ellis}}]{trujillo07}
{Trujillo}, I., {Conselice}, C.~J., {Bundy}, K., {et~al.} 2007, \mnras, 382,
  109

\bibitem[{{Trujillo} {et~al.}(2011){Trujillo}, {Ferreras}, \& {de La
  Rosa}}]{trujillo11}
{Trujillo}, I., {Ferreras}, I., \& {de La Rosa}, I.~G. 2011, \mnras, 938

\bibitem[{{Trujillo} {et~al.}(2006){Trujillo}, {F{\"o}rster Schreiber},
  {Rudnick}, {Barden}, {Franx}, {Rix}, {Caldwell}, {McIntosh}, {Toft},
  {H{\"a}ussler}, {Zirm}, {van Dokkum}, {Labb{\'e}}, {Moorwood},
  {R{\"o}ttgering}, {van der Wel}, {van der Werf}, \& {van
  Starkenburg}}]{trujillo06}
{Trujillo}, I., {F{\"o}rster Schreiber}, N.~M., {Rudnick}, G., {et~al.} 2006,
  \apj, 650, 18

\bibitem[{{Valentinuzzi} {et~al.}(2010{\natexlab{a}}){Valentinuzzi}, {Fritz},
  {Poggianti}, {Cava}, {Bettoni}, {Fasano}, {D'Onofrio}, {Couch}, {Dressler},
  {Moles}, {Moretti}, {Omizzolo}, {Kj{\ae}rgaard}, {Vanzella}, \&
  {Varela}}]{vale10a}
{Valentinuzzi}, T., {Fritz}, J., {Poggianti}, B.~M., {et~al.}
  2010{\natexlab{a}}, \apj, 712, 226

\bibitem[{{Valentinuzzi} {et~al.}(2010{\natexlab{b}}){Valentinuzzi},
  {Poggianti}, {Saglia}, {Arag{\'o}n-Salamanca}, {Simard},
  {S{\'a}nchez-Bl{\'a}zquez}, {D'onofrio}, {Cava}, {Couch}, {Fritz}, {Moretti},
  \& {Vulcani}}]{vale10b}
{Valentinuzzi}, T., {Poggianti}, B.~M., {Saglia}, R.~P., {et~al.}
  2010{\natexlab{b}}, \apjl, 721, L19

\bibitem[{{van de Sande} {et~al.}(2011){van de Sande}, {Kriek}, {Franx}, {van
  Dokkum}, {Bezanson}, {Whitaker}, {Brammer}, {Labb{\'e}}, {Groot}, \&
  {Kaper}}]{vandesande11}
{van de Sande}, J., {Kriek}, M., {Franx}, M., {et~al.} 2011, \apjl, 736, L9+

\bibitem[{{van der Wel} {et~al.}(2009{\natexlab{a}}){van der Wel}, {Bell}, {van
  den Bosch}, {Gallazzi}, \& {Rix}}]{vanderwel09}
{van der Wel}, A., {Bell}, E.~F., {van den Bosch}, F.~C., {Gallazzi}, A., \&
  {Rix}, H.-W. 2009{\natexlab{a}}, \apj, 698, 1232

\bibitem[{{van der Wel} {et~al.}(2008){van der Wel}, {Holden}, {Zirm}, {Franx},
  {Rettura}, {Illingworth}, \& {Ford}}]{vanderwel08esize}
{van der Wel}, A., {Holden}, B.~P., {Zirm}, A.~W., {et~al.} 2008, \apj, 688, 48

\bibitem[{{van der Wel} {et~al.}(2009{\natexlab{b}}){van der Wel}, {Rix},
  {Holden}, {Bell}, \& {Robaina}}]{vanderwel09ba}
{van der Wel}, A., {Rix}, H., {Holden}, B.~P., {Bell}, E.~F., \& {Robaina},
  A.~R. 2009{\natexlab{b}}, \apjl, 706, L120

\bibitem[{{van Dokkum}(2005)}]{vandokkum05}
{van Dokkum}, P.~G. 2005, \aj, 130, 2647

\bibitem[{{van Dokkum} {et~al.}(2008){van Dokkum}, {Franx}, {Kriek}, {Holden},
  {Illingworth}, {Magee}, {Bouwens}, {Marchesini}, {Quadri}, {Rudnick},
  {Taylor}, \& {Toft}}]{vandokkum08}
{van Dokkum}, P.~G., {Franx}, M., {Kriek}, M., {et~al.} 2008, \apjl, 677, L5

\bibitem[{{van Dokkum} {et~al.}(2009){van Dokkum}, {Labb{\'e}}, {Marchesini},
  {Quadri}, {Brammer}, {Whitaker}, {Kriek}, {Franx}, {Rudnick}, {Illingworth},
  {Lee}, \& {Muzzin}}]{vandokkum09}
{van Dokkum}, P.~G., {Labb{\'e}}, I., {Marchesini}, D., {et~al.} 2009, \pasp,
  121, 2

\bibitem[{{van Dokkum} {et~al.}(2010){van Dokkum}, {Whitaker}, {Brammer},
  {Franx}, {Kriek}, {Labb{\'e}}, {Marchesini}, {Quadri}, {Bezanson},
  {Illingworth}, {Muzzin}, {Rudnick}, {Tal}, \& {Wake}}]{vandokkum10}
{van Dokkum}, P.~G., {Whitaker}, K.~E., {Brammer}, G., {et~al.} 2010, \apj,
  709, 1018

\bibitem[{{Vergani} {et~al.}(2008){Vergani}, {Scodeggio}, {Pozzetti}, {Iovino},
  {Franzetti}, {Garilli}, {Zamorani}, {Maccagni}, {Lamareille}, {Le F{\`e}vre},
  {Charlot}, {Contini}, {Guzzo}, {Bottini}, {Le Brun}, {Picat}, {Scaramella},
  {Tresse}, {Vettolani}, {Zanichelli}, {Adami}, {Arnouts}, {Bardelli},
  {Bolzonella}, {Cappi}, {Ciliegi}, {Foucaud}, {Gavignaud}, {Ilbert},
  {McCracken}, {Marano}, {Marinoni}, {Mazure}, {Meneux}, {Merighi}, {Paltani},
  {Pell{\`o}}, {Pollo}, {Radovich}, {Zucca}, {Bondi}, {Bongiorno},
  {Brinchmann}, {Cucciati}, {de la Torre}, {Gregorini}, {Perez-Montero},
  {Mellier}, {Merluzzi}, \& {Temporin}}]{vergani08}
{Vergani}, D., {Scodeggio}, M., {Pozzetti}, L., {et~al.} 2008, \aap, 487, 89

\bibitem[{{Weinzirl} {et~al.}(2011){Weinzirl}, {Jogee}, {Conselice},
  {Papovich}, {Chary}, {Bluck}, {Gr{\"u}tzbauch}, {Buitrago}, {Mobasher},
  {Lucas}, {Dickinson}, \& {Bauer}}]{weinzirl11}
{Weinzirl}, T., {Jogee}, S., {Conselice}, C.~J., {et~al.} 2011, \apj, 743, 87

\bibitem[{{Wetzel} \& {White}(2010)}]{wetzel10}
{Wetzel}, A.~R. \& {White}, M. 2010, \mnras, 403, 1072

\bibitem[{{White} {et~al.}(2005){White}, {Clowe}, {Simard}, {Rudnick}, {De
  Lucia}, {Arag{\'o}n-Salamanca}, {Bender}, {Best}, {Bremer}, {Charlot},
  {Dalcanton}, {Dantel}, {Desai}, {Fort}, {Halliday}, {Jablonka}, {Kauffmann},
  {Mellier}, {Milvang-Jensen}, {Pell{\'o}}, {Poggianti}, {Poirier},
  {Rottgering}, {Saglia}, {Schneider}, \& {Zaritsky}}]{white05}
{White}, S.~D.~M., {Clowe}, D.~I., {Simard}, L., {et~al.} 2005, \aap, 444, 365

\bibitem[{{Williams} {et~al.}(2011){Williams}, {Quadri}, \&
  {Franx}}]{williams11}
{Williams}, R.~J., {Quadri}, R.~F., \& {Franx}, M. 2011, \apjl, 738, L25

\bibitem[{{Williams} {et~al.}(2010){Williams}, {Quadri}, {Franx}, {van Dokkum},
  {Toft}, {Kriek}, \& {Labb{\'e}}}]{williams10}
{Williams}, R.~J., {Quadri}, R.~F., {Franx}, M., {et~al.} 2010, \apj, 713, 738

\bibitem[{{Xu} {et~al.}(2012){Xu}, {Zhao}, {Scoville}, {Capak}, {Drory}, \&
  {Gao}}]{xu12}
{Xu}, C.~K., {Zhao}, Y., {Scoville}, N., {et~al.} 2012, \apj, 747, 85

\end{thebibliography}
\bibliographystyle{aa}

\end{document}